\documentclass[useAMS,usenatbib]{mn2e}
\usepackage{color}
\usepackage{amsmath}
\usepackage{txfonts}
\usepackage{graphicx}
\usepackage{tablefootnote}

\title[IPTA Stochastic Analysis]{From Spin Noise to Systematics: Stochastic Processes in the First International Pulsar Timing Array Data Release}
\author[L. Lentati et al.]{\parbox{\textwidth}{L. Lentati$^{1}$\thanks{E-mail: ltl21@cam.ac.uk}, R. M. Shannon$^{2,3}$, W. A. Coles$^{4}$, J.~P.~W. Verbiest$^{5,6}$, R. van Haasteren$^{7}$,  J. A. Ellis$^{7}$, R.~N. Caballero$^{6}$, R. N. Manchester$^{2}$,
Z. Arzoumanian$^{8}$,
S. Babak$^{9}$,
C. G. Bassa$^{10}$,
N.~D.~R. Bhat$^{3}$,
P. Brem$^{9}$,
M. Burgay$^{11}$,
S. Burke-Spolaor$^{12}$,
D. Champion$^{6}$,
S. Chatterjee$^{13}$,
I. Cognard$^{14, 15}$,
J. M. Cordes$^{13}$,
S. Dai$^{16, 2}$,
P. Demorest$^{12}$,
G. Desvignes$^{6}$,
T. Dolch$^{13, 17}$,
R. D. Ferdman$^{18}$,
E. Fonseca$^{19}$,
J. R. Gair$^{20}$,
M. E. Gonzalez$^{21}$,
E. Graikou$^{6}$,
L. Guillemot$^{14,15}$,
J. W. T. Hessels$^{12,22}$,
G. Hobbs$^{2}$,
G.~H. Janssen$^{10}$,
G. Jones$^{24}$,
R. Karuppusamy$^{6}$,
M. Keith$^{25}$,
M. Kerr$^{2}$,
M. Kramer$^{6}$,
M. T. Lam$^{13}$,
P. D. Lasky$^{26}$,
A. Lassus$^{6}$,
P. Lazarus$^{6}$,
T. J. W. Lazio$^{7}$,
K. J. Lee$^{27}$,
L. Levin$^{28, 25}$,
K. Liu$^{6}$,
R. S. Lynch$^{29}$,
D. R. Madison$^{30}$,
J. McKee$^{25}$,
M. McLaughlin$^{28}$,
S. T. McWilliams$^{28}$,
C. M. F. Mingarelli$^{31,6}$,
D. J. Nice$^{32}$,
S. Os{\l}owski$^{5,6}$,
T. T. Pennucci$^{33}$,
B. B. P. Perera$^{25}$,
D. Perrodin$^{11}$,
A. Petiteau$^{34}$,
A. Possenti$^{11}$,
S. M. Ransom$^{30}$,
D. Reardon$^{2, 26}$,
P. A. Rosado$^{35}$,
S. A. Sanidas$^{22}$,
A. Sesana$^{36}$,
G. Shaifullah$^{6,5}$,
X. Siemens$^{37}$,
R. Smits$^{10}$,
I. Stairs$^{19}$,
B. Stappers$^{25}$,
D. R. Stinebring$^{38}$,
K. Stovall$^{39}$,
J. Swiggum$^{28,37}$,
S. R. Taylor$^{7}$,
G. Theureau$^{14,15,40}$,
C. Tiburzi$^{6,5}$,
L. Toomey$^{2}$,
M. Vallisneri$^{7}$,
W. van Straten$^{35}$,
A. Vecchio$^{36}$,
J.-B. Wang$^{41}$,
Y. Wang$^{42}$,
X. P. You$^{43}$,
W. W. Zhu$^{6}$,
X.-J. Zhu$^{44}$}}\vspace{0.4cm}

\begin{document}

\maketitle

\label{firstpage}

\begin{abstract}
We analyse the stochastic properties of the 49 pulsars that comprise the first International Pulsar Timing Array (IPTA) data release. We use Bayesian methodology, performing model selection to determine the optimal description of the stochastic signals present in each pulsar. In addition to spin-noise and dispersion-measure (DM) variations, these models can include timing noise unique to a single observing system, or frequency band. We show the improved radio-frequency coverage and presence of overlapping data from different observing systems in the IPTA data set enables us to separate both system and band-dependent effects with much greater efficacy than in the individual PTA data sets. For example, we show that PSR J1643$-$1224 has, in addition to DM variations, significant band-dependent noise that is coherent between PTAs which we interpret as coming from time-variable scattering or refraction in the ionised interstellar medium. Failing to model these different contributions appropriately can dramatically alter the astrophysical interpretation of the stochastic signals observed in the residuals. In some cases, the spectral exponent of the spin noise signal can vary from 1.6 to 4 depending upon the model, which has direct implications for the long-term sensitivity of the pulsar to a stochastic gravitational-wave (GW) background. By using a more appropriate model, however, we can greatly improve a pulsar's sensitivity to GWs. For example, including system and band-dependent signals in the PSR J0437$-$4715 data set improves the upper limit on a fiducial GW background by $\sim 60\%$ compared to a model that includes DM variations and spin-noise only.
\end{abstract}

\section{Introduction}

\begin{figure*}
\begin{center}$
\begin{array}{cc}
\hspace{-0.9cm}
\includegraphics[width=95mm]{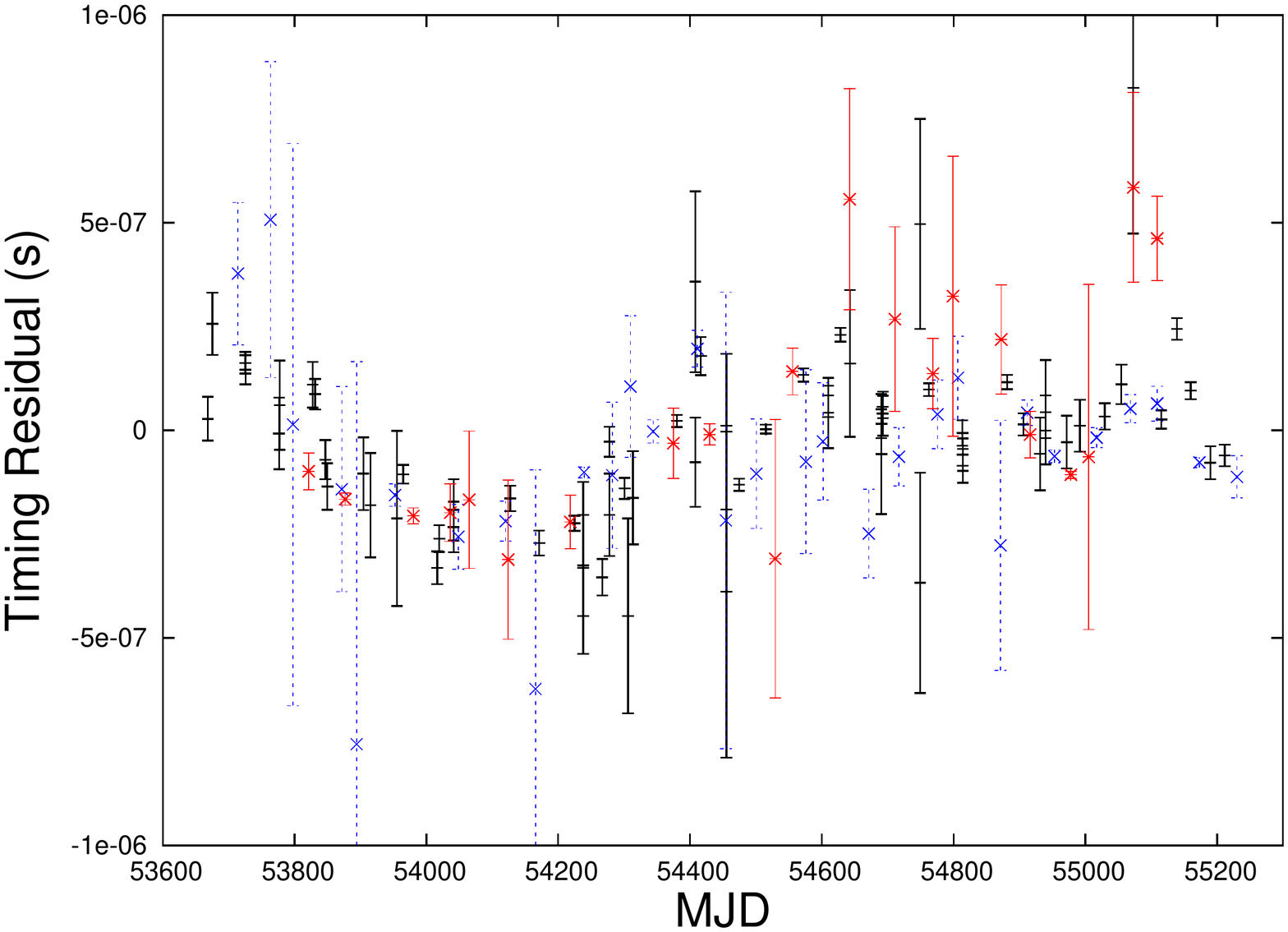} &
\hspace{-1.3cm}
\includegraphics[width=95mm]{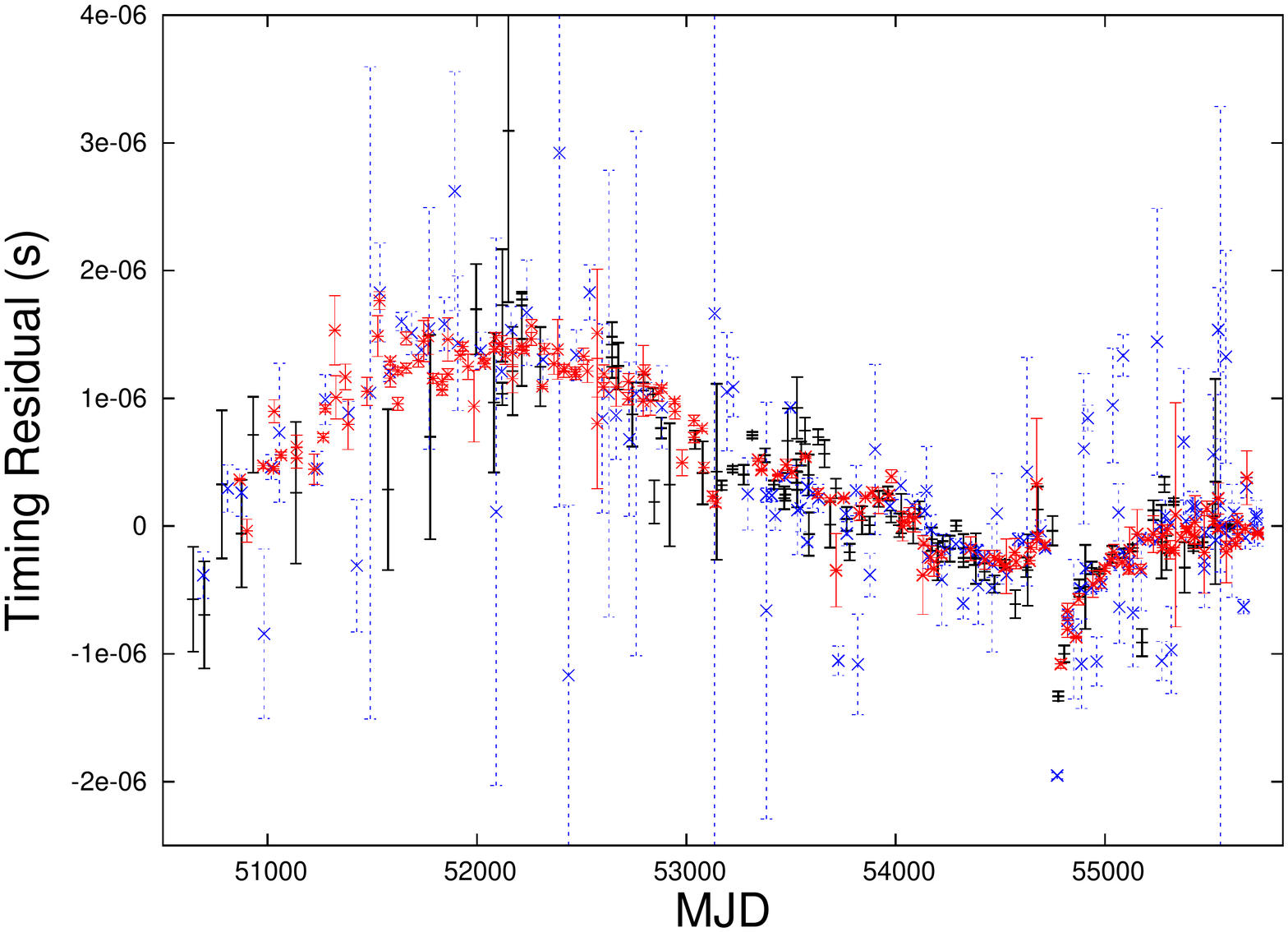}
\end{array}$
\end{center}
\caption{Timing residuals for subsections of the PSR J1909$-$3744 (left) and PSR J1713+0747 (right) IPTA data sets for observations between 1200 and 1700 MHz, where all three PTAs have overlapping data.  Timing solutions are obtained from the model in bold in Table~\ref{Table:evidenceValues}.  For clarity, a weighted average of the residuals has been performed across 50 day epochs separately for each observing system. The different PTAs are indicated by colour; PPTA (black), EPTA (blue), and NANOGrav (red).  Error bars in all cases are given by $(\sum{\sigma_i^{-2}})^{-1/2}$, with $\sigma_i$ the formal uncertainty for ToA $i$ provided with the data set, without any modification from white-noise parameters, and where the sum is performed over all ToAs that fall within the averaging window.  Finally, no power-law noise processes have been subtracted from the data. While all three PTAs can be seen to track the broad features of the data sets, statistically significant outliers are present.  In this paper we attempt to determine the optimal models for such data sets, and to determine the relative contributions of different noise processes.  These processes can include intrinsic spin noise, extrinsic DM variations including non stationary `events' such as the discontinuity seen in the PSR J1713+0747 data set at MJD~54757, and additionally, terms due to excess system- or band-dependent noise.}
\label{Fig:MoneyPlot}
\end{figure*}

The recent direct detection of gravitational-waves (GWs)  from the merger of a pair of black holes \citep{PhysRevLett.116.061102} marks a turning point in experimental physics. The entire GW spectrum is expected to probe cosmological and astrophysical phenomena ranging from quantum fluctuations in the very early universe at frequencies of $10^{-16}$~Hz, to merging binary neutron stars and stellar black holes at frequencies of $10^2$~Hz \citep{2002gr.qc.....4090C}.    Multiple experiments worldwide are or will be designed to observe different regions of this spectrum.  These include Cosmic Microwave Background polarization tests \citep{2015PhRvL.114j1301B}, space-based interferometers such as the evolving Laser Interferometer Space Antenna (eLISA,\citealt{2013arXiv1305.5720C}), and a network of ground-based interferometers consisting of LIGO \citep{2010CQGra..27h4006H}, Virgo \citep{2015CQGra..32b4001A} and KAGRA \citep{2012CQGra..29l4007S}. Additionally, the high precision with which the time of arrival (ToA) of electromagnetic pulses from millisecond pulsars (MSPs) can be measured provides a window into the nano-hertz GW Universe, for which the principal source is expected to be merging supermassive black hole binaries with masses of $10^8-10^{10}$~M$_{\odot}$.

In particular, by using a set of MSPs, referred to as a pulsar timing array \citep[PTA,][]{1990ApJ...361..300F}, the cross correlation of the GW signal between pulsars in the array \citep{1983ApJ...265L..39H} makes it possible to discriminate between GWs and other potential sources of noise in the data (e.g.,  \citealt{2015MNRAS.453.2576L}).

It was with this goal in mind that the International Pulsar Timing Array (IPTA, \citealt{2016arXiv160203640V}; \citealt{2013CQGra..30v4010M, 2010CQGra..27h4013H}) was formed as a collaboration between the three main existing PTAs:

\begin{itemize}
\item the European Pulsar Timing Array (EPTA, \citealt{2013CQGra..30v4009K}),
\item the North American Nanohertz Observatory for Gravitational Waves (NANOGrav, \citealt{2013ApJ...762...94D}) and,
\item the Parkes Pulsar Timing Array (PPTA, \citealt{2013PASA...30...17M}) in Australia.
\end{itemize}
Recently, the first IPTA data release was completed \citep{2016arXiv160203640V}, combining observations from the three PTAs for a total of 49 pulsars. In this paper we investigate the properties of the stochastic signals present in those pulsars.

A detailed review of pulsar timing can be found in, for example,  \cite{2004hpa..book.....L}.  Briefly, the ToAs for a given pulsar are recorded by an observatory as a series of discrete observations made over a period of many years.  Before any analysis can be performed, these arrival times are corrected for the motion of the Earth by transforming them  into a common frame of reference, that of the Solar System Barycentre.

At this stage a deterministic `timing model' for the pulsar is fitted to the ToAs which characterises the pulsar's astrometric and timing properties, such as its position, and rotational frequency.  This can be performed using the \textsc{Tempo}\footnote{http://tempo.sourceforge.net/}, and \textsc{Tempo2} \citep{2006MNRAS.369..655H, 2006MNRAS.372.1549E} pulsar-timing packages, or more recently, using the Bayesian pulsar timing package \textsc{TempoNest}\footnote{https://github.com/LindleyLentati/TempoNest} (\citealt{2014MNRAS.437.3004L}, see e.g., Desvignes et al. (submitted) for the use of  \textsc{TempoNest} for pulsar timing, and e.g,  \citealt{2015arXiv151009194C, 2015Sci...349.1522S}  for its use in noise characterisation).  After subtracting the timing model from the ToAs we are left with the `timing residuals', which contain any effects not accounted for by the timing model.

In many pulsars, these residuals show time-correlated structure that deviates significantly from what could be expected from instrumental noise alone.  One possible origin for this structure is intrinsic `spin-noise' that arises from rotational irregularities of the neutron star itself.  This is generally regarded as an achromatic, stochastic noise process with a red power spectrum, and while large statistical studies have been performed in the past (e.g., \citealt{2010MNRAS.402.1027H}), much remains to be understood about spin noise.    It is precisely this uncertainty that makes studies into the properties of spin noise so valuable, as most models for a stochastic GW background predict that this too will induce a red spectral signal in the timing residuals \citep{1995ApJ...446..543R, 2003ApJ...583..616J, 2003ApJ...590..691W, 2004ApJ...611..623S}.

While `normal' pulsars (with spin periods of $\sim 1~\mathrm{s}$) have been systematically observed to be affected by steep-spectrum spin noise, some of the most stable MSPs show no sign of any intrinsic spin-noise processes at the level of 100~ns, even after a decade of observation (e.g., \citealt{2015Sci...349.1522S, 2009MNRAS.400..951V}).  The low level of the intrinsic noise, coupled with the relative scarcity of statistically robust analysis on large samples of MSPs, makes characterising and predicting the properties of spin noise in MSPs difficult. This is of particular importance given that the strength and properties of the intrinsic spin-noise will certainly affect the timeline for detection of GWs using a PTA (e.g., \citealt{2010ApJ...725.1607S}).

In addition to this intrinsic spin-noise process, variations in the pulse ToAs result from  processes extrinsic to the pulsar, e.g.,  the passage of the pulse through the Earth's atmosphere, or even from asteroid belts surrounding the pulsar \citep{2013ApJ...766....5S}.  The dominant source of extrinsic noise for MSPs in the radio band, however, is typically due to variations in the dispersion measure (DM)  introduced as our line of sight to the pulsar through the ionised interstellar medium (IISM) changes with time (e.g, \citealt{2013MNRAS.429.2161K}).

Early after pulsars were first discovered, temporal variations in the DM were observed \citep{1977ApJ...214..214I}.  As our line of sight through the IISM changes  so does the observed column density of electrons along that line of sight.  Observations over long time spans have shown that these variations are largely consistent with those expected for an IISM characterized by a Kolmogorov turbulence spectrum.  This results in a noise process in the residuals with spectral exponent $\gamma = 8/3$ such that the variations are larger on longer time scales, see e.g., \citealt{1995ApJ...443..209A}.  Deviations from this simple model however have also been observed, with recent analyses suggesting there can be discrete changes in the DM variation on short time scales \citep{2015arXiv150607948C, 2013MNRAS.429.2161K}.

Simultaneous analysis of DM variations with the spin-noise is essential for robust estimation of the characteristics of the intrinsic noise processes.  However, in the IPTA data set, early data are often available at only a single frequency.  If those data were considered in isolation, we would be unable to distinguish between the two sources of noise.  Different approaches to DM correction have been applied by different PTAs in the past.  For example, in \cite{2013ApJ...762...94D} a set of independent parameters are applied that represent the amplitude of the DM variations at each measurement epoch, while in \cite{2013MNRAS.429.2161K} a linear interpolation is performed using some sampling interval.  Neither method is able to perform a statistically robust extrapolation into epochs where there is no multi-frequency coverage, and the former in addition requires near-simultaneous measurement of ToAs at different frequencies across the entire data set, which is not the case in the IPTA data set.  We therefore apply the DM correction method presented in \cite{2014MNRAS.437.3004L},  which makes the assumption that the dominant part of the DM signal is described by a time-stationary process. Thus, in our Bayesian analysis, constraints placed on the signal from epochs where there is multi-frequency coverage are automatically and robustly applied to epochs where there is only single-frequency data, without any need for bootstrapping of models (see e.g., \cite{2014MNRAS.441.2831L} for descriptions of extrapolating and interpolating time domain stochastic signals).  Given the observation of discrete changes in the DM variations, however, we also compare models that include non-stationary DM `events'. We describe our model for these in Section \ref{Section:AdditionalDM} and show the impact that ignoring these events can have on parameter estimation in Section \ref{Section:DMEvents}.

Finally, in addition to DM variations, we also show in Sections \ref{Section:SystemNoise} and \ref{Section:BandNoise} the importance of including `system', and `band' noise.  Henceforth, as defined in \cite{2016arXiv160203640V}, a system refers to a unique combination of observing telescope, recording system and receiver (or centre frequency).  System noise represents possible calibration errors or instrumental effects that might exist in a single observing system or telescope.  Band noise then models signals that exists in a particular frequency band.  This might have its origins in the IISM as a result of processes that are not coherent between different bands, or that do not scale in amplitude with the inverse square of the observing frequency.  Alternatively, it might result from sources of radio-frequency-interference (RFI) that are present in the same frequency band independent of the observing site (e.g., due to satellites, or digital broadcasts).

In Fig. \ref{Fig:MoneyPlot} we show examples of timing residuals for subsections of the IPTA data set for two pulsars:  PSR J1909$-$3744 (left) and PSR J1713+0747 (right).  In particular, we show the residuals for observations between 1200 and 1700~MHz, where all three PTAs have overlapping data.  The goal of this paper is to determine an optimal description of such data, exploiting not just the  significant overlap in time between PTAs that have observed with different telescopes, and calibrated using different techniques, but also the broad frequency coverage present in the IPTA data set.  We will show explicitly that not only does the IPTA data set enable us to separate out these extra effects with much greater efficacy, but that failing to do so can dramatically alter the interpretation of the signals observed in the residuals, in the most extreme cases revealing an apparent detection of spin noise to be a purely systemic effect.

In Section \ref{Section:Models}, we provide a description of the models we employ for the different components of the deterministic and stochastic signals in each pulsar in the IPTA data set, and in Section \ref{Section:AnalysisMethods} we give a brief overview of Bayesian analysis techniques.  In Section ~\ref{Section:Dataset} we give a brief description of the IPTA data set.  Sections \ref{Section:Results} to \ref{Section:SpinNoise} contain the results of our analysis, and finally in Section \ref{Section:Conclusions} we offer some concluding remarks.

\section{Deterministic and Stochastic Models}
\label{Section:Models}

 The key goal of this paper is to estimate the properties of stochastic signals that affect the pulse ToAs for each pulsar in the IPTA data set.  The key difficulty in this process is that each ToA will be affected by contributions from all intrinsic and extrinsic astrophysical processes -- both deterministic and stochastic --  in addition to potential system noise that might affect only those ToAs that come from a particular telescope or observing system.  As a consequence, all contributions to the total signal must be \emph{simultaneously} estimated in the analysis in order to be able to draw meaningful conclusions from the results.

For each pulsar our measured data will consist of a set of $N_d$ observed pulse ToAs.  Adopting the same notation for our signal model as in \citealt{2015MNRAS.453.2576L}, we write the vector that contains the ToAs $\mathbf{d}$, as the sum of a number of components:

\begin{equation}
\label{Eq:TotalSignal}
\mathbf{d} = \mathbf{\tau}^\mathbf{TM} + \mathrm{\tau}^\mathbf{WN} + \mathrm{\tau}^\mathbf{SN} + \mathrm{\tau}^\mathbf{DM}  + \mathrm{\tau}^\mathbf{Sys} + \mathrm{\tau}^\mathbf{BN}\,.
\end{equation}
In Eq. \ref{Eq:TotalSignal} we have:

\begin{itemize}
\item $\mathbf{\tau}^\mathbf{TM}$, the deterministic pulse timing model.
\item $\mathrm{\tau}^\mathbf{WN}$, the stochastic uncorrelated contribution to the noise due to both instrumental thermal noise, and white noise intrinsic to the pulsar.
\item $\mathrm{\tau}^\mathbf{SN}$, the stochastic time-correlated contribution due to achromatic red spin noise.
\item $\mathrm{\tau}^\mathbf{DM}$, the stochastic time-correlated contribution due to the dispersion of radio pulses travelling through the interstellar medium (consisting of time-stationary behaviour  and/or discrete events) and through the Solar wind.
\item $\mathrm{\tau}^\mathbf{Sys}$, the stochastic contribution due to time-correlated instrumental effects unique to a single observing system.
\item $\mathrm{\tau}^\mathbf{BN}$, the stochastic contribution due to time-correlated `Band Noise' unique to a particular observing frequency.
\end{itemize}
As in \citealt{2015MNRAS.453.2576L}, we first note that an additional term, $\tau^\mathbf{GW}$, could be added to account for the influence of GWs;  in the present work we take this term to be zero except for the specific case of J0437$-$4715 in Section \ref{Section:BandNoise0437}, in which we obtain upper limits for this term for a set of different models.  Secondly, as before we note that, with the exception of our model for DM events, all stochastic contributions are assumed to be zero-mean Gaussian processes.

Each of these terms then enters into our Bayesian analysis which we describe in Section \ref{Section:AnalysisMethods}, in which we will use the evidence to determine the optimal set of model components required to describe the data.

\subsection{The timing model}

We begin by incorporating the deterministic evolution in the pulse ToAs due to the pulsar timing model into our analysis. To do this we first construct a linear approximation to timing model which allows us to marginalise over the timing model parameters analytically. This linear approximation is obtained as in the $\textsc{Tempo2}$ timing package, which we describe in brief below.  We first define the length $m$ vector that contains the particular set of deterministic timing model parameters for a given pulsar as $\mathbf{\epsilon}$.  We then write the arrival times predicted by that set of parameter values as $\mathbf{\tau}(\mathbf{\epsilon})$.

Given these arrival times, we can define the vector of `timing residuals' that result from subtracting the theoretical ToA for each pulse from our observed ToA at the Solar System Barycentre:

\begin{equation}
\mathbf{d}_{\mathrm{TR}} = \mathbf{d} - \mathbf{\tau}(\mathbf{\epsilon}).
\end{equation}
The linear approximation is obtained using an initial estimate of the $m$ timing model parameters $\mathbf{\epsilon_{\mathrm{0}}}$.  Any variation from those initial estimates can then be described using the $m$ parameters $\mathbf{\delta\epsilon}$ such that:

\begin{equation}
\delta\epsilon_i = \epsilon_i - \epsilon_{\mathrm{0}i}.
\end{equation}
Therefore, any changes in the timing residuals that result from deviations in the linear timing model parameters $\mathbf{\delta\epsilon}$ can be written as:
\begin{equation}
\mathbf{\delta t} = \mathbf{d}_{\mathrm{TR}} -  \mathbfss{M}\mathbf{\delta\epsilon},
\end{equation}
where $\mathbfss{M}$ is the $N_{\mathrm{d}}\times m$ `design matrix' which describes how the timing residuals depend on the parameters $\mathbf{\delta\epsilon}$.

As stated previously, the use of the linear timing model allows us to marginalise analytically over the parameters $\mathbf{\delta\epsilon}$.  When performing this marginalisation the matrix $\mathbfss{M}$ is numerically unstable. Using the approach advocated in \cite{2014PhRvD..90j4012V} we take the singular value decomposition of $\mathbfss{M}$, to form the set of matrices $\mathbfss{U}\mathbfss{S}\mathbfss{V}^T$. The matrices $\mathbfss{U}$ and $\mathbfss{V}^T$ contain the left-singular and right-singular vectors of our original matrix $\mathbfss{M}$, while \mathbfss{S} is a diagonal matrix containing the singular values of $\mathbfss{M}$ themselves.  Here $ \mathbfss{U}$ is an $N_d\times N_d$  matrix, which we divide into two components:
\begin{equation}
\mathbfss{U} = \big(\mathbfss{G}_\mathbfss{c}, \mathbfss{G}\big),
\end{equation}
where $\mathbfss{G}$ is a $N_d\times(N_d-m)$ matrix, and $\mathbfss{G}_\mathbfss{c}$ is the $N_d\times m$ complement.  The matrix $\mathbfss{G}_\mathbfss{c}$ contains a set of orthonormal basis vectors that includes the same information as $\mathbf{M}$ but is numerically stable.  As such, we use $\mathbfss{G}_\mathbfss{c}$ instead of $\mathbfss{M}$  in the linear model.

\subsection{White noise}
\label{Section:White}

Next we consider the stochastic white-noise component, $\tau^{\mathrm{WN}}_i$. This model is divided into three components.  The first two components, referred to as EFAC and EQUAD, are common parameterisations of the white noise in pulsar timing analyses.  The third component, referred to as ECORR, has been applied more recently in NANOGrav data analysis (See e.g., \citealt{2014ApJ...794..141A, 2015arXiv150803024A}).  We now describe each of these components briefly below:

\begin{itemize}
  \item When the ToAs are formed through the cross-correlation of a profile template with the integrated pulse profile from that observation an estimate of the uncertainty on that ToA is also obtained.  The EFAC parameter, defined separately for each observing system, accounts for possible errors that arise in the cross-correlation process due to, e.g., profile variations.  The EFAC parameter multiplies all the ToA uncertainties for a given pulsar, associated with a particular system. \\
  \item The second model component, EQUAD,  represents an additional source of time-independent noise.  This could have its origins in some physical process, for example, as a result of stochastic shape variations in the integrated pulse profile due to averaging over a finite number of single pulses when forming each ToA \citep[see e.g.,][]{2014MNRAS.443.1463S}. If this were the case then the value of EQUAD should not be dependent on the observing system that recorded it.  However,  as the integration times for ToAs from different observing epochs can vary, and because these integration times are not available for all early observations in the IPTA data set, such an interpretation is not always possible.  As for the EFAC parameter, we therefore include an EQUAD parameter per observing system in our analysis.  In principle one might wish to add multiple terms in quadrature, such as a non-Gaussian term that could describe the impact of non-stationary RFI such as in \cite{2014MNRAS.444.3863L}, however due to the increase in dimensionality that results we do not take that approach here. \\
  \item The final white-noise component we consider, ECORR,  is only applicable to the NANOGrav data, for which many ToAs are present for a single observing epoch.  ECORR then represents a jitter-like effect that is fully correlated between all the ToAs in a given epoch, and uncorrelated between different epochs.
\end{itemize}
The first two components, EFAC and EQUAD, are typically defined by modifying the uncertainty $\sigma_{i}$, defining $\hat{\sigma}_{i}$ such that:

\begin{equation}
\hat{\sigma}_{i}^2 = \alpha_{i}^2\sigma_i^2 + \beta_{i}^2
\end{equation}
where $\alpha_i$ and $\beta_i$ represent the EFAC and EQUAD parameters applied to ToA $i$ respectively.  While in principle we could incorporate the effect of non-Gaussianity in the uncorrelated noise in our analysis, as in \cite{2014MNRAS.444.3863L}, this is not an approach we consider here.  However, even when significant non-Gaussianity was observed in PSR J0437$-$4715 the effect on the timing and stochastic results was found to be minimal, so we do not anticipate it will affect our results significantly.

Our model for ECORR is incorporated into our analysis by first defining the $N_d \times N_e$ matrix $\mathbfss{U}_\mathbfss{e}$, where as before $N_d$ is the total number of ToAs in the data set, and $N_e$ is the number of unique observing epoch/observing system combinations in our ECORR model.  For the purposes of this work we consider an epoch to be an interval of ten seconds, however given the average observing cadence for NANOGrav data is 4 to 6 weeks, the exact value is not important, as long as it is less than this value.

The matrix $\mathbfss{U}_\mathbfss{e}$ takes either a value of 1 or 0, depending on whether a particular ToA $i$ falls within a particular epoch/system combination $j$, i.e.:

\begin{equation}
  U_{e,ij}
  \begin{cases}
    1, & \text{if ToA $i$ falls in epoch/system combination $j$} \\
    0, & \text{otherwise.}
  \end{cases}
\end{equation}
Note that by construction, $U_{e,ij}$ will always be zero for all ToAs not from NANOGrav.

We then define the $N_e$ length vector of free parameters $\mathbf{a}$ that represent the time shift at each epoch, such that we can write the signal due to the ECORR model parameters as:

\begin{equation}
\tau^{\mathrm{ECORR}} = \mathbfss{U}_\mathbfss{e}\mathbf{a}.
\end{equation}

We then define the $N_e \times N_e$ matrix $\mathbf{\Psi}^{(\mathrm{ECORR})}$, which describes the variance, $\mathcal{J}$,  in the signal parameters $\mathbf{a}$, such that:

\begin{equation}
\langle a_ia_j\rangle = \Psi^{(\mathrm{ECORR})}_{ij} = \mathcal{J}_i\delta_{ij},
\end{equation}
where we include in our model one $\mathcal{J}$ per NANOGrav observing system as a free parameter to be fitted for.

\subsection{Spin noise}
\label{Section:Spin}

In order to define our model for spin noise, we will use the `time-frequency' method described in \cite{2014MNRAS.437.3004L} (henceforth L14) which we describe in brief below. Here the timing noise is decomposed into a set of Fourier basis vectors where for each pulsar the model includes the set of frequencies $n/T$, with $T$ the total observation time for the pulsar, and where $n$ runs from 1 to some maximum $n_{\mathrm{c}}$.  The model thus includes $2n_{\mathrm{c}}$ basis vectors, representing the sine and cosine at each frequency in the model.  In our analysis we take $n_{\mathrm{c}}$ to be the integer such that $T/n$ is closest to a period of 60 days, which was found to be sufficient in an analysis of the EPTA 2015 data release \citep{2015arXiv151009194C}.  In principle we note that one might wish to marginalise over the value of this cut off numerically as part of the analysis, however that is not an approach that we have taken here.  We list the value used for each pulsar in Table \ref{Table:PulsarInfo}.

As in L14 we take the lowest frequency in our spin noise model to be $1/T$.   This approximation is possible because the quadratic term present in the timing model that describes the pulse spin frequency and spin down significantly diminishes our sensitivity to longer periods.  The efficacy of this quadratic as a proxy to low-frequency spin noise does, however, begin to decrease for extremely steep spectrum spin noise ($\gamma > 6$).  We will show in the case of PSR J1939+2134 that our parameter estimates for the power law properties of the spin noise are completely consistent between models that use the quadratic in the pulsar timing model as a proxy to the low-frequency variations in the data, and a model that explicitly parameterises those low frequencies using the methods in van Haasteren \& Valisneri (2015).

In our analysis of the IPTA data set, as in L14, we consider a two-parameter power-law model in frequency, such that the power $\varphi$ at a Fourier frequency $f$ is given by:

\begin{equation}
\label{Eq:RedPowerLaw}
\varphi(f, A_{\mathrm{SN}}, \gamma_{\mathrm{SN}}) = \frac{A_{\mathrm{SN}}^2}{12\pi^2}\left(\frac{1}{1\mathrm{yr}}\right)^{-3} \frac{f^{-\gamma_{\mathrm{SN}}}}{T},
\end{equation}
where $A_{\mathrm{SN}}$ and $\gamma_{\mathrm{SN}}$ are the amplitude and spectral exponent of the power-law.  In L14 a more general analysis was also performed, where the power at each frequency in the model is a free parameter.  Because of the large increase in dimensionality, however, this is not an approach we pursue in this work.  The Fourier coefficients are then marginalised over analytically using the model power spectrum as a prior, a process that is described in detail in L14.

\subsection{Dispersion measure variations}

For a detailed description of the effects of the IISM on pulsar timing data see, e.g. \cite{1990puas.book.....L}. In brief, the plasma in the IISM, as well as in the Solar wind and the ionosphere, results in time-variable delays in the propagation of the pulse signal between the pulsar and the observatory.  This manifests in the timing residuals as an additional time-correlated signal.

Unlike spin noise, however, the magnitude of the DM variations are dependent upon the observing frequency.  Given a set of observations over a wide enough band width, we can therefore use this additional information to decouple DM variations from spin-noise.

We will consider our model for DM variations as the sum of several different components.  First we consider the time-stationary stochastic component of the signal, which we henceforth refer to as `DM noise' and is incorporated into our analysis as in L14.  We describe this method in brief below.

\subsubsection{DM Noise}
\label{SubSection:DMNoise}

To include DM noise in our model we define the vector $\mathbf{D}$, of length $N_d$ for a given pulsar, as:

\begin{equation}
\label{Eqn:DMScale}
D_i = 1/(K\nu^2_{(o,i)})
\end{equation}
for observation $i$ with observing frequency $\nu_{(o,i)}$, where the dispersion constant $K$ is given by:
\begin{equation}
\label{Eq:DMScale}
K \equiv 2.41 \times 10^{-16}~\mathrm{Hz^{-2}~cm^{-3}~pc~s^{-1}}.
\end{equation}

As in L14, we decompose our DM signal into a series of Fourier modes, and determine the set of frequencies to be included as for the red spin noise.  Now, however, our basis vectors are scaled using Eq. \ref{Eqn:DMScale} in order to incorporate the frequency dependence of the signal.

As before we consider a two-parameter power-law model, with an equivalent form to Eq. \ref{Eq:RedPowerLaw}, however without the factor $12\pi^2$ for the DM noise, i.e. we have:

\begin{equation}
\label{Eq:DMPowerLaw}
\varphi(f, A_{\mathrm{DM}}, \gamma_{\mathrm{DM}}) = A_{\mathrm{DM}}^2\left(\frac{1}{1\mathrm{yr}}\right)^{-3} \frac{f^{-\gamma_{\mathrm{DM}}}}{T}.
\end{equation}

In contrast to spin noise, where the spin-down quadratic in the timing model acts as a proxy for the low-frequency ($f < 1/T$) fluctuations in our data, the dependence of DM on observing frequency leaves us sensitive to low-frequency power in the DM noise.  This power must be accounted for in the model, either by including low-frequency Fourier modes in the model, or by including a quadratic in DM as an approximation.  In L14 the quadratic approach was used, and we follow this method in our analysis here.  This quadratic model is defined as:

\begin{equation}
\label{Eq:DMQuad}
 Q^{\mathrm{DM}}(t_i)= q_0 t_iD_i + q_1 (t_iD_i)^2
\end{equation}
with $q_{0,1}$ free parameters to be fitted for, and $t_i$ the barycentric arrival time for ToA $i$.  This can be achieved by adding the first and second time-derivatives of the DM, $DM1$ and $DM2$, into the timing model for the pulsar.  These parameters are equivalent to $q_0$ and $q_1$ in Eq. \ref{Eq:DMQuad}, and this is the approach we also take.

\subsubsection{Additional DM terms}
\label{Section:AdditionalDM}

We also consider two possible extensions to our model for the DM variations in each pulsar timing data set.  Firstly, yearly DM variations as observed by \cite{2013MNRAS.429.2161K}, are described by a two-parameter model with amplitude $A_\mathrm{yrDM}$ and phase $\phi_\mathrm{yrDM}$ given by:

\begin{equation}
\tau^\mathrm{yrDM}_i = A_\mathrm{yrDM}\sin\left(2\pi\mathrm{yr}^{-1} t_i + \phi_\mathrm{yrDM}\right)D_i.
\end{equation}
Secondly, we include a DM `event' model that accounts for sudden changes in the DM that are not well described by the time-stationary processes described thus far.

To model the DM events we use shapelet basis functions.  A complete description of shapelets can be found in \cite{2003MNRAS.338...35R}, with astronomical uses in e.g, \cite{2004AJ....127..625K,2015MNRAS.447.2159L,2003MNRAS.338...48R}.  Here we only describe what is needed for our DM-event model.

In one dimension, shapelets are described by the set of basis functions:

\begin{equation}
B_n(t, \Lambda) \equiv \left[\Lambda2^nn!\sqrt{\pi}\;\right]^{-1/2} H_n\left(\frac{t-t_0}{\Lambda}\right)\;\exp\left(-\frac{(t-t_0)^2}{2\Lambda^2}\right),
\end{equation}
where $t_0$ is the reference point of the event, $\Lambda$ is the scale factor which is a free parameter in our analysis, $n$ is a non-negative integer, and $H_n$ is the $n$th Hermite polynomial.  The 0th order shapelet is therefore simply given by a Gaussian ($H_0(x) = 1$), while higher order shapelets are described by a Gaussian multiplied by the relevant Hermite polynomial.

We can then represent a function $f(t)$ as the sum:

\begin{equation}
\label{Eq:shapefunction}
f(t, \mathbf{\zeta}, \Lambda) = \sum_{i\mathrm{=0}}^{n_{\mathrm{max}}} \zeta_iB_i(t;\Lambda),
\end{equation}
where $\zeta_i$ are shapelet amplitudes, and $n_{\mathrm{max}}$ is the number of shapelet terms included in the model.

We can modify Eq. \ref{Eq:shapefunction} to form our DM-event model by multiplying for each ToA $i$ at time $t_i$, the corresponding element from the vector in Eq. \ref{Eqn:DMScale}, resulting in:

\begin{equation}
\tau^{\mathrm{DMEvent}}(t_i) = f(t_i, \mathbf{\zeta}, \Lambda)D_i.
\end{equation}

Finally, while we do not consider it a free parameter in our analysis, we also incorporate a simple spherically-symmetric, time-stationary model for the Solar-wind density.  This assumes a quadratic decrease with Solar distance given by:

\begin{equation}
\mathrm{DM}_\odot = 4.85\times 10^{-6} n_0 \frac{\theta}{\sin{\theta}}~\mathrm{cm}^{-3}~\mathrm{pc},
\end{equation}
with $\theta$ the pulsar-Sun-observatory angle, and $n_0$ the electron density at 1~AU from the sun in units of cm$^{-3}$.  We use the default value for $n_0$ in \textsc{Tempo2}, which is 4~cm$^{-3}$.  In principle one would want to fit for $n_0$ as a part of the analysis, however in this work we assume any deviation from this value can be described using our DM model.

\subsection{System and Band Noise}
\label{Section:SystemandBandModels}

The final two noise components that we will consider in our model we refer to as `system' and `band' noise.  These are additional timing noise processes that are applied to, respectively,  a specific observing system, or to all ToAs within a particular frequency band.

In the case of system noise, in principle one might consider defining a separate stochastic noise process in the same vein as the spin noise described in Section \ref{Section:Spin} for every observing system used in a particular data set, much as we already have done for the white-noise parameters EFAC and EQUAD.  In practice this is not computationally tractable, as every time-correlated signal added increases the size of the matrix operations required as part of the analysis.  We instead define a system search parameter, which can take a value between 1 and the number of systems present in a particular data set. We then sample the three-dimensional parameter space of this system index, along with the power-law amplitude and spectral exponent of the noise process applied to the system.  We can then use the evidence (see Section \ref{Section:AnalysisMethods}) to determine how many such terms are required to model the data, with the assumption that all systems are a priori equally likely to have such excess noise.

For the band noise, the number of wide ($\sim$1~GHz) frequency bands is significantly less than the number of observing systems, and so we do not incorporate a `band search' parameter, but instead simply define three bands, ($\nu < 1$ GHz, $1 < \nu < 2$ GHz, $\nu > 2$ GHz) and include an additional power-law noise process in each.  For some pulsars we then subdivide these values into smaller bands, however this is discussed on a case-by-case basis in Section \ref{Section:BandNoise}.

We note here that, if only a single observing system exists for each frequency band, the system-noise, and band-noise terms will be completely covariant.  In this case, when we perform model selection between these two cases, we will naturally be unable to distinguish one from the other.  For many pulsars in the IPTA data set, multiple observing systems operate within the same frequency band.  This breaks the degeneracy between these two competing models, and so enables us to distinguish between band and system noise with much greater efficacy.

\section{Bayesian Methods}
\label{Section:AnalysisMethods}

In our analysis we will we make use of Bayesian methods, which provide a means of estimating a set of parameters $\Theta$ in a model or hypothesis $H$ given the data, $D$  (see e.g., \cite{2014bmc..book.....H} for a description of Bayesian inference, and its application to a range of problems in different astrophysical fields).  Central to all Bayesian analysis is Bayes' theorem, which states that:

\begin{equation}
\mathrm{Pr}(\Theta \mid D, H) = \frac{\mathrm{Pr}(D\mid \Theta, H)\mathrm{Pr}(\Theta \mid H)}{\mathrm{Pr}(D \mid H)},
\end{equation}
where $\mathrm{Pr}(\Theta \mid D, H) \equiv \mathrm{Pr}(\Theta)$ is the posterior probability distribution of the parameters,  $\mathrm{Pr}(D\mid \Theta, H) \equiv L(\Theta)$ is the likelihood, $\mathrm{Pr}(\Theta \mid H) \equiv \pi(\Theta)$ is the prior probability distribution, and $\mathrm{Pr}(D \mid H) \equiv \mathcal{Z}$ is known as the evidence.

In order to discriminate between different models, $H_0$ and $H_1$, in a Bayesian analysis we must consider the \textit{odds ratio}, $R$:

\begin{equation}
R=  \frac{\mathcal{Z}_1}{\mathcal{Z}_0}\frac{\mathrm{Pr}(H_1)}{\mathrm{Pr}(H_0)},
\label{Eq:Rval}
\end{equation}
where $\mathrm{Pr}(H_1)/\mathrm{Pr}(H_0)$ is the {\it a priori} probability ratio for the two models.  In the work that follows we will take the prior probability ratio of different models to be one, in which case $R$ reduces to the `Bayes factor'.

The Bayes factor relates to the probability of one model compared the other as:

\begin{equation}
P = \frac{R}{1+R}.
\end{equation}
In our analysis we deal with the log Bayes factor, which is just the difference in the log evidence for two competing models. While different interpretations of the Bayes factor exist (e.g., \citealt{bayesRef}) in our analysis we will require an increase in the log evidence of three to prefer one model to another, corresponding to a threshold probability of approximately 95\%.

\begin{table*}
\caption{Details of the individual pulsar data sets.  We define $\sigma_\mathrm{w}$ as the weighted RMS of the residuals, after subtracting out the maximum likelihood time-correlated signals for the optimal model determined by our analysis. N$_{\mathrm{sys}}$ is the number of observing systems associated with each PTA, and finally n$_{\mathrm{c}}$ is the number of Fourier frequencies included in the time-correlated noise processes (see Section \ref{Section:Spin} for details). Note that PSRs J1824−2452A, J1857+0943, J1939+2134 and J1955+2908 also have B1950 names, which are, respectively, PSRs B1821-24A, B1855+09, B1937+21 and B1953+29.}
\begin{tabular}{lrrrrrrr}
\hline\hline
PSR J-Name &	Timespan&	Frequency	&	$\sigma_\mathrm{w}$ &		&	N$_{\mathrm{sys}}$	&	&	n$_{\mathrm{c}}$	\\
(J2000)	&	(yrs)	&	(MHz)		&	($\mu$s)	&	EPTA	&	PPTA	&	NANOGrav	&		\\
\hline
\hline
J0030+0451	 	&	12.7	&	420 -	2628	&	1.5	&	6	&	--	&	2	&	78	\\
J0034$-$0534	 &	11.1	&	324 -	1410	&	4.4	&	5	&	--	&	--	&	68	\\
J0218+4232		 &	15.2	&	324 -	2048	&	6.7	&	13	&	--	&	--	&	93	\\
J0437$-$4715	 &	14.9	&	690 -	3117	&	0.1	&	--	&	14	&	--	&	91	\\
J0610$-$2100	 &	4.5	&	1366 -	1630	&	5.2	&	3	&	--	&	--	&	28	\\
J0613$-$0200	 &	13.7	&	324 -	3101	&	1.1	&	14	&	14	&	2	&	84	\\
J0621+1002		 &	14.3	&	324 -	2636	&	7.2	&	10	&	--	&	--	&	88	\\
J0711$-$6830	 &	17.1	&	689 -	3102	&	2.0	&	--	&	13	&	--	&	105	\\
J0751+1807		 &	15.3	&	1353 -	2695	&	3.3	&	9	&	--	&	--	&	94	\\
J0900$-$3144	 &	4.5	&	1366 -	2206	&	2.8	&	5	&	--	&	--	&	28	\\
J1012+5307		 &	14.4	&	324 -	2636	&	1.6	&	15	&	--	&	2	&	88	\\
J1022+1001		 &	15.2	&	324 -	3102	&	2.1	&	10	&	11	&	--	&	93	\\
J1024$-$0719	 &	15.9	&	689 -	3102	&	1.5	&	9	&	13	&	--	&	97	\\
J1045$-$4509	 &	17.0    &	689 -	3102	&	3.7	&	--	&	13	&	--	&	104	\\
J1455$-$3330	 &	7.4	&	760 -	1699	&	3.8	&	4	&	--	&	2	&	46	\\
J1600$-$3053	 &	9.9	&	689 -	3104	&	0.7	&	4	&	12	&	2	&	61	\\
J1603$-$7202	 &	15.3	&	689 -	3102	&	1.8	&	--	&	26	&	--	&	94	\\
J1640+2224		 &	15.0	&	420 -	2636	&	2.0	&	8	&	--	&	2	&	92	\\
J1643$-$1224	 &	17.8	&	689 -	3102	&	1.8	&	9	&	13	&	2	&	109	\\
J1713+0747	 	 &	21.2	&	689 -	3102	&	0.2	&	14	&	15	&	14	&	130	\\
J1721$-$2457	 &	10.3	&	1335 -	1412	&	25.7	&	3	&	--	&	--	&	63	\\
J1730$-$2304	 &	17.8	&	689 -	3102	&	1.6	&	7	&	13	&	--	&	109	\\
J1732$-$5049	 &	8.0	&	689 -	3101	&	2.5	&	--	&	10	&	--	&	49	\\
J1738+0333		 &	4.9	&	1366 -	1628	&	2.6	&	2	&	--	&	--	&	30	\\
J1744$-$1134	 &	17.0	&	324 -	3102	&	0.8	&	9	&	13	&	2	&	104	\\
J1751$-$2857	 &	5.7	&	1398 -	1411	&	2.4	&	1	&	--	&	--	&	35	\\
J1801$-$1417	 &	4.8	&	1396 -	1698	&	2.0	&	3	&	--	&	--	&	30	\\
J1802$-$2124	 &	4.7	&	1366 -	2048	&	2.9	&	4	&	--	&	--	&	29	\\
J1804$-$2717	 &	5.9	&	1395 -	1520	&	4.4	&	2	&	--	&	--	&	36	\\
J1824$-$2452A	 &	5.8	&	689 -	3101	&	0.6	&	--	&	10	&	--	&	36	\\
J1843$-$1113	 &	8.7	&	1335 -	1630	&	1.0	&	5	&	--	&	--	&	53	\\
J1853+1303		 &	7.0	&	1370 -	2378	&	1.1	&	2	&	--	&	2	&	43	\\
J1857+0943		 &	26.0	&	420 -	3102	&	1.3	&	10	&	11	&	2	&	159	\\
J1909$-$3744	 &	9.0	&	565 -	3106	&	0.2	&	3	&	16	&	2	&	55	\\
J1910+1256		 &	6.9	&	1366 -	2378	&	1.4	&	2	&	--	&	2	&	43	\\
J1911+1347		 &	4.9	&	1366 -	1408	&	5.2	&	1	&	--	&	--	&	30	\\
J1911$-$1114	 &	5.7	&	1398 -	1520	&	0.7	&	3	&	--	&	--	&	35	\\
J1918$-$0642	 &	10.5	&	792 -	1520	&	1.5	&	5	&	--	&	2	&	64	\\
J1939+2134		 &	27.1	&	689 -	3101	&	0.3	&	12	&	14	&	--	&	165	\\
J1955+2908		 &	5.8	&	1386 -	2378	&	5.0	&	3	&	--	&	2	&	36	\\
J2010$-$1323	 &	5.0	&	1366 -	2048	&	2.0	&	5	&	--	&	--	&	31	\\
J2019+2425	 	&	6.8	&	1366 -	1520	&	8.8	&	3	&	--	&	--	&	42	\\
J2033+1734	 	&	5.5	&	1368 -	1520	&	13.3	&	3	&	--	&	--	&	34	\\
J2124$-$3358	 &	17.6	&	689 -	3102	&	3.0	&	5	&	13	&	--	&	108	\\
J2129$-$5721	 &	15.4	&	689 -	3102	&	1.2	&	--	&	12	&	--	&	94	\\
J2145$-$0750	 &	17.5	&	324 -	3142	&	1.0	&	12	&	14	&	2	&	107	\\
J2229+2643	 	&	5.8	&	1355 -	2638	&	3.8	&	6	&	--	&	--	&	36	\\
J2317+1439	 	&	14.9	&	317 -	2638	&	1.0	&	8	&	--	&	2	&	91	\\
J2322+2057		 &	5.5	&	1395 -	1698	&	7.0	&	4	&	--	&	--	&	34	\\
\hline
\end{tabular}
\label{Table:PulsarInfo}
\end{table*}

We perform our analysis using either the \textsc{MultiNest} algorithm \citep{2008MNRAS.384..449F, 2009MNRAS.398.1601F}, or the recently introduced \textsc{PolyChord} \citep{2015arXiv150201856H}.  Both these algorithms make use of nested sampling \citep{2004AIPC..735..395S}  which allows for efficient calculation of the evidence and also produces posterior distributions for the parameters being sampled.

Which algorithm we make use of depends upon the dimensionality of the pulsar data set being analysed, as in high dimensions ($\gtrsim$ 50), the number of samples required by \textsc{MultiNest} increases exponentially (See Fig. 4 in \citealt{2015arXiv150201856H}).

In the IPTA data set some pulsars have significantly more than 50 dimensions in their model.  For example, the PSR J1713+0747 data set is comprised of observations taken from a total of 43 observing systems, of which 14 are from NANOGrav telescopes.  Simply considering the white-noise parameters alone (EFAC, EQUAD, ECORR) this results in a dimensionality of 100 before any additional terms describing spin noise or DM variations have been added to the model.  It is therefore not computationally feasible to use the \textsc{MultiNest} algorithm to compute the evidence for these larger data sets.

The \textsc{PolyChord} algorithm, however, scales with dimensionality $d$ as $d^3$ at worst, making it the ideal choice for the more complex data sets.    We have updated \textsc{TempoNest} to use the \textsc{PolyChord}  algorithm, and this was used in all analyses where the dimensionality is greater than 50. For smaller-dimensional problems we find \textsc{MultiNest} is more efficient. The evidence values and parameter estimates from both samplers are consistent. The updated version of \textsc{TempoNest} is publicly available as part of \textsc{Tempo2}.

Regardless of the sampler used, unless otherwise stated we use priors that are uniform in the log of the parameter for all noise amplitudes with the exception of the EFAC parameter, and the shapelet amplitudes that describe our DM-event model for which we use a prior that is uniform in the amplitude.   Finally we use priors that are uniform in spectral exponent for all power-law noise models and in the phase of the yearly DM model.  These priors are chosen to be uninformative in all cases, so as to result in conservative detections of the noise processes under investigation.  Finally,  when  marginalising over the timing model analytically we use a uniform prior on the amplitude of these parameters.  In principle one could include further prior information, such as constraints on system jumps obtained from additional information not present in the IPTA data set, however, as stated, incorporating less prior information will only result in our analysis being more conservative.

\section{The Data set}
\label{Section:Dataset}

The first IPTA data release includes a total of 49 pulsars, of which 14 are solitary and 35 are in binary orbits.  For full details refer to \cite{2016arXiv160203640V}. Here we will give only a brief overview and provide details relevant to the analysis that follows.

The IPTA data release combines observations previously released by the three PTAs independently in Demorest et al. (2013), Manchester et al. (2013) and Desvignes et al. (submitted). The EPTA data set contains observations from the four largest radio telescopes in Europe: the Effelsberg Radio Telescope in Germany,  the Lovell Radio Telescope at the Jodrell Bank Observatory in the UK, the Nan{\c c}ay Radio Telescope in France, and the Westerbork Synthesis Radio Telescope in The Netherlands.  The PPTA data release contains data from the 64-m Parkes radio telescope, and the NANOGrav observations make use of the 100-m Robert C. Byrd Green Bank Telescope, and the 305-m William E. Gordon Telescope of the Arecibo Observatory.   In addition, for PSRs J1857+0943 and J1939+2134, publicly available data taken with the Arecibo radio telescope from Kaspi et al. (1994) have been included, extending the timing baseline for these two pulsars backwards to 1986 and 1982, respectively.  Finally for PSR J1713+0747, archival data previously used in \cite{2015ApJ...809...41Z} is also included in the IPTA data set.

In Table \ref{Table:PulsarInfo}, we list some properties of the pulsars in the first IPTA data release relevant to the stochastic analysis performed here.  Namely, we list the timespan for each pulsar, the frequency range covered by all observations, the weighted rms scatter of the residuals, $\sigma_\mathrm{w}$, obtained after subtracting the maximum-likelihood time-correlated stochastic signals from the optimal model determined by our analysis, the number of observing systems per PTA, and finally the number of Fourier-frequencies included in our model, $n_c$ (see Section \ref{Section:Spin}), such that we sample timescales as short as 60 days.  The weighted rms is calculated using the ToA error bars, after modifying them with the EFAC and EQUAD parameters obtained from our analysis.

The large number of telescopes and observing systems in the IPTA data release presents numerous challenges when attempting to perform a robust statistical analysis on the data, many of which are discussed in \cite{2016arXiv160203640V}.  In Table \ref{Table:PulsarInfo} we draw attention to one particular aspect of this challenge, namely the number of unique observing systems (N$_{\mathrm{sys}}$)  present for the different pulsars.  These are listed separately for the EPTA, PPTA and NANOGrav contributions to the data set.  For each system we include the two white-noise parameters, EFAC and EQUAD, and in addition for each NANOGrav system we include a separate ECORR parameter.

As we will discuss in Section \ref{Section:SystemNoise}, the existence of a wealth of overlapping data from different telescopes analysed using different data reduction pipelines, allows us to separate  system noise with much greater efficacy compared to the individual data sets.  Similarly, the much greater multi-frequency coverage afforded by the IPTA data set, spanning in some instances from 300~MHz to 3~GHz, allows us to better separate DM variations from other effects, and to explore deviations from the standard $\nu_{\mathrm{obs}}^2$ paradigm for band-dependent noise, which we discuss further in Section \ref{Section:BandNoise}.

\section{Results}
\label{Section:Results}

Table \ref{Table:evidenceValues} lists the relative evidence values for different models applied to the 49 pulsars in the IPTA data set.  In each case we indicate the combination of spin noise, DM noise, system noise and band noise that is included in the model by marking the included components with a cross.

For each pulsar we indicate the model most favoured by the evidence by giving the evidence value corresponding to that model in bold. We note again here that we only perform model selection between different time-correlated signals, and so all models include as a minimum an EFAC and an EQUAD per system, and an ECORR for each NANOGrav system. The timing model parameters included in the analysis are the same as those in \cite{2016arXiv160203640V}, and the relevant white noise parameters are included for all systems.  As discussed in Section \ref{Section:AnalysisMethods}, we require a change in the log evidence of 3 to warrant the inclusion of extra parameters.   As such, in many cases the model that has the highest evidence is not considered the favoured model, as the increase in the log evidence is less than 3.  Similarly, in some cases multiple models will have their log evidence values in bold, as different models of similar `complexity' (i.e. models including the same number of components) will have log evidence values such that the difference does not exceed 3.

\begin{table*}
\caption{Relative log evidence values for different models. Crosses indicate which components have been included in the model. Values in bold reflect the `optimal' models as supported by the evidence (see Section \ref{Section:Results} for details).  As stated in Section~\ref{SubSection:DMNoise} we do not perform evidence comparisons for the quadratic in DM used to model the low-frequency fluctuations. $^1$  Indicates the data set has low-frequency DM variations with greater than 2$\sigma$ significance using the model shown in bold.}
\begin{tabular}{ccccccccccccc}
\hline\hline
Spin Noise    & -       & x        & -        & -       & x       &    x   &    -     &    -   &  x     &  x     & -      & x      \\
DM Noise & -       & -        & x        & -       & x       &    -   &    x     &    x   &  x     &  x     & x      & x      \\
System Noise  & -       & -        & -        & x       & -       &    x   &    x     &    -   &  x     &  -     & x      & x      \\
Band Noise    & -       & -        & -        & -       & -       &    -   &    -     &    x   &  -     &  x     & x      & x      \\
\hline
Pulsar Name   &         &          &          &         &         &        &          &         &        &        &        &        \\
J0030+0451    &   0.0   &\bf{6.0}  &\bf{5.6}  &   2.8   &   7.3   &  5.9   &   5.3    &  -      & -      & -      & -      & -      \\
J0034$-$0534$^1$    &\bf{0.0} &   0.3    &     1.5  &   0.3   &   -     & -      &    -     &    -    & -      & -      & -      & -      \\
J0218+4232    &	0.0     &26.0	   &\bf{124.1}&	35.4    &124.2    &48.7    & 123.7    &  -      & -      &-       & -      & -      \\
J0437$-$4715    &	-       &-	   &0.0       &	  -  & 85.9       &-       & -        &  -      & 240.0  &-       &\bf{270.1} &270.6  \\
J0610$-$2100    &\bf{0.0} &   -0.1   &     0.1  &   0.1   &   -     & -      &    -     &    -    & -      & -      & -      & -      \\
J0613$-$0200    & -       &   0.0    &     3.4  &   -     &   8.0   & -      &    -     &   -     &\bf{21.9}& -      & -      & 19.3   \\
J0621+1002    &   0.0   &\bf{117.9}&110.9     &   67.5  &   117.7 &  117.3 &   111.6  &  -      & -      & -      & -      & -      \\
J0711$-$6830    &  0.0    &1.0       &\bf{5.2}  & -0.8    &   5.0   &  1.5   &  4.3     &  1.3    & -      & -      & -      & -      \\
J0751+1807    &   0.0   &\bf{7.3}  &\bf{7.4}  &   1.8   &   7.5   &  7.1   &   7.2    &  -      & -      & -      & -      & -      \\
J0900$-$3144    &  0.0    &\bf{9.1}  &\bf{9.9}  &\bf{7.2} &  9.3    &  8.5   &  9.1     &  -      & -      & -      & -      & -      \\
J1012+5307    & 0.0     &\bf{8.3}  &4.0       & 1.1     &  9.4    & 8.5    &   -      &  -      & -      & -      & -      & -      \\
J1022+1001    & 0.0     & 10.2     &17.4      & 16.2    & 17.5    &-       &\bf{32.3} &  26.5   & -      & -      & 31.6   & -      \\
J1024$-$0719$^1$     & -       &\bf{0.0}  &-59.7     &  -148.5 &   -1.5  &  -1.7  &   -    &  -     & -       & -          & -      & -      \\
J1045$-$4509    & 0.0     &190.5     &\bf{339.2}&   94.5  &   340.6 &  -     &   339.0&  -     & -       & -          & -      & -      \\
J1455$-$3330$^1$     &  0.0    &   0.6    &     0.1  & \bf{3.8}&   -     & 2.4    &    2.1 &    -   & -       & -          & -      & -      \\
J1600$-$3053    & 0.0     &89.8	   &116.8     &	61.3    &119.9    &  -     & 143.9  & 138.0  & 142.7   &   137.5     & \bf{161.5}      & 160.5     \\
J1603$-$7202    & 0.0     &19.3      &\bf{43.6}      &   28.8  &   44.1  &  -     &   50.3 &  50.3  & -       & -      & -      & -      \\
J1640+2224$^1$     &\bf{0.0} &0.6       &2.3       &   0.4   &   2.0   &  0.6   &   2.0  &  -     & -       & -      & -      & -      \\
J1643$-$1224    & 0.0     &  -       &    -     &    -    &   0.0   &  22.8  &   21.2 &  -     & 43.9    & -      & \bf{80.4}   & 80.8   \\
J1713+0747    & -       &  -       &    0.0   &     -   &   \bf{37.0}  &  -     &   -    &  -     & 160.2   & 50.1   & -      & 160.3  \\
J1721$-$2457    &\bf{0.0} &   0.0    &     0.2  &   0.0   &   -     & -      &    -   &    -   & -       & -      & -      & -      \\
J1730$-$2304$^1$     & 0.0     &-0.9	   &-1.3      &\bf{5.1} &	 -& 4.1    &   4.3  &    -   & -       & -      & -      & -      \\
J1732$-$5049    &  0.0    &6.7       &\bf{9.9}  &3.6      &  9.4    &  6.4   &  9.6   &  9.8   & -       & -      & -      & -      \\
J1738+0333    &\bf{0.0} &   -0.5   &     -0.3 &   -0.2  &   -     & -      &    -   &    -   & -       & -      & -      & -      \\
J1744$-$1134    & 0.0     &   9.0    &     10.0 &   19.0  &   10.4  &\bf{35.6}&\bf{33.5}&    -   & -     & -      & -      & -      \\
J1751$-$2857    &\bf{0.0} &   -0.2   &     0.0  &   0.0   &   -     & -      &    -   &    -   & -       & -      & -      & -      \\
J1801$-$1417    &  0.0    &\bf{12.8} &\bf{13.1} &\bf{12.0}&  13.3   &  12.9  &  13.3  &  -     & -       & -      & -      & -      \\
J1802$-$2124    &  0.0    &\bf{61.7} &\bf{61.7} &\bf{60.1}&  62.1   &  61.6  &  61.7  &  -     & -       & -      & -      & -      \\
J1804$-$2717    &\bf{0.0} &   -0.3   &     -0.1 &   0.0   &   -     & -      &    -   &    -   & -       & -      & -      & -      \\
J1824$-$2452A   &  -      &0.0	   &5.5	      &  -76.7  &\bf{27.5}& 6.2    &  9.0   & 27.4   & 29.5    & -      & -      & -      \\
J1843$-$1113    &  0.0    &\bf{17.7} &\bf{16.7} &9.6      &  17.8   &  17.4  &  16.3  &  -     & -       & -      & -      & -      \\
J1853+1303    &\bf{0.0} &   -0.6   &     -0.4 &   0.4   &   -     & -      &    -   &    -   & -       & -      & -      & -      \\
J1857+0943    &   -     &  0.0     &     \bf{34.0} &   -     &   32.6  & -      &   32.81&    34.1     & -       & -      & -      & -      \\
J1909$-$3744    &  -      &    -     &\bf{0.0}  &   -     &  -2.0   &  -     &  -3.6  &  -2.4  & -4.5    & -      & -      & -      \\
J1910+1256    &  0.0    &\bf{8.6}  &\bf{8.2}  &4.6      &  8.0    &  6.7   &  8.8   &  -     & -       & -      & -      & -      \\
J1911$-$1114    &\bf{0.0} &   -0.3   &     -0.2 &   0.0   &   -     & -      &    -   &    -   & -       & -      & -      & -      \\
J1911+1347    &\bf{0.0} &   -0.6   &     -0.4 &   -0.5  &   -     & -      &    -   &    -   & -       & -      & -      & -      \\
J1918$-$0642$^1$     &\bf{0.0} &   -4.0   &     -0.3 &   -8.6  &   -     & -      &    -   &    -   & -       & -      & -      & -      \\
J1939+2134    &   -     &  -       & -        &   -     &   0.0   & -      &   -    &  -     & 89.7    & 11.7   & -      & \bf{105.7}  \\
J1955+2908    &\bf{0.0} &   -0.3   &     -0.1 &   0.7   &   -     & -      &    -   &    -   & -       & -      & -      & -      \\
J2010$-$1323$^1$     &\bf{0.0} &   -0.2   &     0.0  &   -0.1  &   -     & -      &    -   &    -   & -       & -      & -      & -      \\
J2019+2425    &\bf{0.0} &   -0.5   &     -0.3 &   -0.2  &   -     & -      &    -   &    -   & -       & -      & -      & -      \\
J2033+1734    &\bf{0.0} &   -0.4   &     -0.1 &   -0.1  &   -     & -      &    -   &    -   & -       & -      & -      & -      \\
J2124$-$3358$^1$     &\bf{0.0} &0.8       &-0.2      &   0.4   &   -     &  -     &   -    &  -     & -       & -      & -      & -      \\
J2129$-$5721$^1$     &\bf{0.0} &0.0       &0.0       &   -0.5  &   -     &  -     &   -    &  -     & -       & -      & -      & -      \\
J2145-0750    &    -    &0.0       &   -20.8  &   -87.0 &   3.2   &  53.5  &   29.7 &  -     &\bf{56.9}& -      & -      & 56.0   \\
J2229+2643$^1$     &\bf{0.0} &   -0.5   &     0.0  &   0.2   &   -     & -      &    -   &    -   & -       & -      & -      & -      \\
J2317+1439    &   0.0   &48.2      &\bf{125.1}&   40.0  &   124.2 &  71.5  &   124.9&  -     & -       & -      & -      & -      \\
J2322+2057    &\bf{0.0} &   -0.4   &     -0.1 &   0.0   &   -     & -      &    -   &    -   & -       & -      & -      & -      \\
\hline
\end{tabular}
\label{Table:evidenceValues}
\end{table*}

For example, the PSR J0030+0451 data set has a maximum value for the log evidence of 7.3 for a model that includes both DM noise and spin noise. This, however,  is only 1.3 greater than a model that includes spin noise alone, and only 1.7 greater than a model that includes DM noise alone.  Thus we conclude that the timing model must include either spin noise or DM noise, but likely due to a lack of quality overlapping multi-frequency data, we cannot distinguish between these two models from this data set, and so both are bold in the table.

\begin{table*}
\caption{Relative log evidence values for  additional DM model components.  Bold font in column 3 identifies the two pulsars for which significant DM events were identified.  For both pulsars, the remaining model parameters are given by the emboldened column in Table \ref{Table:evidenceValues}.}
\begin{tabular}{ccc}
\hline\hline
Pulsar Name   &    Yearly DM     &    DM Event          \\
\hline
J0437$-$4715    & 269.1 		 & 268.3  		\\
J0218+4232    & 123.7  		 & 123.5  		\\
J0613$-$0200    & 22.5		 &  18.3		\\
J1022+1001    & 30.1  		 & 30.3 	        \\
J1045$-$4509    & 339.8  		 & 339.3  		\\
J1600$-$3053    & 159.1  		 & 161.2	        \\
J1603$-$7202    &     47.9         & \bf{53.3}   	        \\
J1643$-$1224    &      80.1        &  79.1  	        \\
J1713+0747    &      158.7       &  \bf{195.1}	        \\
J1732$-$5049    & 9.8   		 & 11.7   		\\
J1824$-$2452A   &     28.7         &  26.5               \\
J1857+0943    &     33.7         &      32.9		\\
J1909$-$3744    & -1.4  		 & -2.6  	        \\
J1939+2134    &  104.9 		 & 103.6  	        \\
J2145$-$0750    & 55.8             &   54.4              \\
J2317+1439    & 124.5  		 & 125.2  		\\
\hline
\end{tabular}
\label{Table:EMevidenceValues}
\end{table*}

Table \ref{Table:EMevidenceValues} then lists the evidence values when including either additional non-stationary DM events, or yearly DM variations.  We only consider these additional DM components for those pulsars with data sets that support DM noise in their optimal model, and where the DM noise can be clearly distinguished from spin noise, and other system- or band-dependent effects.  Only two data sets support the inclusion of non-stationary DM events, those for PSRs J1713+0747 and J1603$-$7202.  For these two the emboldened model in Table  \ref{Table:evidenceValues} reflects the set of model components that optimally describes the data when also including the DM-event model, and as such might not be the highest number in that row.  In both cases however in Table \ref{Table:EMevidenceValues} the optimal model is greater than any non-event model listed in Table \ref{Table:evidenceValues}.

In total we find that:
\begin{itemize}
\item For 19 pulsars, the data support no time-correlated timing noise components.  Typically these are shorter data sets, with all but six of these pulsars having less than 7 years of observations.  Notably, however,  PSRs J1640+2224, J2124$-$3358, and J2129$-$5721 all have time spans of greater than 15 years and $\sigma_\mathrm{w}$ of less than 3$\mu$s.  These three data sets do, however, have significant low-frequency DM variations.\\

\item For a further 20 pulsars, the data support a single time-correlated noise component,  of which for seven we are unable to distinguish between spin noise and DM noise.  With the exception of PSR J0030+0451, however,  these are all data sets for which there is no data at less than $1400$~MHz, or no high-precision high-frequency  ($\ge$ 3~GHz) data,  stressing the importance of including broad frequency coverage if we wish to disentangle timing noise that is intrinsic to the pulsar (or GWs) from that induced by variable propagation effects in the IISM.\\

\item Of the remaining ten pulsar data sets, four support a combination of two noise components, five support three components and only the PSR J1939+2134 data set supports all four types of time-correlated signal in the model.\\
\end{itemize}

For clarity we note that there are ten pulsars that do not show evidence for the DM noise model (i.e., higher order DM variations, see Section~\ref{SubSection:DMNoise}) which do still have significant low-frequency DM variations which are modeled by the DM quadratic we include in the timing model (see \cite{2016arXiv160203640V}). These pulsars are indicated by a superscript~1 on pulsar name in Table~\ref{Table:evidenceValues}.

Of particular note are those pulsars whose data support system noise or band noise in their model.  For example, when considering only spin noise and DM noise for PSR J1600$-$3053, we find that a model that includes both has an increase of more than three in the log evidence compared to a model that includes only DM noise, which would lead us to conclude that this pulsar suffered from spin noise.  When including the possibility of both system and band noise, however,  we find no evidence for spin noise in the data set.

In previous analyses of the NANOGrav five-year data set, included as a subset of the IPTA data set,  \cite{2013arXiv1311.3693P} found that PSR J1643$-$1224 suffered from red timing noise.  In our analysis described in Section \ref{Section:BandNoise} we show that this data set has significant additional frequency-dependent noise that is coherent between different frequency bands observed by different PTAs, but scales more steeply than either $\nu_\mathrm{o}^0$ or $\nu_\mathrm{o}^2$ as would be expected from either spin noise or DM noise respectively.  This allows us to interpret this timing noise as likely coming from time-variable scattering or refraction in the IISM, as opposed to being due to spin noise intrinsic to the pulsar.

\begin{figure}
\begin{center}$
\begin{array}{c}
\hspace{-1.0cm}
\includegraphics[width=100mm]{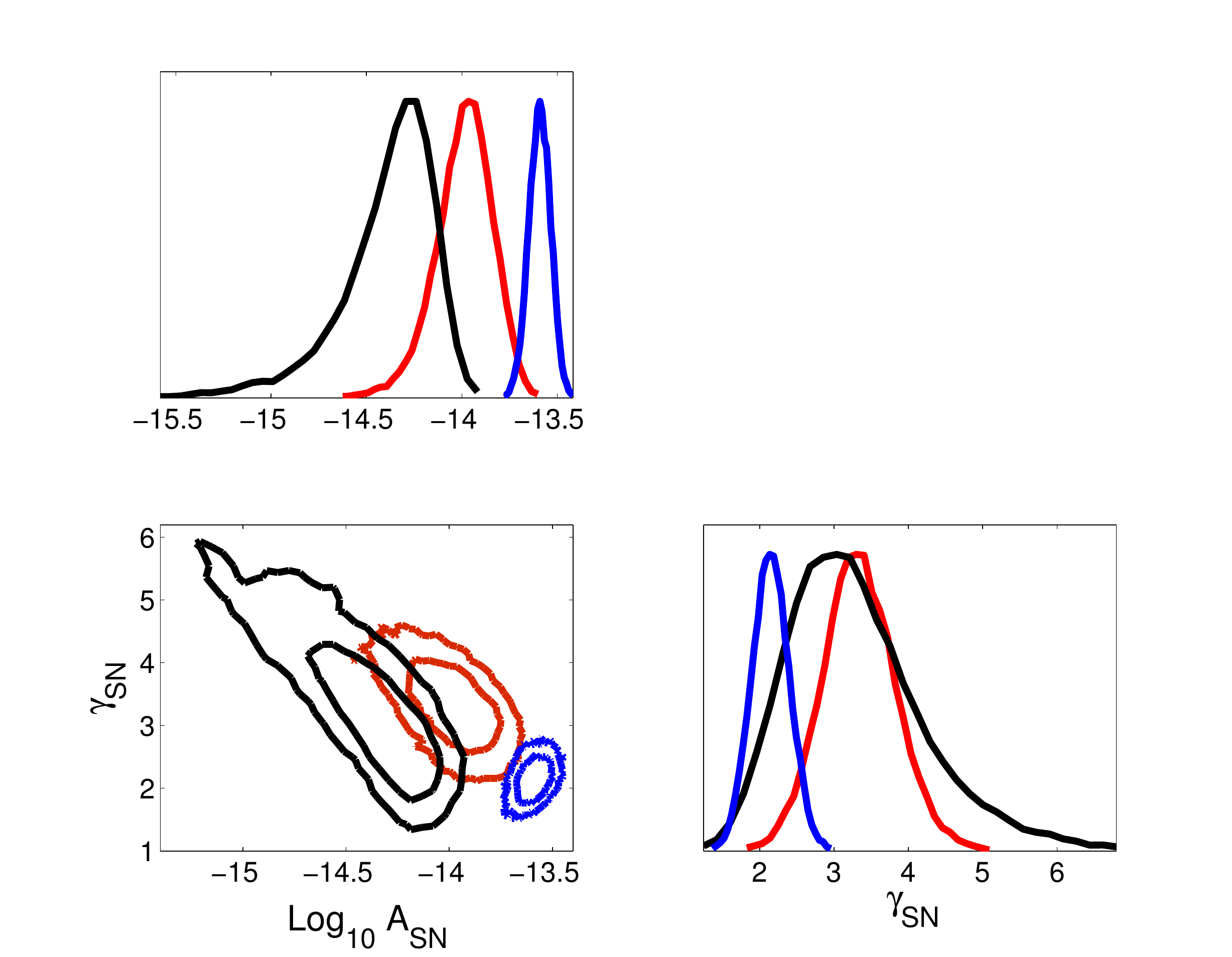}
\end{array}$
\end{center}
\caption{One- and two-dimensional marginalised posterior distributions for the spin-noise amplitude and spectral exponent in the PSR J0437$-$4715 data set when including: Only spin noise and DM noise (blue lines), additional system noise (red lines), and both additional system noise, and band noise terms in the 600-800~MHz and 1200-1600~MHz bands (black lines).  The inferred parameter estimates change significantly depending upon the chosen model, demonstrating the importance of determining the optimal model when attempting to draw astrophysical conclusions from pulsar timing data.  In this plot, and all triangular plots henceforth, the contours represent one- and two-$\sigma$ confidence intervals.  In addition, the axis labels on the bottom row apply to all plots above it, and similarly, all axis labels on the left apply to all plots to the right of it.}
\label{Fig:J0437SpinNoiseComp}
\end{figure}

\begin{table*}
\caption{System and band-noise model parameters. We denote the integrated power for each model as $\sigma_\mathrm{Sys}$, and $\sigma_\mathrm{BN}$ for the system-noise, and band-noise processes respectively  (see Section \ref{Section:SystemNoise} for details).}
\begin{tabular}{ccccc}
\hline\hline
\multicolumn{5}{c}{System-noise model parameter estimates} \\
\hline
Pulsar & System  & Log$_{10}$ A$_{\mathrm{Sys}}$ & $\gamma_{\mathrm{Sys}}$ & Log$_{10} \sigma_{\mathrm{Sys}}$ \\
\hline\hline
J0613$-$0200  &  Nan{\c c}ay 1400~MHz & -14.8   $\pm$  0.8   &  4.9  $\pm$  1.4  &  -6.08  $\pm$  0.16 \\
J1022+1001  &  Nan{\c c}ay 1400~MHz & -12.68  $\pm$  0.07  &  1.8  $\pm$  0.3  &  -5.58  $\pm$  0.14 \\
J1455$-$3330  &  Nan{\c c}ay 1400~MHz & -13.5   $\pm$  1.0   &  3.6  $\pm$  1.5  &  -5.8   $\pm$  0.8 \\
J1600$-$3053  &  Nan{\c c}ay 1400~MHz & -13.28  $\pm$  0.10  &  1.9  $\pm$  0.4  &  -6.23  $\pm$  0.14 \\
J1643$-$1224  &  Nan{\c c}ay 1400~MHz & -12.60  $\pm$  0.07  &  1.7  $\pm$  0.3  &  -5.55  $\pm$  0.17 \\
J1730$-$2304  &  Nan{\c c}ay 1400~MHz & -12.66  $\pm$  0.14  &  1.6  $\pm$  0.5  &  -5.63  $\pm$  0.14 \\
J1744$-$1134  &  Nan{\c c}ay 1400~MHz & -13.36  $\pm$  0.16  &  2.3  $\pm$  0.4  &  -5.99  $\pm$  0.13 \\
J2145$-$0750  &  Nan{\c c}ay 1400~MHz & -12.71  $\pm$  0.06  &  1.7  $\pm$  0.2  &  -5.66  $\pm$  0.11 \\
J0437$-$4715  &  Parkes CPSR2 1400~MHz	    & 	  -13.34 $\pm$  0.07	 &   0.9  $\pm$  0.3	  &   	-5.49  $\pm$  0.08 		\\
J0437$-$4715  &  Parkes CPSR2 legacy 1400~MHz	    & 	 -13.51  $\pm$  0.17	 &   2.8  $\pm$  0.7	  &   	-4.8  $\pm$  0.3	        \\
J1939+2134  &   Nan{\c c}ay DDS 1400~MHz	&	-12.95  $\pm$  0.05  &  1.8  $\pm$  0.2  &  -5.78  $\pm$  0.15	\\
J1939+2134  &	Parkes CPSR2  1400~MHz &	-13.7  $\pm$  0.3  &  1.7  $\pm$  0.7  &  -6.5  $\pm$  0.2 \\
\hline
\hline
\multicolumn{5}{c}{Band-noise model parameter estimates} \\
\hline
Pulsar & Band  & Log$_{10}$ A$_{\mathrm{BN}}$ & $\gamma_{\mathrm{BN}}$ & Log$_{10} \sigma_{\mathrm{BN}}$ \\
\hline\hline
J0437$-$4715  & 0-1000~MHz    	    &    $-$13.32  $\pm$  0.16	 &   0.6  $\pm$  0.4	  &  $-$5.46  $\pm$  0.16       \\
J0437$-$4715  & 1000-2000~MHz          &	 $-$13.83  $\pm$  0.14	 &   0.8  $\pm$  0.4	  &      $-$5.96  $\pm$  0.16		\\
J0437$-$4715  & $>$ 2000~MHz	    & 	  $-$14.4  $\pm$  0.3	 &   3.3  $\pm$  0.9	  &  	$-$5.5  $\pm$  0.3 	        \\
J1600$-$3053 &  0-730~MHz       &   $-$12.7 $\pm$ 0.15          &    2.2 $\pm$ 0.6       &      $-$5.4  $\pm$  0.3                         \\
J1643$-$1224  & 0-730~MHz    &    $-$12.02  $\pm$  0.06     &  2.2  $\pm$  0.3  	  &  $-$4.7  $\pm$  0.2  \\
J1643$-$1224  & 750-890~MHz    &    $-$12.19  $\pm$  0.07     &  1.5  $\pm$  0.3       &  $-$5.2  $\pm$  0.16 \\
J1939+2134  &  0-800~MHz	    &	$-$12.85  $\pm$  0.10      &  2.2  $\pm$  0.3  	  &  $-$5.44  $\pm$  0.19 \\
J1939+2134  & 2000-2500~MHz	    &   $-$13.6  $\pm$  0.3        &  2.1  $\pm$  0.6   	  &  $-$6.2  $\pm$  0.3 \\
\hline
\end{tabular}
\label{Table:SystemAndBandNoiseParams}
\end{table*}

As a final example, in Fig. \ref{Fig:J0437SpinNoiseComp} we show one- and two-dimensional marginalised posterior distributions for the spin-noise amplitude and spectral exponent for PSR J0437$-$4715 when including; a) only spin noise and DM noise, b) additional system noise, and c) additional system noise, and band noise terms in the 600-800~MHz and 1200-1600~MHz bands.  When performing a standard analysis including only spin noise and DM noise we find a relatively flat spectral exponent for the spin-noise model, with $\gamma_{\mathrm{SN}} \sim 1-2$.  When including system noise terms the spin-noise model drops significantly in amplitude, and becomes steeper, with $\gamma_{\mathrm{SN}} \sim 3-4$, more in line with the steep timing noise observed in young pulsars and the MSP PSR J1939+2134.  Finally, when including band noise in the model, we find that the spin noise is again consistent with smaller amplitudes and steeper spectral exponents.   That the inferred parameter estimates change significantly depending upon the chosen model clearly demonstrates the importance of performing model comparisons when attempting to draw astrophysical conclusions from pulsar timing data.

In the following sections we discuss in more detail the results for the different models used in our analysis, including system noise in Section \ref{Section:SystemNoise}, band noise in Section \ref{Section:BandNoise}, DM noise, and non-stationary DM events in Section \ref{Section:DMEvents} and finally spin noise in Section \ref{Section:SpinNoise}.

\section{System Noise}
\label{Section:SystemNoise}

We find that for ten pulsars in the IPTA data release, the data support system noise in addition to, or in favour of other time-correlated stochastic signals.  The inferred properties of these signals are given in Table \ref{Table:SystemAndBandNoiseParams} which lists the log amplitude, spectral exponent, and total integrated power with 1-$\sigma$ confidence intervals in each case.  Note that, while the one sigma confidence interval is mathematically well defined regardless of the shape of the posterior, for highly non-Gaussian probability distributions the significance of a parameter can only be determined using the full posterior, and the relative evidence  for models with and without that parameter.  We calculate the integrated power over only the Fourier frequencies included in the power law model, from $1/T$ to $n_c/T$, with $T$ the time span of the pulsar, and $n_c$ the number of frequencies included in the model. The mean and uncertainty are derived directly from the posterior distributions for the amplitude and spectral exponent of the noise processes.

\begin{figure*}
\begin{center}$
\begin{array}{cc}
\hspace{-1cm}
\includegraphics[width=100mm]{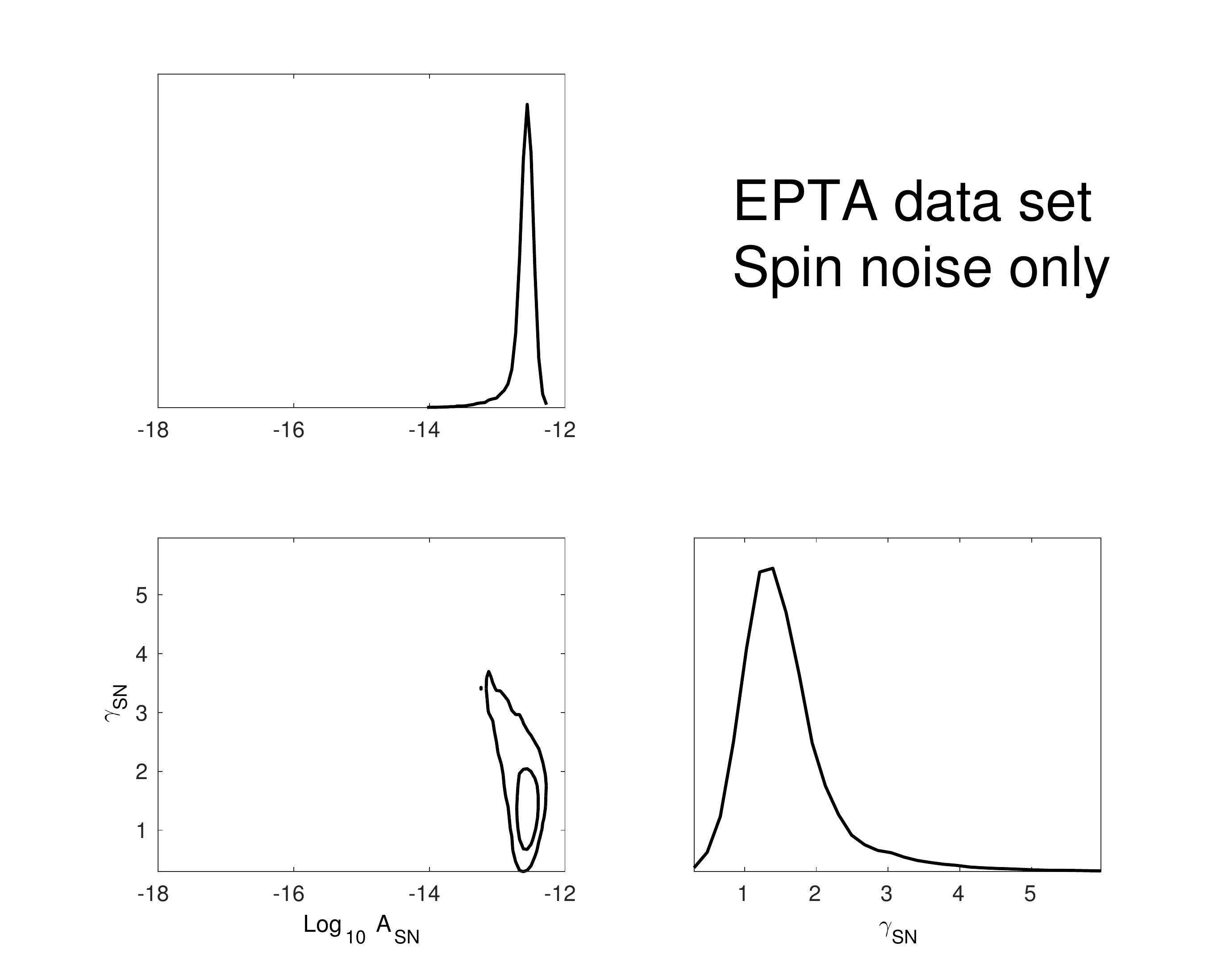} &
\hspace{-1cm}
\includegraphics[width=100mm]{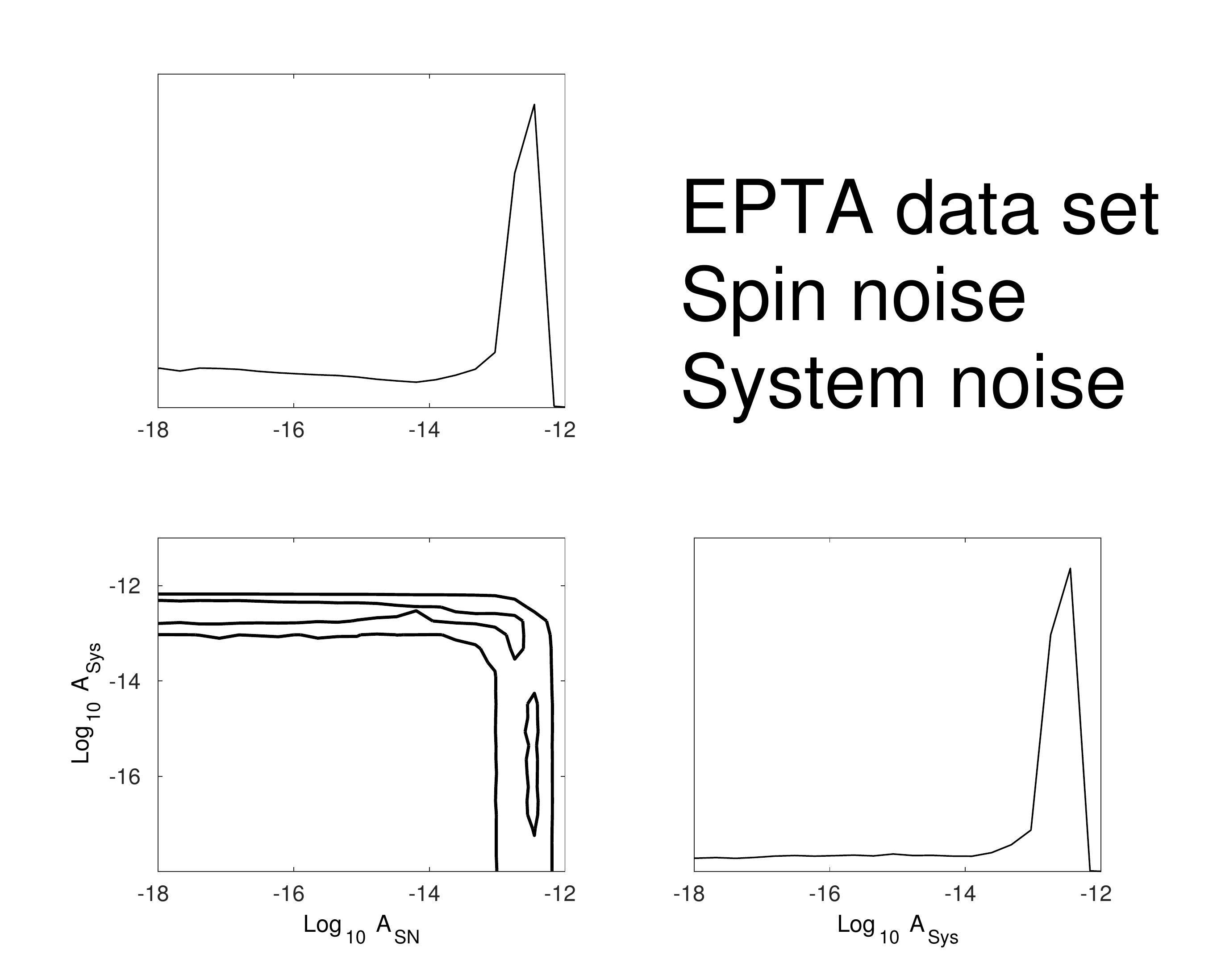} \\
\hspace{-1cm}
\includegraphics[width=100mm]{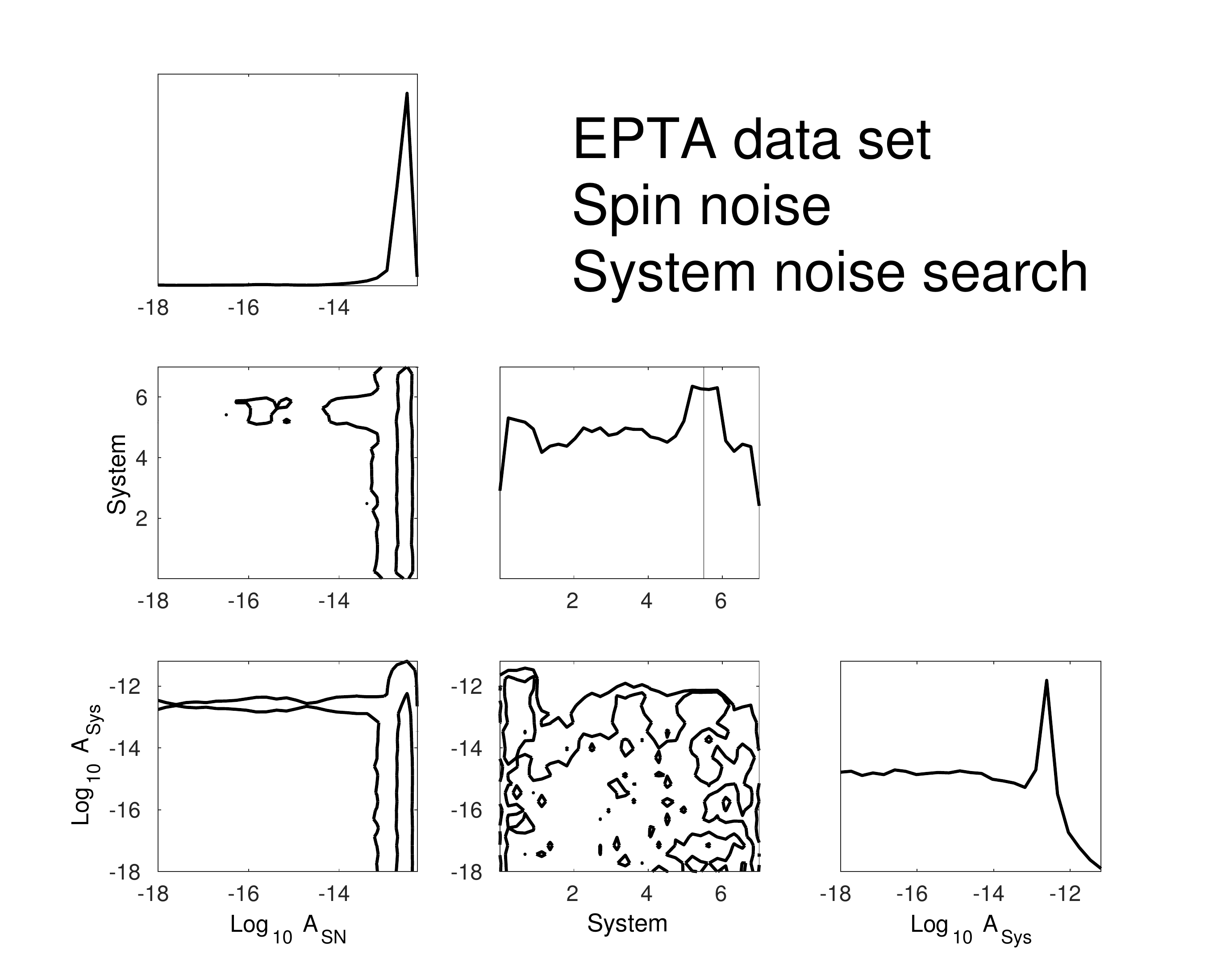} &
\hspace{-1cm}
\includegraphics[width=100mm]{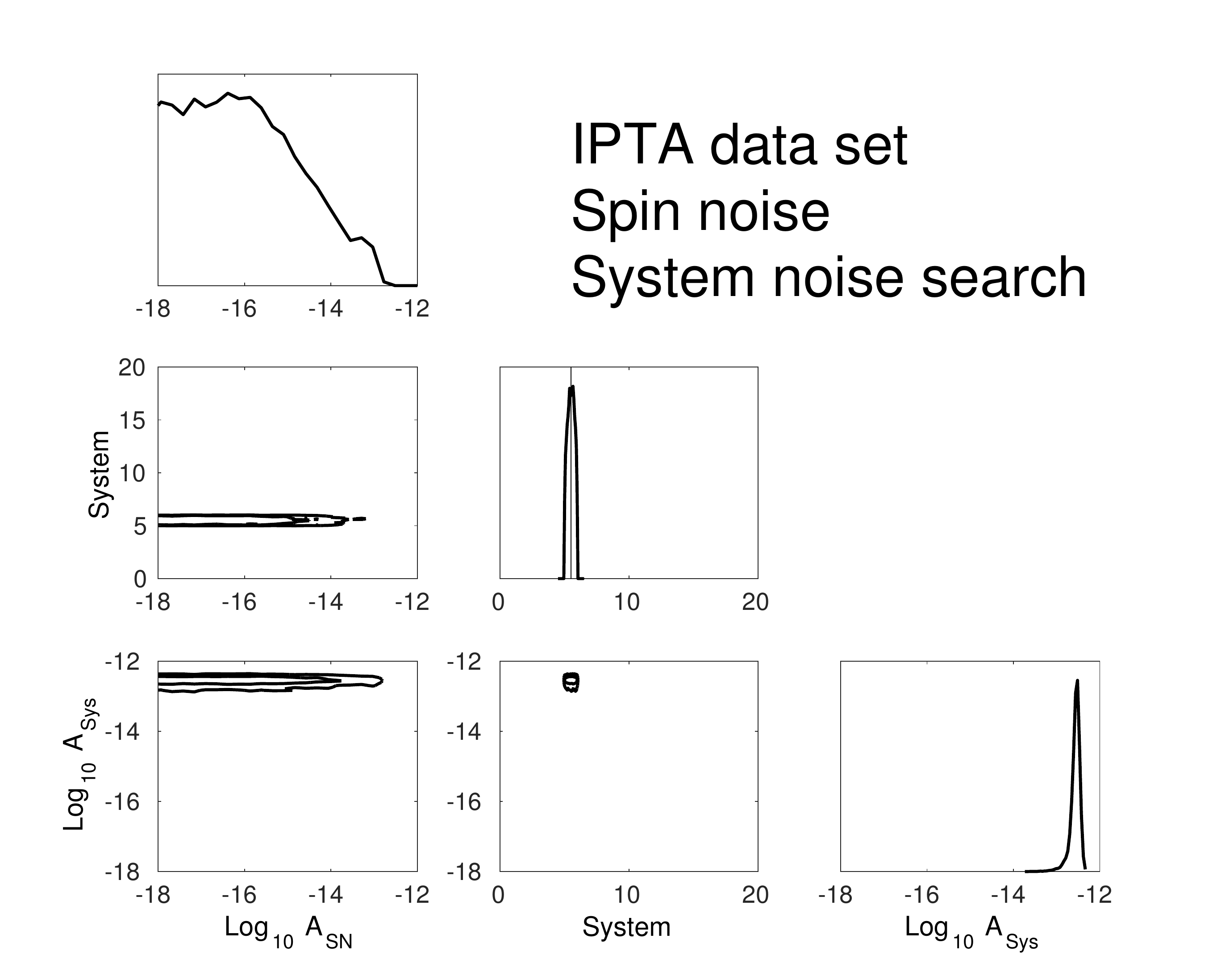} \\
\end{array}$
\end{center}
\caption{Each of the four sets of panels shows one- and two-dimensional marginalised posterior distributions for stochastic parameter estimates from different
models, either for the full PSR J1730$-$2304 IPTA data set, or for the EPTA data set alone. The top-left panel shows posteriors for the amplitude and spectral exponent
of a red spin-noise process fitted to the EPTA data set, when this is the only time-correlated component included in the model. We find a difference in the log evidence for this model compared to a white-noise-only model of 8 which, if we were
to consider only these two models, would lead us to conclude that there was strong support for the existence of spin noise in this pulsar. The top-right panel
shows the posteriors for the amplitude of the spin-noise process, and the amplitude of an additional system-noise process simultaneously applied to the
Nan{\c c}ay 1400~MHz data. The Nan{\c c}ay 1400~MHz data set contributes over half the ToAs and, based on the formal ToA uncertainties, over 90\% of the weight to
the EPTA PSR J1730$-$2304 data set. Consequently these two model components are highly covariant.  If we assume {\it a priori} that we expect additional noise in
the Nan{\c c}ay 1400~MHz data, we cannot distinguish between `true' spin noise and system noise in this data set. The bottom-left panel shows the posteriors
for a similar analysis to the top-right panel, however this time we have also fitted for the system to which we apply the additional noise process. In this case there
is a significant penalty associated with searching over the system, however we still see a peak in the posterior associated with system noise in the Nan{\c c}ay
1400~MHz data, indicated in this plot as index 5 for the system parameter, and marked with a vertical line.  However, if we assume the same prior probability for system noise in all systems, we cannot claim that there is any significant system noise in the EPTA data set. Finally, the bottom-right panel shows
the same set of parameters for the same model as the bottom-left panel, this time applied to the full IPTA data set which includes additional data
from the PPTA. In this case, even though we have doubled the number of systems that we have to search over, and thus increased the penalty for performing the search,
the Nan{\c c}ay 1400~MHz data are sufficiently inconsistent with the PPTA data that we can separate the system noise from the spin noise in this pulsar.}
\label{Fig:J1730GroupNoise}
\end{figure*}

In Fig. \ref{Fig:J1730GroupNoise} we explicitly demonstrate the advantages of working with the IPTA data set in terms of isolating system-dependent timing noise using the PSR J1730$-$2304 data set.  When performing a timing analysis of the subset of the data set provided by the EPTA, fitting for white-noise parameters and an additional spin-noise model, we find a significant detection of the spin-noise process.  The difference in the $\log$ evidence for the more complex model that includes spin noise, compared to the white noise only model is eight, with $\log_{10} A_{\mathrm{SN}} =  -12.66  \pm  0.14$ and $\gamma_{\mathrm{SN}} =   1.6  \pm  0.5$.

If we fit for a model that includes both spin noise and an additional system-dependent time-correlated signal applied to only the Nan{\c c}ay 1400~MHz data, we find that the spin-noise amplitude is highly covariant with the system noise amplitude, as shown in the top-right panel.  This clearly indicates that the EPTA data alone are not sufficient to discriminate between these two models, because the Nan{\c c}ay 1400~MHz data contributes over 90\% of the weight to the EPTA J1730$-$2304 data set.

The effect of this imbalance can also be seen if we adopt the reasonable prior that each of the seven observing systems present in the EPTA data set is equally likely to suffer from  system noise, rather than assuming {\it a priori} that there is system noise present in the Nan{\c c}ay 1400~MHz data.  In this case we search for the system to which we apply the system noise term, as described in Section \ref{Section:SystemandBandModels}, and find that there is a much greater probability that the timing noise present in the data set should be attributed to spin noise, rather than system noise.  This is shown in the bottom-left panel of Fig. \ref{Fig:J1730GroupNoise}.  In comparison to the top-left panel, however, we find there is now some probability that the spin-noise amplitude is consistent with zero, as we would expect. This tail on the amplitude of the spin noise corresponds to a peak in the posterior probability distribution associated with system five, marked with a vertical line in the one dimensional figure, which is associated with the  Nan{\c c}ay 1400~MHz data.

\begin{figure*}
\begin{center}$
\begin{array}{c}
\hspace{-2cm}
\includegraphics[width=200mm]{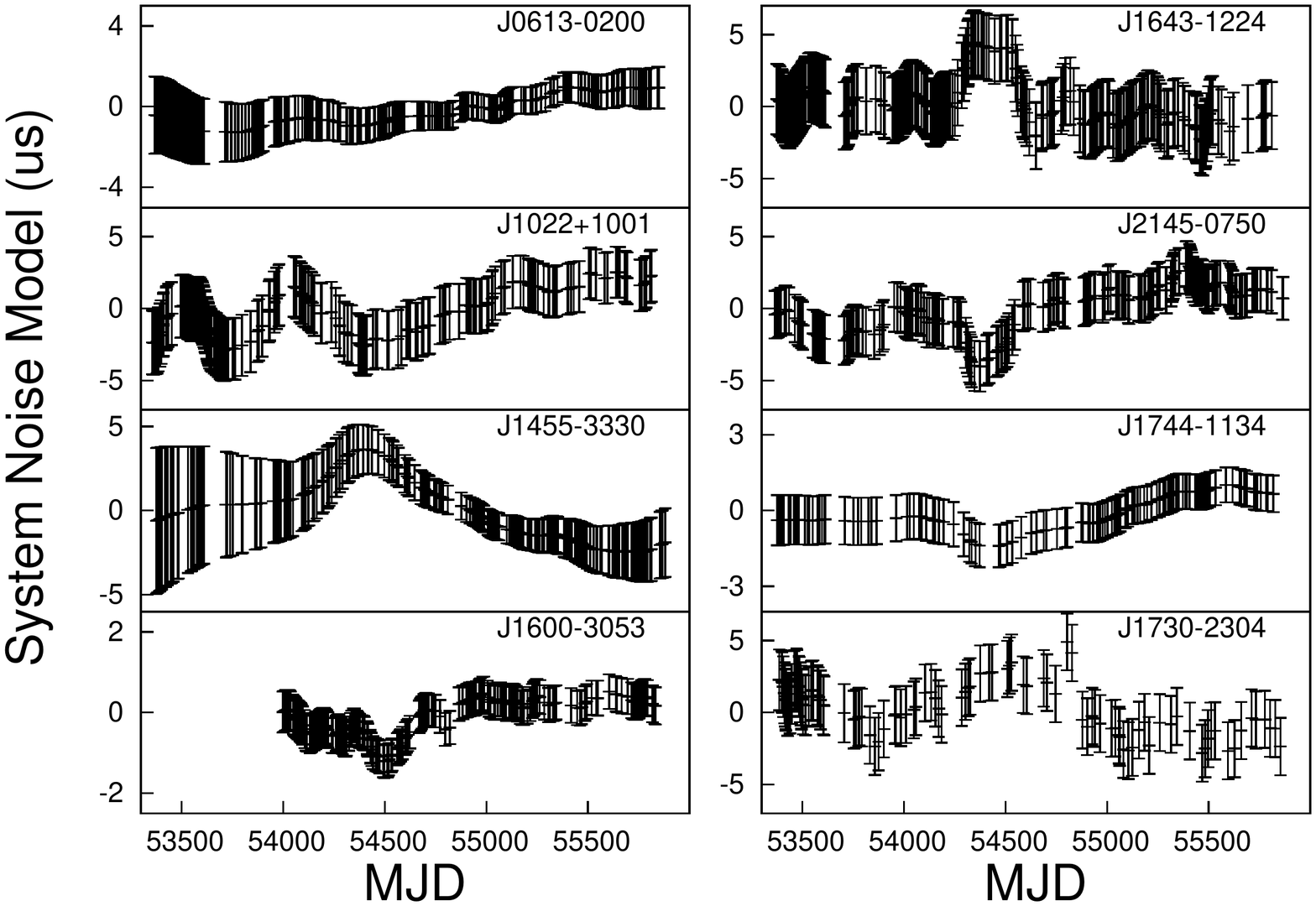}\\
\end{array}$
\end{center}
\caption{Maximum-likelihood signal realisations for the system noise detected in eight pulsars for the Nan{\c c}ay 1400~MHz system.  The bumps and troughs around MJDs $\sim$ 54300-54500 correspond to known times where errors in polarization calibration have affected the ToAs.}
\label{Fig:J1730GroupNoiseSignals}
\end{figure*}

The full IPTA data set, however, includes significant additional PPTA data.  Analysis of this data set (bottom-right panel) shows that the Nan{\c c}ay  1400~MHz data is sufficiently inconsistent with the PPTA data that we can separate the system noise from the spin noise in this pulsar.  The optimal model then includes no support for spin noise in this data set, only system noise in the Nan{\c c}ay  1400~MHz data.

In total we find that eight pulsars have significant system noise in the EPTA Nan{\c c}ay 1400~MHz data.  In Fig. \ref{Fig:J1730GroupNoiseSignals} we show the maximum-likelihood signal realisations and 1-$\sigma$ confidence intervals given the maximum-likelihood power spectrum from the full Bayesian analysis.  In particular, PSRs J1643$-$1224, J2145$-$0750  and J1600$-$3053 show significant bumps or troughs around MJDs $\sim$ 54300-54500, an interval known to be contaminated by polarization calibration errors.

We note that of these eight pulsars, all but PSRs J1455$-$3330 and J0613$-$0200 have also been found previously to be susceptible to polarization calibration errors at the $> 100$~ns level at 1400~MHz in \cite{2013ApJS..204...13V} (henceforth S13).  PSR J1455$-$3330 was not a part of the sample analysed in S13, and PSR J0613$-$0200 was found to be susceptible at a lower level.  We find that the amplitude of the system-noise signal determined by our analysis for these pulsars tracks the rms of the  systematic timing error introduced by polarization calibration errors found in S13. PSR J0613$-$0200 (Log$_{10}$ A$_{\mathrm{Sys}}$ =  -14.8   $\pm$  0.8), and PSR J1744$-$1134 (Log$_{10}$ A$_{\mathrm{Sys}}$ =  -13.36  $\pm$  0.16) being less prone to calibration errors than  PSR J1643$-$1224 (Log$_{10}$ A$_{\mathrm{Sys}}$ =  -12.60  $\pm$  0.07), and PSR J1022+1001 (Log$_{10}$ A$_{\mathrm{Sys}}$ =  -12.68  $\pm$  0.07), for example.

Clearly if the calibration errors are localised to a particular set of observational epochs, a time-stationary power-law model applied to the entire Nan{\c c}ay 1400~MHz data set is not going to be the most optimal model to describe the system noise, as it will down-weight the entire data set in order to model the excess noise over this small time period.  As a final test we therefore investigate the stationarity of the system noise in PSR J1730$-$2304.  In principle we could apply the same approach that we use to model DM events, applying the shapelet basis to model a non-stationary system-noise process.  However, we perform a simpler test, limiting the time period over which this noise term is applied to only those MJDs greater than the date that the polarization errors are known to have occurred ($\sim$ 54200).  We find the difference in the evidence between this model and a model where the system noise is applied to the entire Nan{\c c}ay 1400~MHz data set is $\sim 3$, in favour of the non-stationary model.    Future IPTA data releases will clearly have to develop these system-noise models further, utilising as much prior information as possible about the observations.  In our analysis, however, we can still say that while it is likely not the most suitable model, the stationary description of the system noise included in our analysis is still  preferred to not including system noise, as supported by the evidence in each case.

\begin{figure*}
\begin{center}$
\begin{array}{cc}
\hspace{-0.5cm}
\includegraphics[trim = 30 0 30 30, clip,width=100mm]{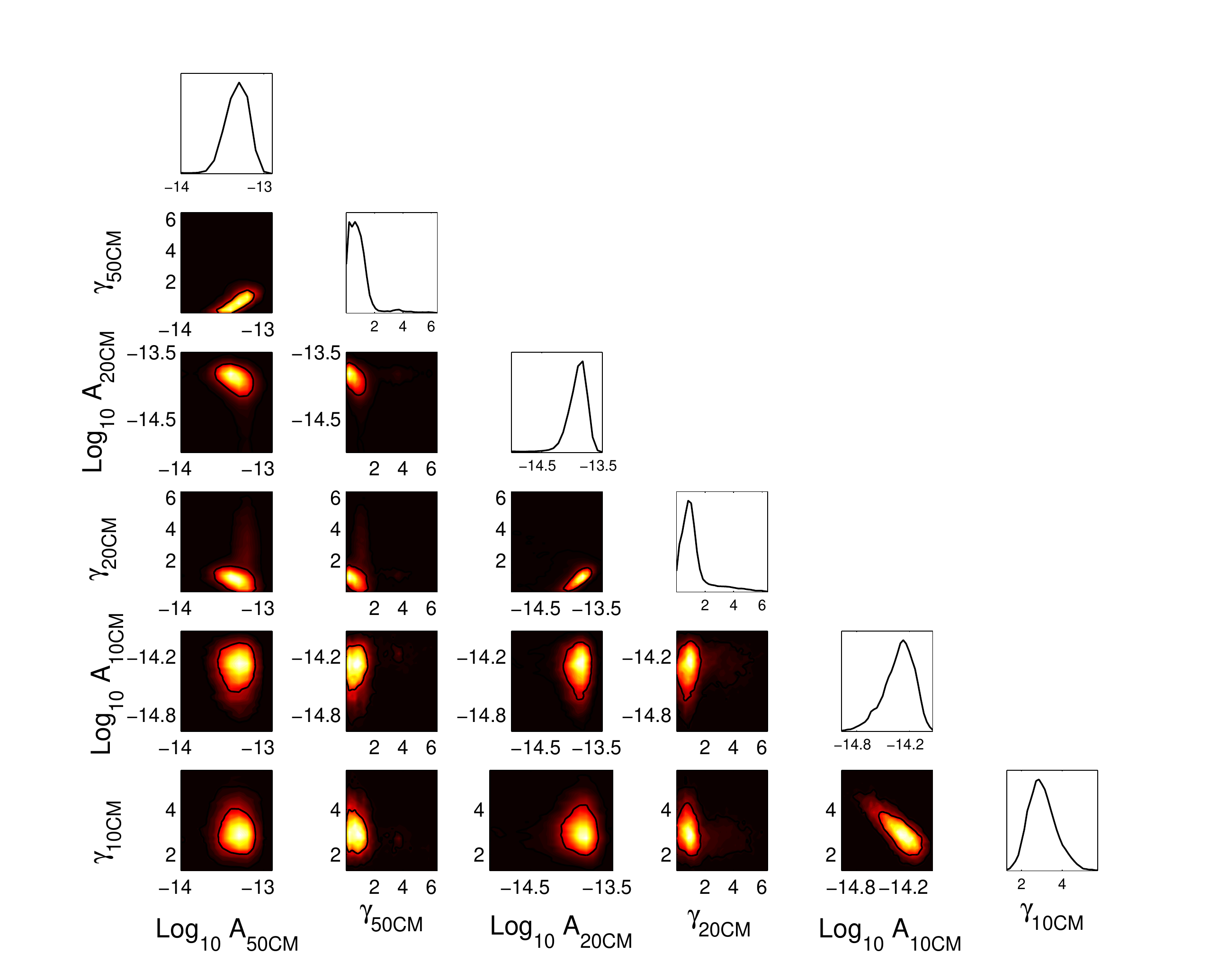} &
\hspace{-0.5cm}
\includegraphics[trim = 30 0 70 50, clip,width=90mm]{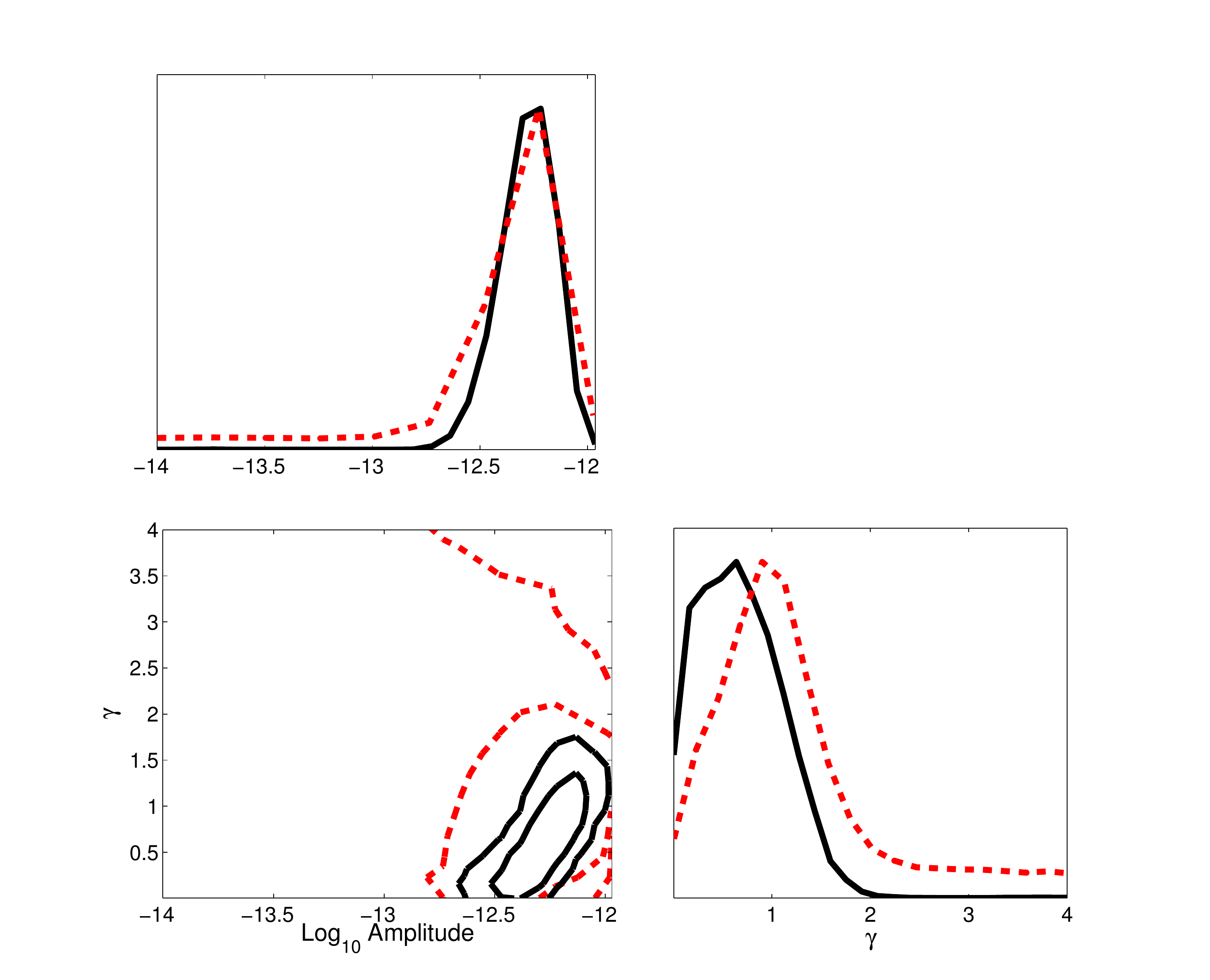} \\
\hspace{-1cm}
\includegraphics[width=100mm]{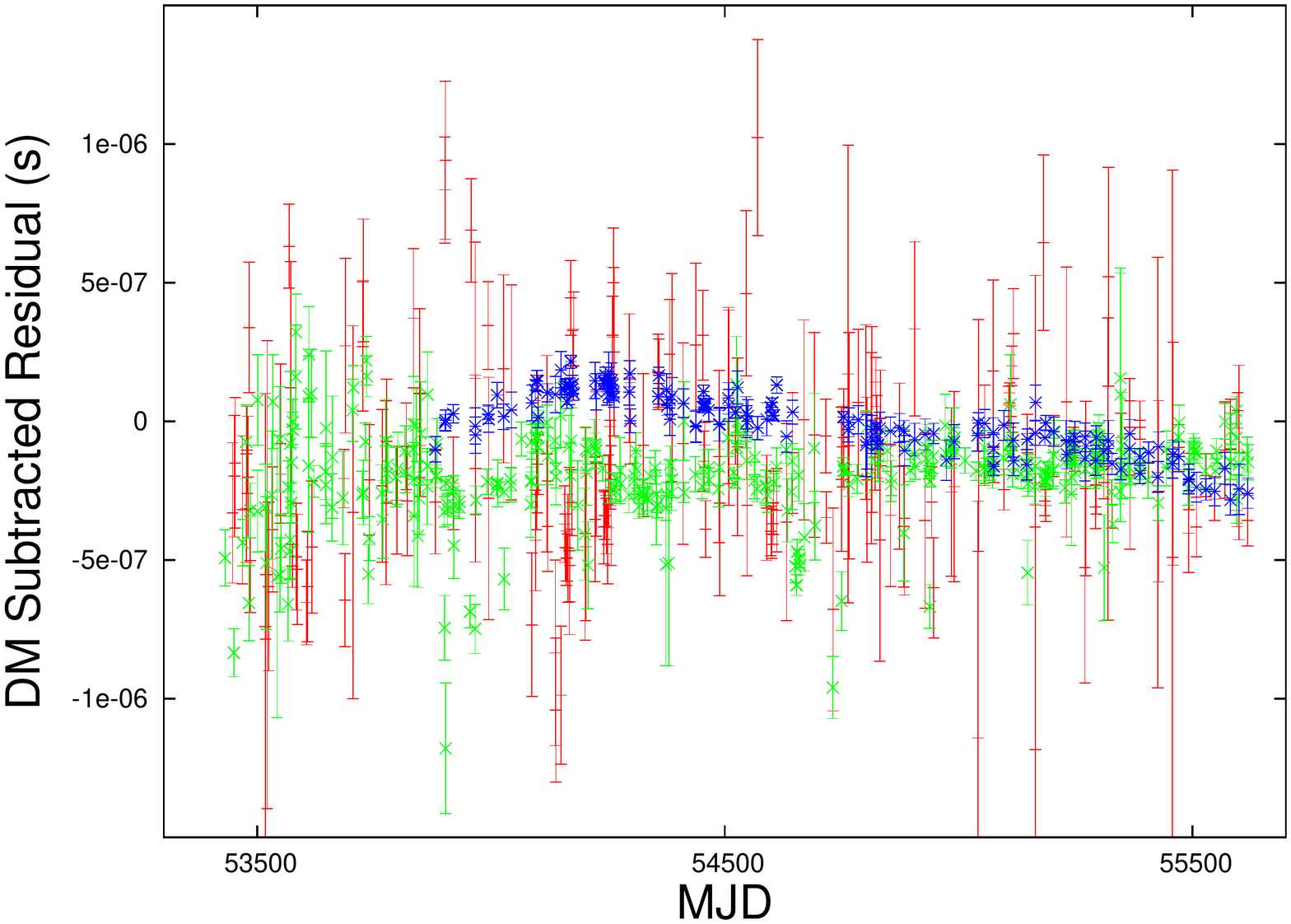} &
\hspace{-1cm}
\includegraphics[width=100mm]{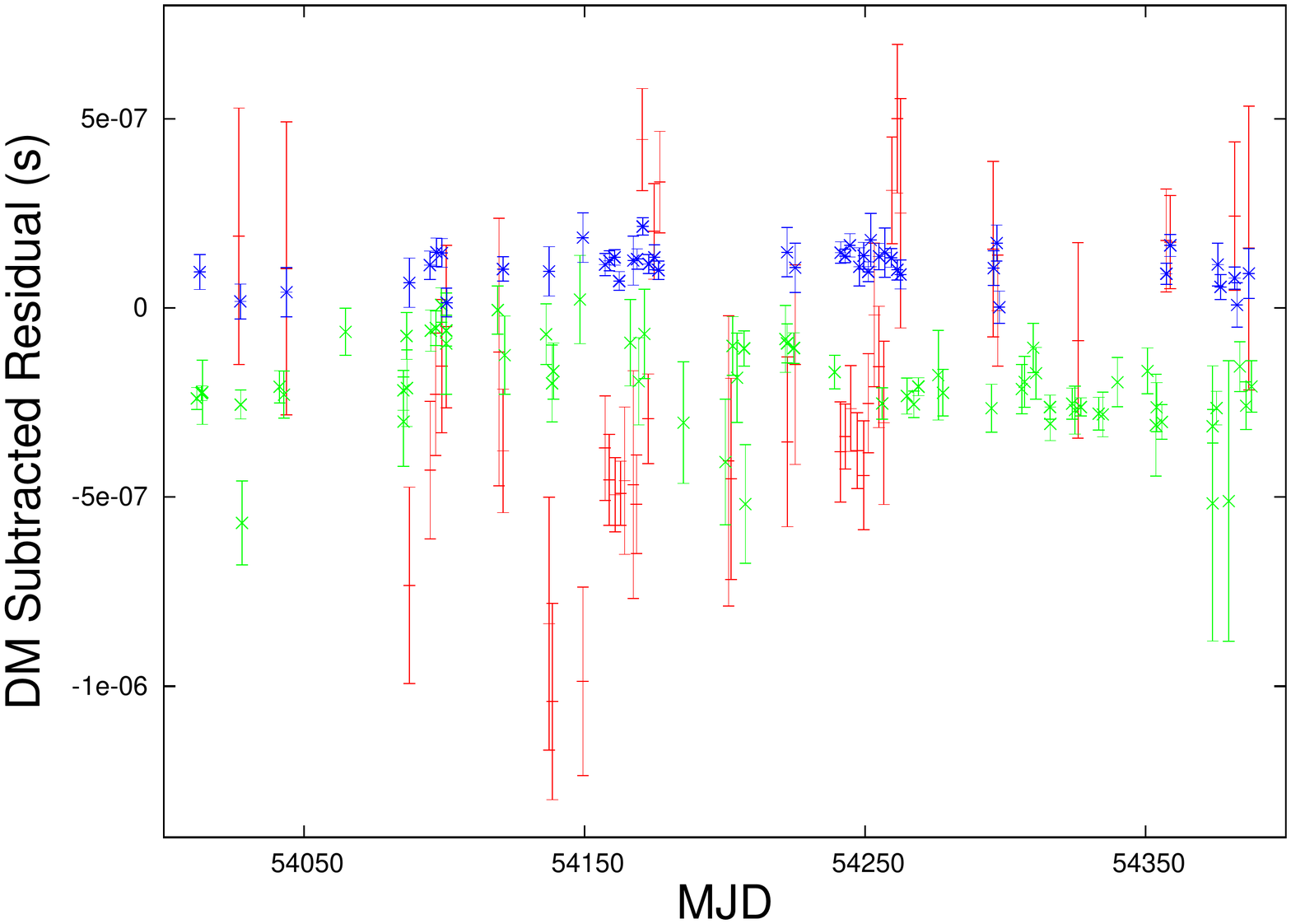} \\
\end{array}$
\end{center}
\caption{(Top left) One- and two-dimensional marginalised posterior parameter estimates for the band noise power-law amplitudes and spectral exponents from the optimal noise model for PSR J0437$-$4715.  The amplitude-$\gamma$ space occupied by the three bands are inconsistent with one another.  Both the 50~cm and 20~cm band-noise terms have flat-spectrum noise models with $\gamma_{\mathrm{BN}} \sim 1$, while the 10~cm band noise is much steeper with $\gamma_{\mathrm{BN}} \sim 3$.  (Top right) If we perform our analysis of the optimal model, but scale the noise in each band as with the DM variations, we find that both the 50~cm (black solid lines) and 20~cm terms (red dashed lines) are completely consistent in $A-\gamma$ space, and the evidence supports describing both with a single amplitude and spectral exponent.  (Bottom left)  DM-subtracted residuals for the whole multi-frequency data span,   and (bottom right) for a $\sim$250-day period. For clarity, residuals have been time averaged in two-day windows, separately for each system group.  Colours indicate observing frequency; 50~cm data (red  + points), 20~cm (green x points) and 10~cm (blue $\ast$ points).}
\label{Fig:J0437BandNoise}
\end{figure*}

\section{Band Noise}
\label{Section:BandNoise}

Out of the 49 pulsars in the IPTA data set, four show significant evidence for the presence of `band noise' (see Section \ref{Section:SystemandBandModels}); J0437$-$4715, J1600$-$3053, J1643$-$1224, and J1939+2134.   The origin of this band noise is as of yet unclear.  In principle, band noise could have its origins in the IISM, either as an  incoherent component of the DM variations that represents photons at different frequencies experiencing different scattering volumes \citep{2015arXiv150308491C}, or as terms that have steeper dependencies on the observing frequency than $\nu^2$, and thus would present themselves as an excess of noise at longer wavelengths.  Alternatively, the excess noise could be the result frequency-dependent calibration errors, which can show correlations over hundreds of days \citep{2013ApJS..204...13V}, or RFI.  This last source of band noise, however,  is likely to be uncorrelated between observations and thus would be modelled by the white-noise parameters in our analysis.  In the following sub-sections we look in greater detail at the four pulsars that have evidence for band noise in order to try and ascertain which of these possibilities could be the origin of the excess noise in these pulsars.

As for the system noise, in Table \ref{Table:SystemAndBandNoiseParams} we list the mean parameter estimates and 1-$\sigma$ uncertainties, along with total integrated power, for the band noise components of the stochastic model.

\subsection{PSR J0437$-$4715}
\label{Section:BandNoise0437}

For PSR J0437$-$4715 the optimal noise model included (in addition to DM noise and the system noise discussed in Section \ref{Section:SystemNoise})  band-noise processes in the 0-1000~MHz, 1000-2000~MHz, and $>$ 2000~MHz bands, which we refer to in this section as the 50~cm, 20~cm and 10~cm bands. In Fig. \ref{Fig:J0437BandNoise} (top-left panel) we show the one- and two-dimensional posterior probability distributions for the power-law amplitudes and spectral exponents for the three band-noise components from the optimal model.  The mean values and 1-$\sigma$ uncertainties for the band-noise models, along with the total integrated power for each term are listed in Table \ref{Table:SystemAndBandNoiseParams}.    We find the evidence supports different amplitudes for the band-noise in each band, with log$_{10}$ values of -13.32  $\pm$  0.16, -13.83  $\pm$  0.14, and -14.4  $\pm$  0.3 for the 50, 20 and 10~cm bands respectively. Both the 50~cm and 20~cm bands have shallow red-spectrum noise models with $\gamma_{\mathrm{BN}} \sim 1$, while the 10~cm band noise is much steeper with $\gamma_{\mathrm{BN}} \sim 3$.

When including an additional spin-noise component in the model we found this to be completely covariant with the 10~cm band noise term, and the difference in evidence for including either the spin-noise model or the 10~cm band-noise term was $\sim 0.4$, indicating the data have no power to discriminate between these two models.  We find that the parameter estimates for both the spin noise or the 10~cm band noise are consistent with one another, and with the timing noise observed in the 10~cm data set analysed in  \cite{2015Sci...349.1522S}.

We find significant evidence that the band-noise signals are incoherent between the 50~cm and 20~cm bands (i.e., the time-domain signals for both processes are inconsistent with one another).  We compare models where we fit for either, (i) a single additional noise process present in both the 50~cm and 20~cm bands with the same (i.e., coherent) signal  and (ii) separate incoherent band-noise processes in the 50~cm and 20~cm bands.  We repeat this process both with a $\nu_o^0$ (equivalent to spin noise) and $\nu_o^2$ (equivalent to DM variations) scaling of the amplitude of the signals with observing frequency. We find a difference in the $\log$ evidence of $\Delta \log \mathcal{Z} = 30.1$ and $\Delta \log \mathcal{Z} = 24.3$ in favour of separate noise processes in the $\nu_o^0$ and $\nu_o^2$ cases respectively.  In Fig. \ref{Fig:J0437BandNoise}  we show DM subtracted residuals for the full multi-frequency data set (bottom-left) and for a ~250-day sub-section.  We subtract the DM by calculating the maximum likelihood signal realisation using the maximum {\it a posteriori} amplitude and spectral exponent for the DM noise process obtained from our Bayesian analysis. In both cases for clarity we have time averaged the residuals in two-day windows, separately for each system. The 50~cm data show high frequency structure that is, even by eye, inconsistent with the 20~cm data, which reflects the magnitude of the increase in the log evidence when allowing the signals to be incoherent between the different bands, despite the increase in dimensionality.

Despite the lack of coherent signals between the 50~cm and 20~cm bands, we find that the spectral properties of the noise in these bands are related. Fig. \ref{Fig:J0437BandNoise} (top-right panel) shows the one- and two-dimensional posterior distributions for the 50~cm and 20~cm  power-law amplitude and spectral exponents from our model where we have separate, incoherent noise terms in each band, but scale the amplitudes in each band with $\nu_o^2$, as for our model for DM variations.  We find the parameter estimates are completely consistent in $A-\gamma$ space, and the evidence supports describing both with a single amplitude and spectral exponent.  Given the expected delay at 20~cm due to scattering in PSR J0437$-$4715 is less than 1~ns \citep{2010ApJ...717.1206C},  and that the DM is relatively low in this pulsar ($\sim$ 2.64~cm$^{-3}$~pc) it is possible that given the amount of data currently available, this is a coincidence and that the origin of the excess band noise could simply be RFI at the telescope site, or the result of polarisation calibration errors.  However, given the high precision of the observations, it is also possible that this could be indicative of small differences in the sampling of the IISM by the different wavelengths emitted by the pulsar.  Without overlapping observations from multiple telescopes at the same frequency, however, it will be difficult to disentangle these different interpretations.

\begin{figure}
\begin{center}$
\begin{array}{c}
\hspace{-1cm}
\includegraphics[width=90mm]{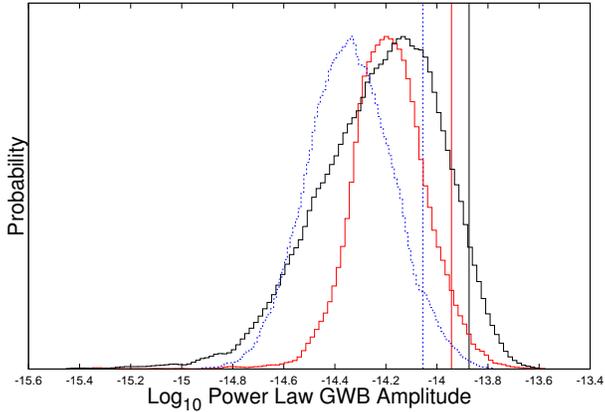} \\
\end{array}$
\end{center}
\caption{One-dimensional marginalised posterior distributions for the amplitude of a steep spin-noise process with $\gamma_{\mathrm{SN}} = 13/3$.  We use this model as a proxy for perturbations that are consistent with those expected from a GW background generated by a population of supermassive black-hole binaries.  A prior that is uniform in the amplitude of the spin-noise is used in order to obtain upper limits.  Different models are assumed for the stochastic signals and different subsets of the PSR J0437$-$4715 data set are analysed.   Black solid lines correspond to an analysis of the full data set, including only spin noise and DM noise in the model, red lines correspond to a spin-noise-only model applied to the 10~cm data only, and finally the blue dashed lines correspond to the optimal model, including system-noise and band-noise terms applied to the full data set.  Vertical lines correspond to 95\% upper limits for each case, which we find to be $1.3\times10^{-14}$, $1.1\times10^{-14}$, and $8.2\times10^{-15}$, respectively.}
\label{Fig:J0437Limits}
\end{figure}

Finally, in Fig. \ref{Fig:J0437Limits} we show the upper limits obtained at a spectral exponent of $\gamma_{SN} = 13/3$ (consistent with the expected spectral exponent for a GWB resulting from inspiraling SMBHBs, c.f. references in the Introduction) for an additional spin-noise process using a uniform prior on the amplitude for three different cases.    As we are only calculating upper limits on the GWB term we do not perform evidence comparison for models with and without this additional parameter.   We show the 95\%  upper limits obtained on the full J0437$-$4715 data set using a simple model that contains only coherent power-law DM noise and spin-noise terms (black line, A$_{95}$ = 1.3$\times10^{-14}$ ), and using the optimal model (blue line, A$_{95}$ = 8.2$\times10^{-15}$), and finally the upper limit obtained using only the 10~cm data, where we include only a spin-noise term in our noise model in addition to the GWB power-law term with fixed spectral exponent (red line, A$_{95}$ = 1.1$\times10^{-14}$).  We find that compared to the simple noise model the 10~cm only data set results in an upper limit that is $\sim 30\%$ lower.  However, when the system- and band-dependent noise terms are modelled more appropriately, the upper limit for the optimal noise model is $\sim 20\%$ lower than the 10~cm only limit, and $60\%$ lower than that for the simple noise model.

\begin{table}
\caption{Relative log evidence values, and upper limits on a stochastic GWB for a simulated PSR J0437$-$0715 data set.}
\begin{tabular}{ccc}
\hline\hline
Model & log evidence & 95\% upper limit          \\
           &                      &  $10^{-16}$ \\
\hline
No noise model & $-$4081.2 & 6162 \\
DM Noise & 0.0 &  5.3 \\
DM Noise, Spin Noise & $-$0.9 & 5.3 \\
DM Noise, Spin Noise, Band Noise & $-$3.0 &  5.3 \\
\hline
\end{tabular}
\label{Table:J0437Sims}
\end{table}

We stress that the improvement in the sensitivity of the data set to gravitational waves as the complexity of the noise model increases is not simply a generic result of including additional parameters in the model.  If further components are added to the model that are not warranted by the data the upper limit on the amplitude of a GWB will either remain constant, or potentially increase.  We show this explicitly by constructing a simulation using the observed time stamps and frequencies from the IPTA PSR J0437$-$4715 data set, with white noise consistent with the formal ToA uncertainties and DM variations with statistical properties consistent with those from the real data.  We then analyse this simulation using four different models which we list in Table~\ref{Table:J0437Sims}.  For the band noise model we include three terms for the $<$1000~MHz, 1000$-$2000~MHz, and $\ge$ 2000~MHz bands.   In each case we perform the analysis twice, once without an additional GWB term, and once with the additional term.  We include in Table~\ref{Table:J0437Sims} the relative log evidences for the models without the GWB term, and the 95\% upper limit on the amplitude of the GWB obtained from that analysis.

We find that the evidence supports the DM-noise only model, and that as expected the upper limit decreases substantially when going from a model with no noise components to the DM-noise model.  However, as additional components are added to the model describing spin noise or band noise, the upper limit decreases no further.  We note that the upper limit is significantly better in the simulation compared to the real data set as we have used the formal ToA uncertainties, rather than the uncertainties modified by the EFAC and EQUAD parameters in constructing our simulation.

\begin{figure*}
\begin{center}$
\begin{array}{cc}
\hspace{-1.1cm}
\includegraphics[width=90mm]{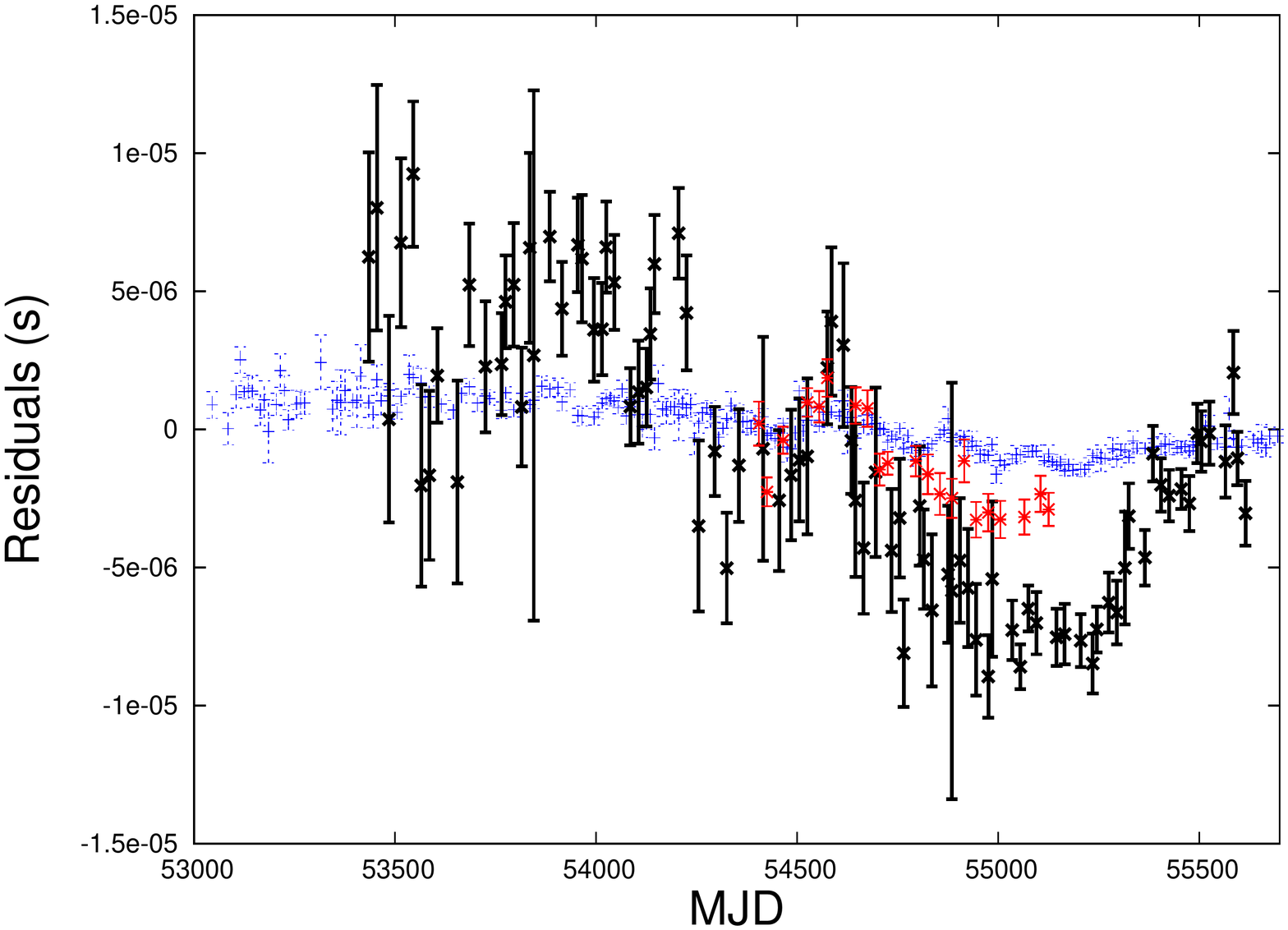} &
\includegraphics[width=90mm]{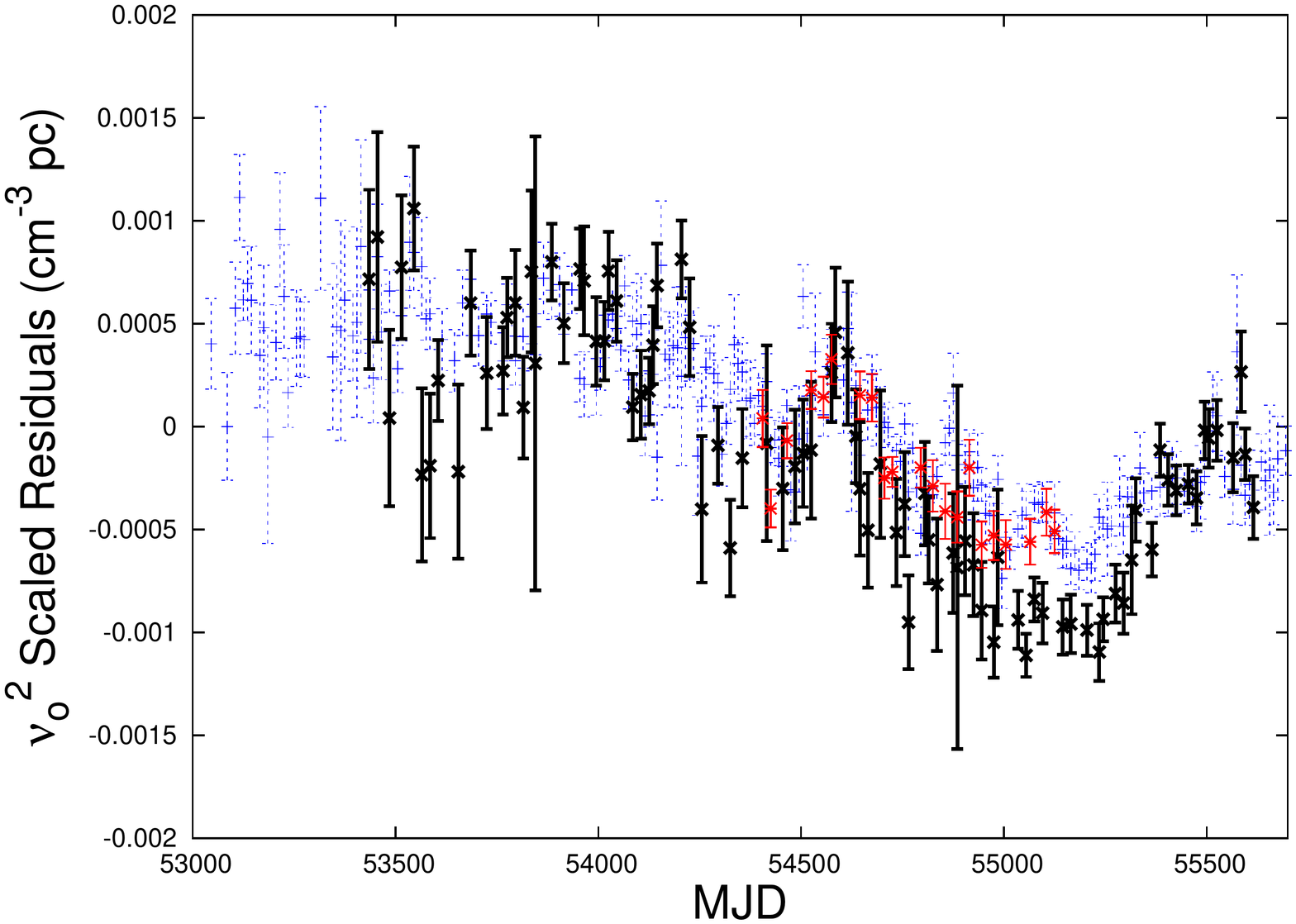} \\
\hspace{-1.1cm}
\includegraphics[width=90mm]{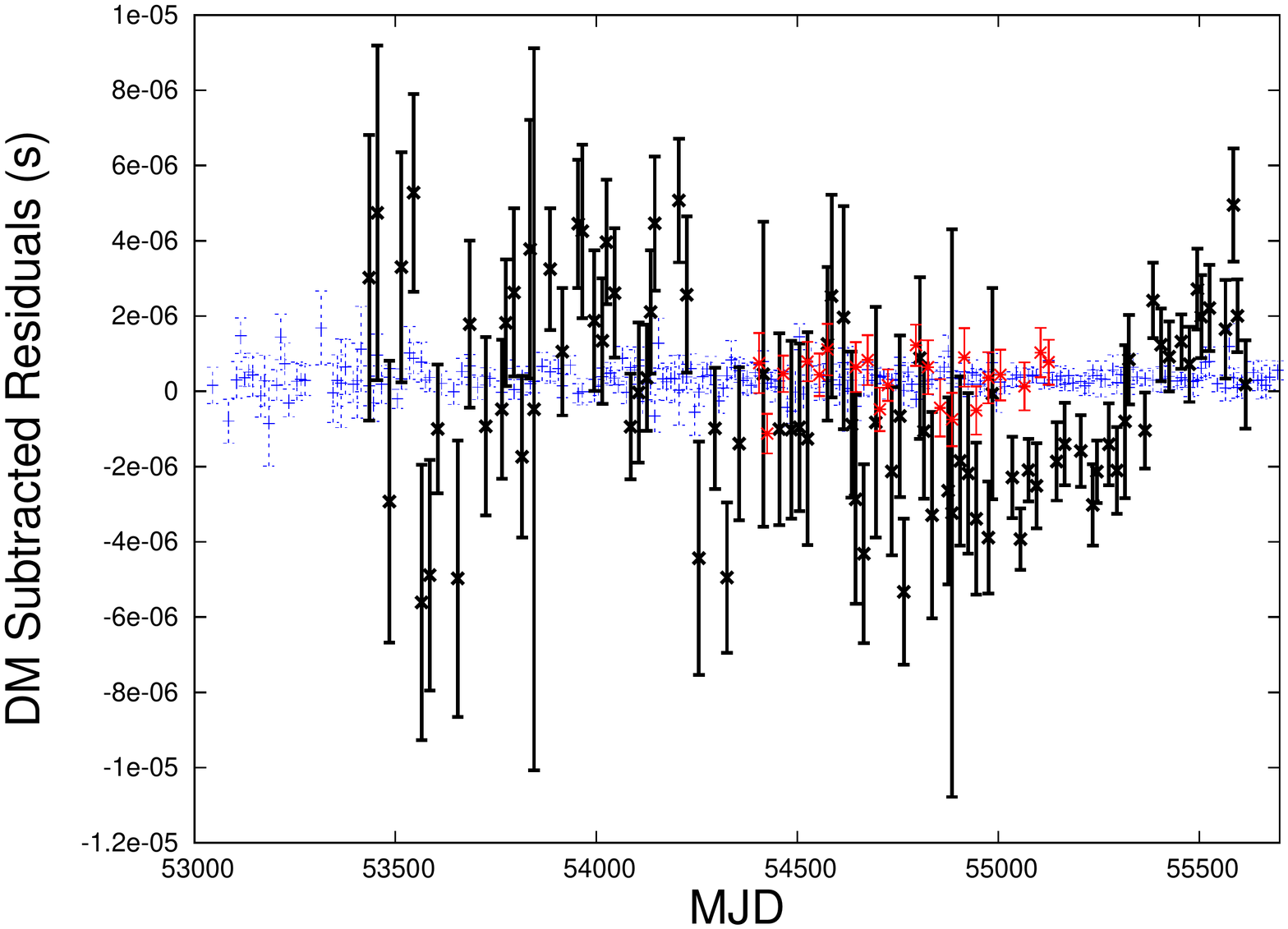} &
\includegraphics[width=90mm]{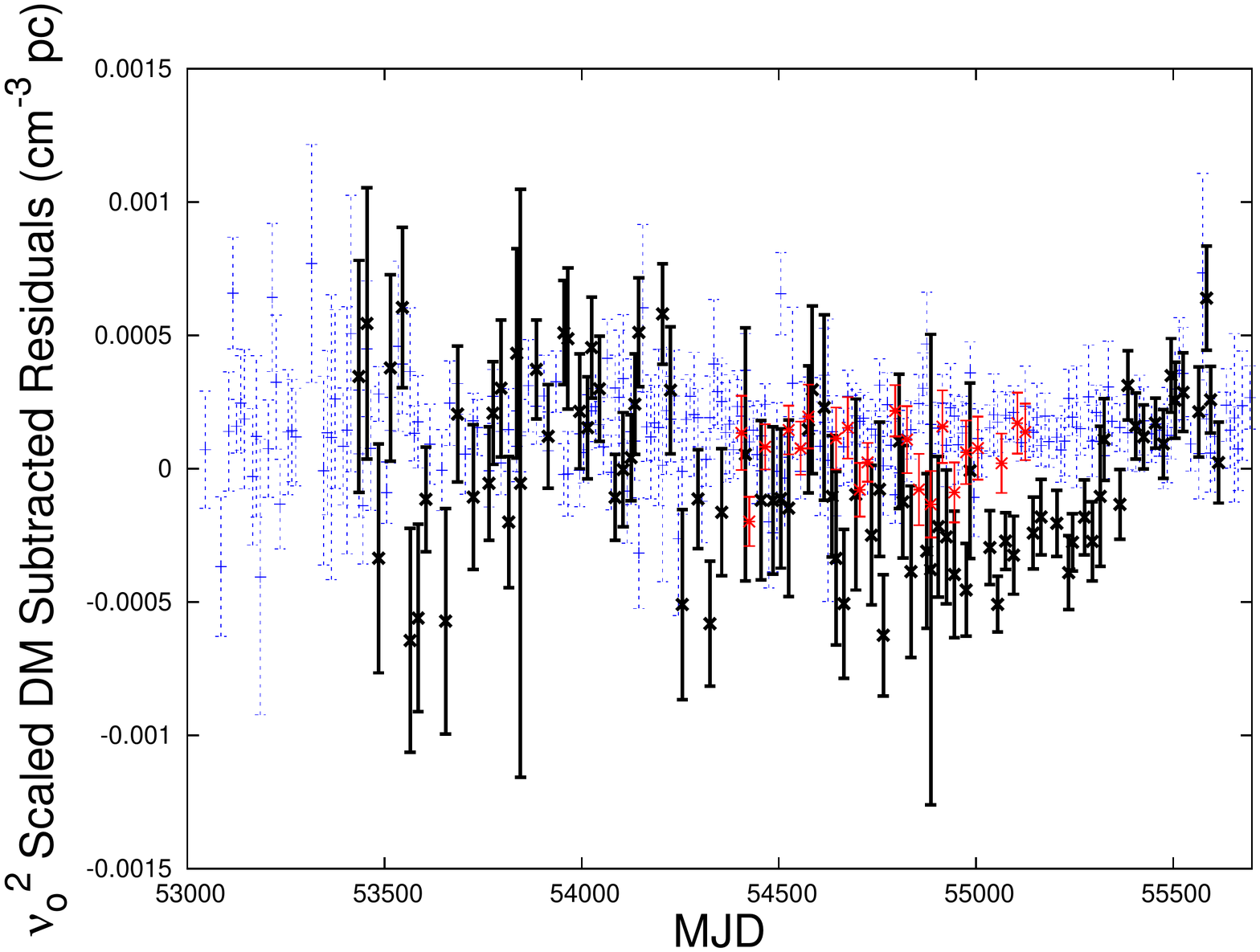} \\
\includegraphics[width=80mm]{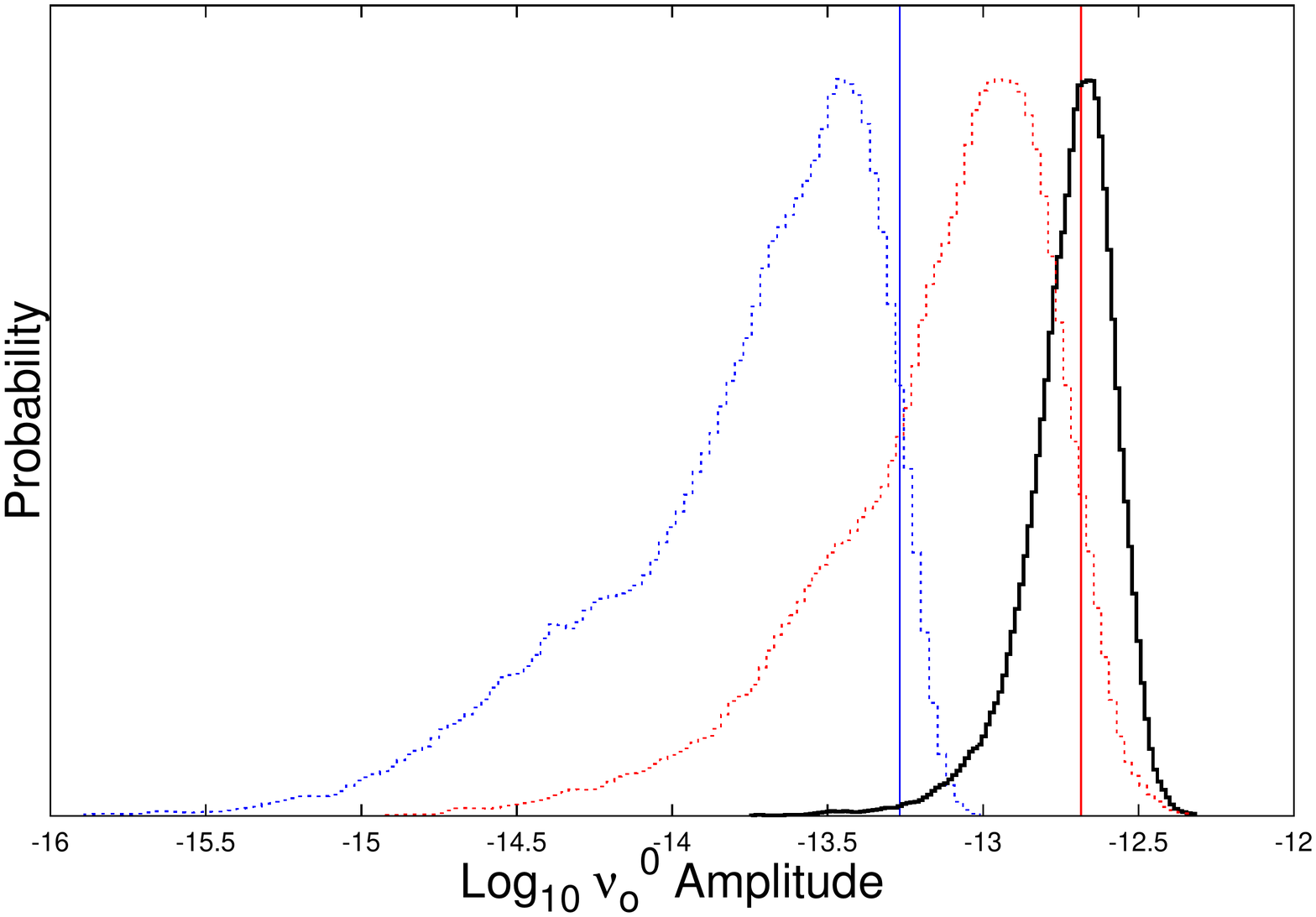} &
\includegraphics[width=80mm]{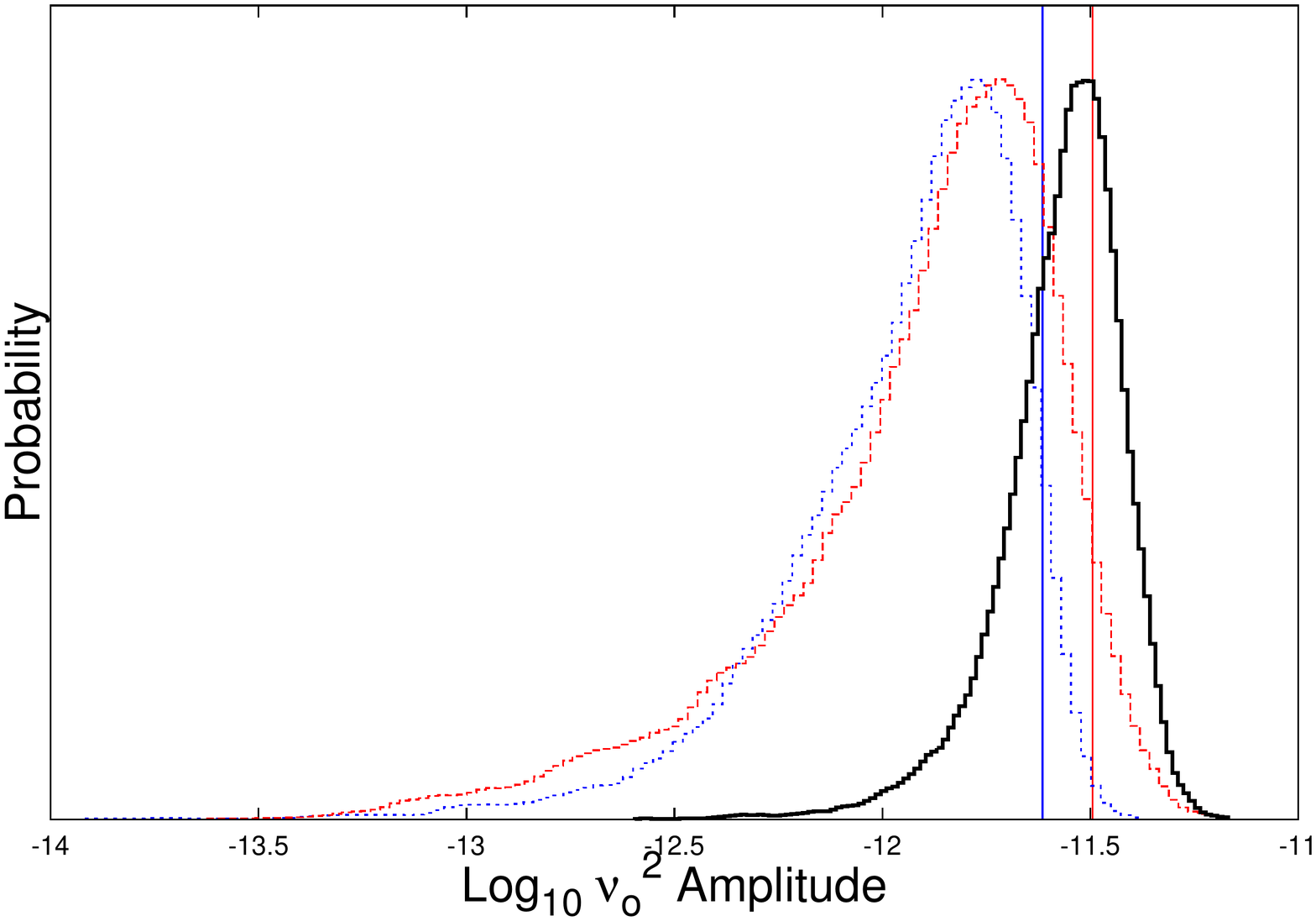} \\
\end{array}$
\end{center}
\vspace{-1cm}
\caption{(Top left) Timing residuals for PSR J1600$-$3053 after subtracting the maximum-likelihood timing model from our optimal model.  Colours represent: The 690-730~MHz PPTA (black points), 780-884MHz NANOGrav (red points) and $1000-2000$~MHz data from all PTAs (blue points). For clarity, the three data sets have been time-averaged over windows of 10~days.  (Top right) As for top-left panel, however the residuals have been scaled by $K\nu_o^2$, with $K$ as defined in Eq. \ref{Eq:DMScale}.   The lower-frequency PPTA data shows more structure than the 1000-2000~MHz data, even after scaling by $K\nu_o^2$.  This additional noise is present in both the Parkes CPSR2 and PDFB3 690-730~MHz data and is statistically consistent between systems.
(Middle) As for top, however after subtracting the maximum likelihood DM variations signal.  The 1000-2000~MHz data shows no significant residual timing noise, however there is still significant structure in the lower-frequency PPTA data.
(Bottom) One-dimensional posterior distributions for the amplitude of the power-law band-noise processes obtained for the PPTA 690-730~MHz data (black line) using a prior uniform in the log of the amplitude parameter, and for the NANOGrav 780-884~MHz data (red line), and 1000-2000~MHz data from all PTAs (blue line) using a prior uniform in the amplitude which we use to obtain 95\% upper limits (vertical lines).  We obtain the upper limits by fitting for a single spectral exponent across all three band-noise terms, but allowing the amplitude in each band to vary, and perform the analysis both in units that scale as $\nu_o^0$ (bottom-left panel) and $\nu_o^2$ (bottom-right panel).}
\label{Fig:J1600BandNoise}
\end{figure*}

\subsection{PSR J1600$-$3053}

For PSR J1600$-$3053 we find that the data support a band-noise model that includes excess noise in the 690-730~MHz PPTA data (which here we treat as a separate band  from the 780-884~MHz NANOGrav data), in addition to DM noise, and system noise.  We note here that, the analysis we perform is a weighted fit.  Thus, while the excess noise is detected in a band with low timing precision (larger uncertainties), these uncertainties are factored into our analysis when calculating our parameter estimates, and evidence values.  That the band noise is significant means that it is inconsistent with the higher-frequency data, even given the larger uncertainties.  Similarly, the evidence would not support a band-noise process in the 690-730~MHz PPTA data if the DM-noise model constrained by data at other frequencies was able to describe this lower-frequency data fully. In Fig. \ref{Fig:J1600BandNoise} (top-left panel) we show the timing residuals for PSR J1600$-$3053 after subtracting the maximum-likelihood timing model from our optimal model, and additionally after subtracting the maximum-likelihood DM noise model (centre-left panel). In the top-right and centre-right panels we show the same thing, with the residuals scaled by $K\nu_o^2$, with $K$ as defined in Eq. \ref{Eq:DMScale}.  This rescaling allows us to visualise a DM-like process in the data, which should be both coherent, and of similar amplitude in all bands.  For clarity, the three data sets have been time-averaged over ten-day intervals.

\begin{figure}
\begin{center}$
\begin{array}{c}
\includegraphics[trim = 30 50 230 50, clip, width=85mm]{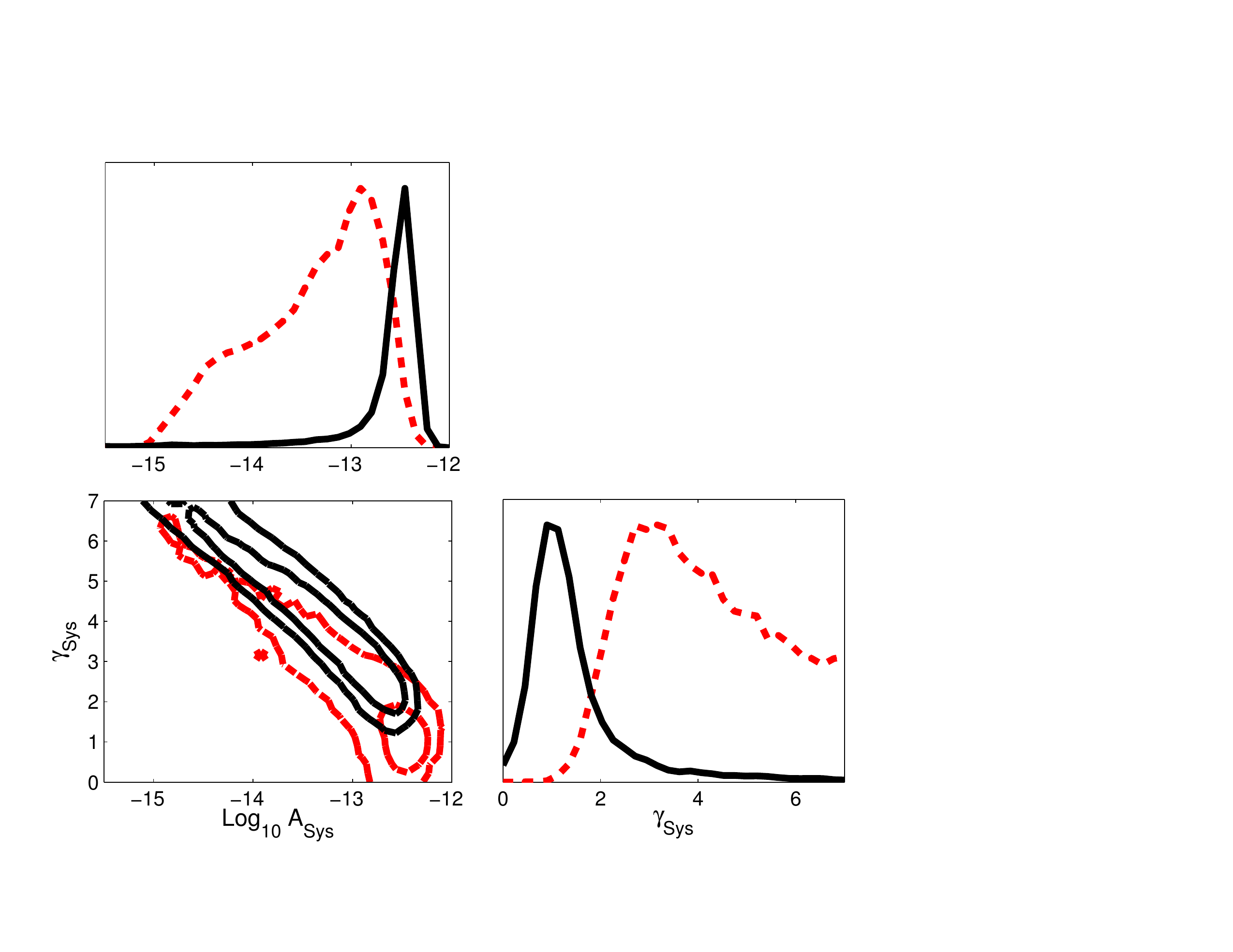} \\
\end{array}$
\end{center}
\caption{One- and two-dimensional posterior parameter estimates for the power-law amplitude and spectral exponent for system-noise terms for PSR J1600$-$3053, fitted simultaneously to the 690-730~MHz Parkes CSPR2 (black solid lines) and Parkes PDFB3 (red dashed lines)  data, instead of a single excess-noise term in the 690-730~MHz band as in the optimal model.  Contours in the two-dimensional plot are at the 1-$\sigma$ and 2-$\sigma$ levels. Both systems have significant detections of the excess noise, which are consistent within statistics.}
\label{Fig:J1600Groups}
\end{figure}

The excess noise in the lower-frequency PPTA data shows more structure than the 1000-2000~MHz data from all PTAs, even after scaling by $K\nu_o^2$.  We test to see if this excess occurs in both observing systems present in the PPTA 690-730~MHz data.    Fig. \ref{Fig:J1600Groups} shows the one- and two-dimensional posterior distributions for the power-law amplitude and spectral exponent for system-noise terms fitted simultaneously to the two Parkes systems that have observed at this band  (the CPSR2 and PDFB3 systems) instead of a single excess noise term in the whole PPTA band.  We perform this analysis using the full PSR J1600$-$3053 IPTA data set, including DM noise, and Nan{\c c}ay 1400~MHz system noise as in the optimal model. Contours in the two-dimensional plot are at the 1-$\sigma$ and 2-$\sigma$ levels. Both Parkes systems have significant detections of the excess noise, and we find that the evidence supports a single power-law model, indicating the two noise terms are statistically consistent.

In Fig. \ref{Fig:J1600BandNoise} (bottom), we show one-dimensional posterior distributions for the amplitude of the power-law band-noise processes obtained for the PPTA 690-730~MHz data  using a prior that is uniform in the log of the amplitude parameter, and for the NANOGrav 780-884~MHz data, and 1000-2000~MHz data from all PTAs using a prior that is uniform in the amplitude which we use to obtain 95\% upper limits.  All band-noise parameters are evaluated simultaneously with the other noise parameters from the optimal model.  We obtain the upper limit fitting for a single spectral exponent across all three band noise terms, but allowing the amplitude in each band to vary, and perform the analysis both in units that scale as $\nu_o^0$ (bottom-left panel) and $\nu_o^2$ (bottom-right panel).   We find that the 1000-2000~MHz data rules out the possibility of a $\nu_o^0$ process that has the same characteristics as the lower frequency noise with a probability of $> 99\%$, however it is unable to rule out a $\nu_o^2$ scaling to high significance.  The upper limits from the NANOGrav 780-884~MHz data are also consistent with the excess noise for both scalings.

If the observed signal were due to scattering by the IISM we would expect a broadening of the pulse profile that varies in time, leading to a change in the observed arrival times relative to the standard template used to form the ToA.  For each observational epoch we have a measurement of the pulse intensity as a function of observing frequency and time, referred to as a dynamic spectrum.  This intensity will fluctuate due to scintillation in the IISM, and the characteristic scale of those fluctuations (referred to as the scintillation bandwidth, $\Delta \nu_d$)  can be used to directly estimate the magnitude of the scattering timescale, $\tau_d$,  observed in our ToAs, as $\Delta \nu_d$ and  $\tau_d$ are the Fourier conjugate of one another.  As $\Delta \nu_d$ scales  as $\nu_o^4$, however,  at low frequencies it will rapidly become unresolved in the dynamic spectra, despite having a large impact on the arrival times.  We therefore use the 3100~MHz PPTA data to estimate  $\Delta \nu_d$, and from that estimate $\tau_d$ in the 690-730~MHz band.  We find $\Delta \nu_d$ to be $\sim$ 10~MHz for the 3100~MHz data, corresponding to a $\tau_d$ of 30~ns at 3100~MHz, which gives a $\tau_d$ of 10~$\mu$s in the 690-730~MHz band.  In Fig. \ref{Fig:J1600BandNoise} we see the peak to peak fluctuations of the 690-730~MHz band noise are $\sim$ 5-10$~\mu$s, consistent with our estimate of the amplitude of the scattering from the high-frequency data.

\subsection{PSR J1643$-$1224}

For PSR J1643$-$1224 we find that the optimal noise model includes band-dependent terms in both the PPTA 690-730~MHz, and NANOGrav 780-884~MHz bands in addition to the DM-noise and system-noise terms discussed previously. In contrast to PSR J0437$-$4715, we find that the evidence supports a coherent signal present in both these bands, however with a different amplitude in each, which is highly suggestive of this excess noise being the result of astrophysical processes, as opposed to RFI or telescope dependent effects.

\begin{figure*}
\begin{center}$
\begin{array}{cc}
\hspace{-1.1cm}
\includegraphics[width=80mm]{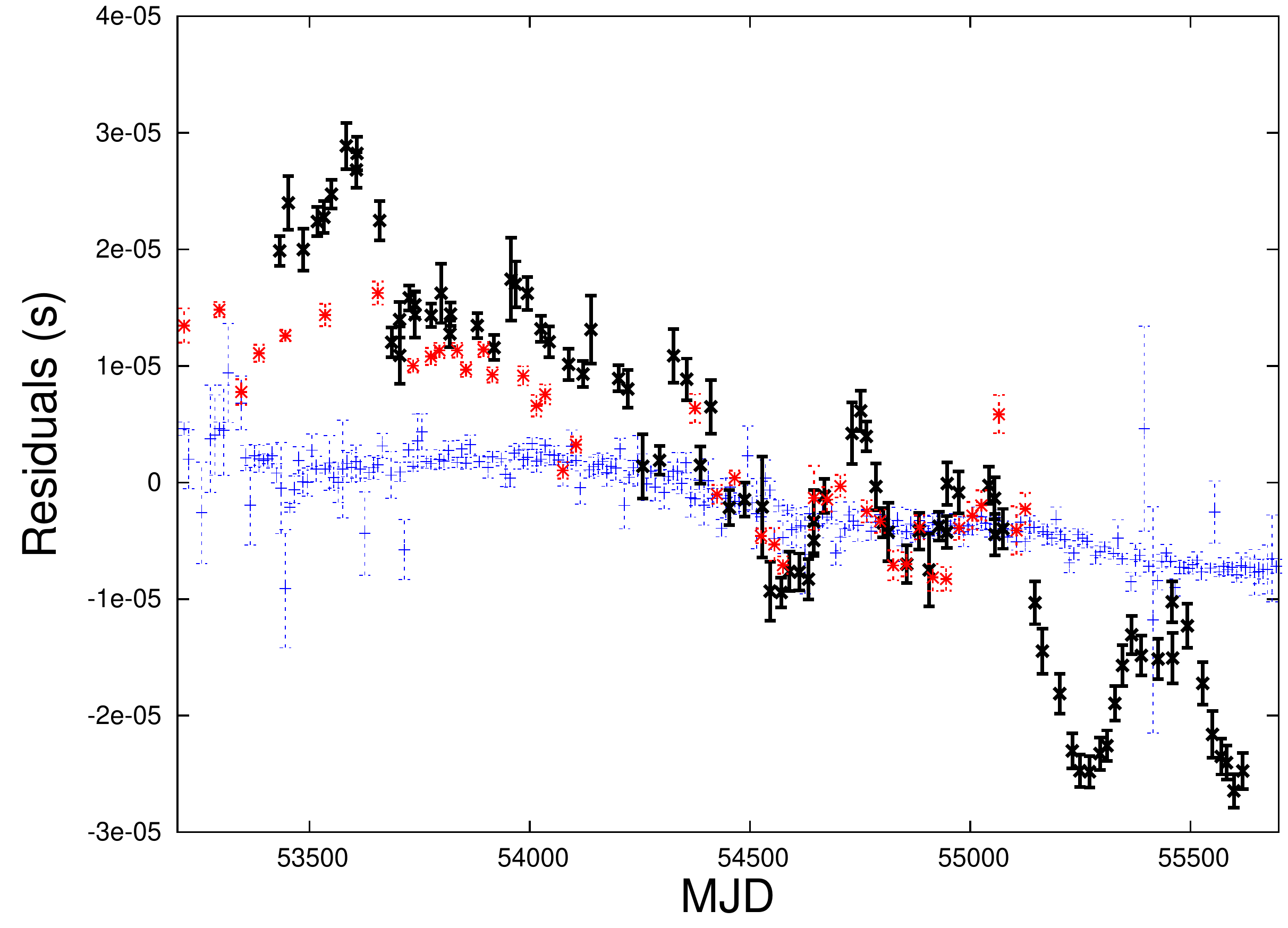} &
\includegraphics[width=80mm]{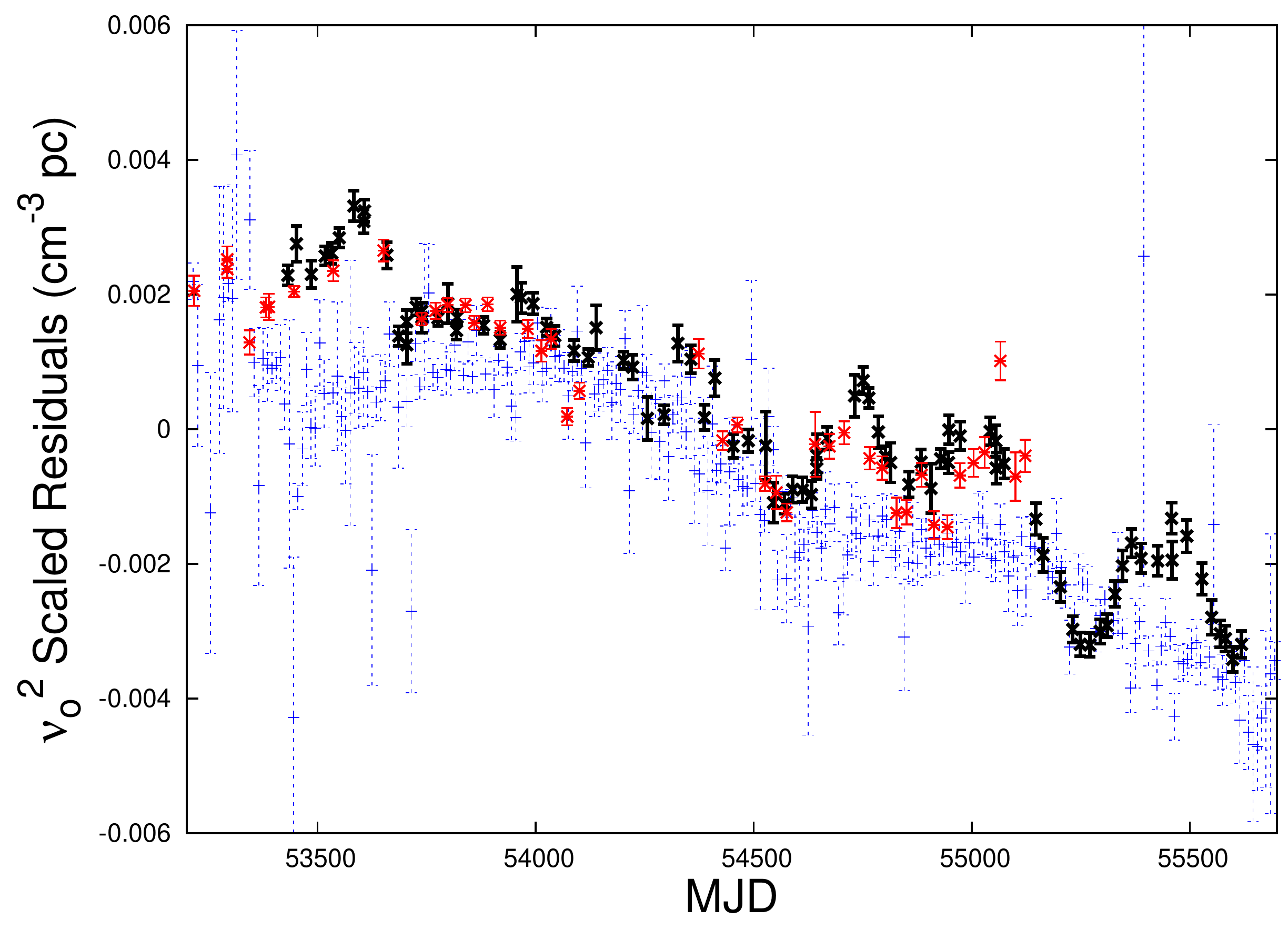} \\
\hspace{-1.1cm}
\includegraphics[width=80mm]{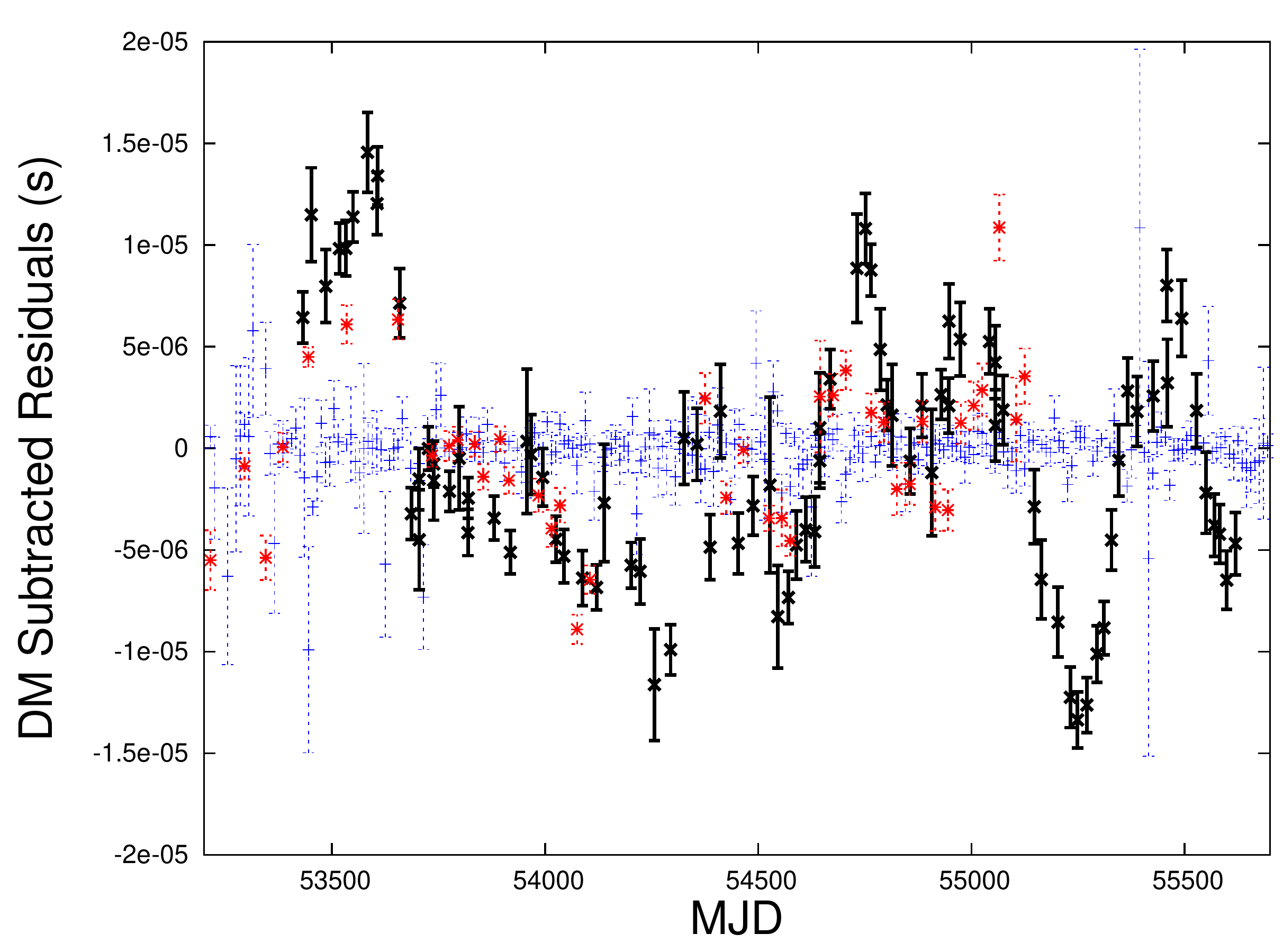} &
\includegraphics[width=80mm]{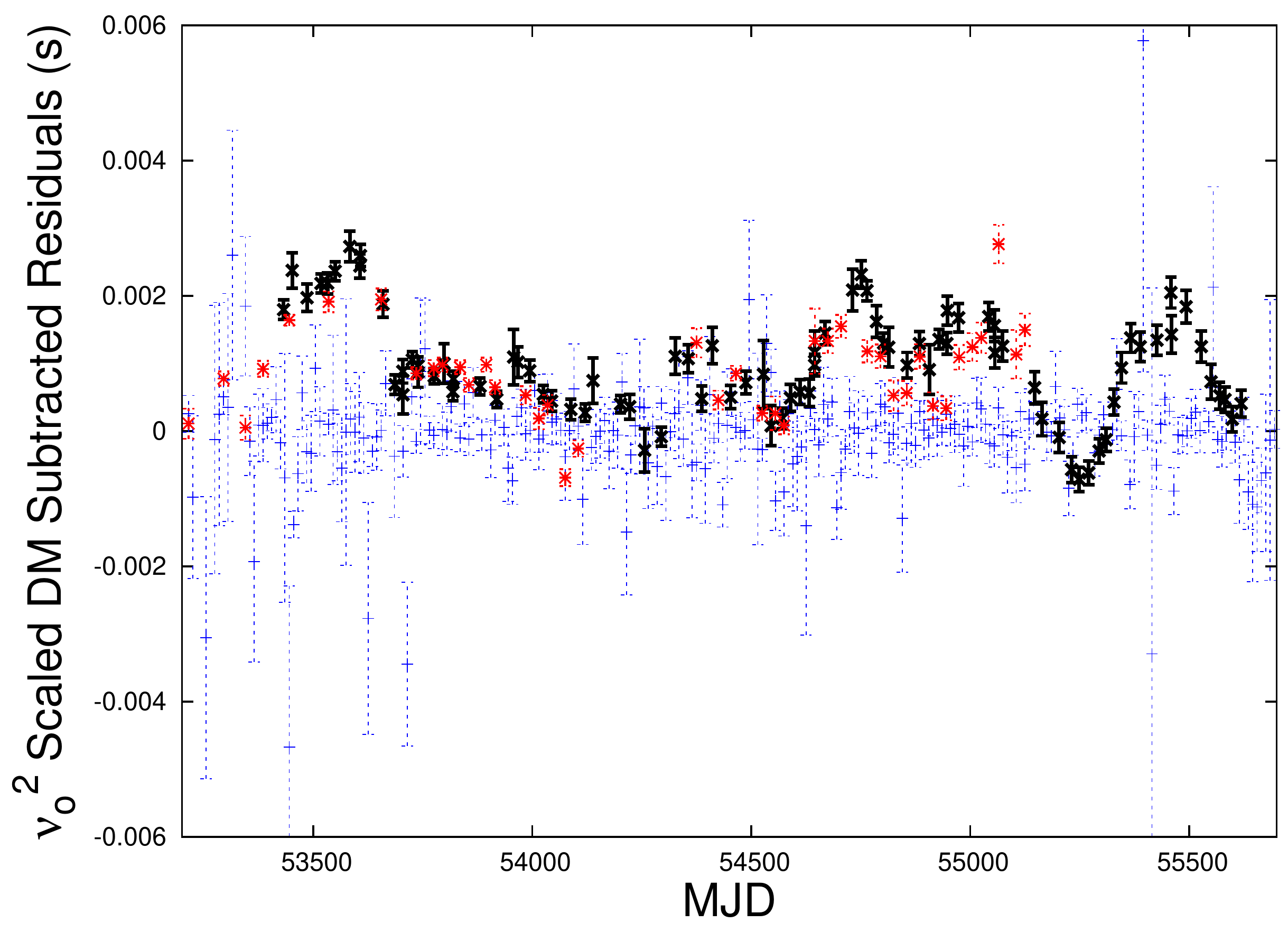} \\
\includegraphics[width=85mm]{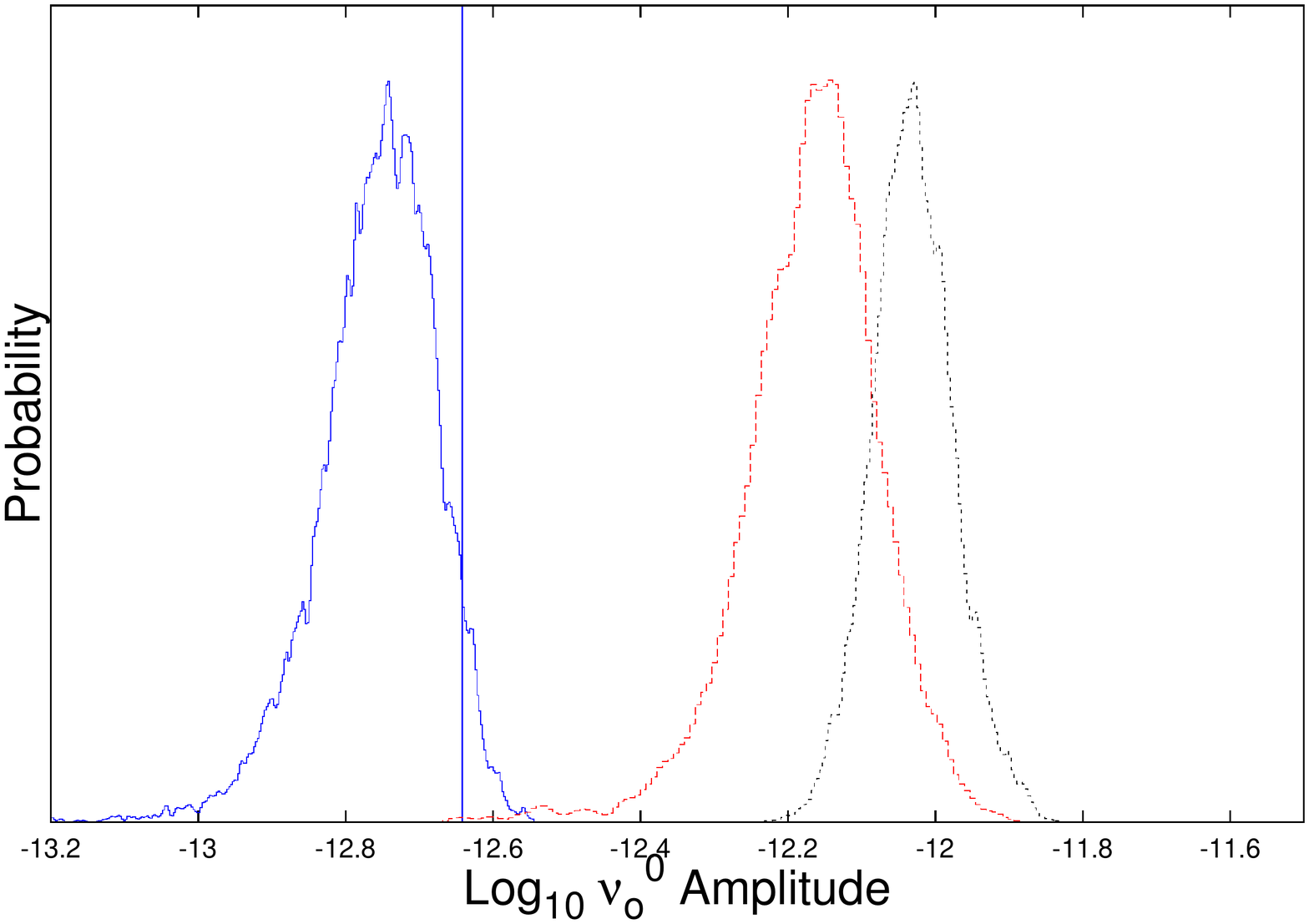} &
\includegraphics[width=85mm]{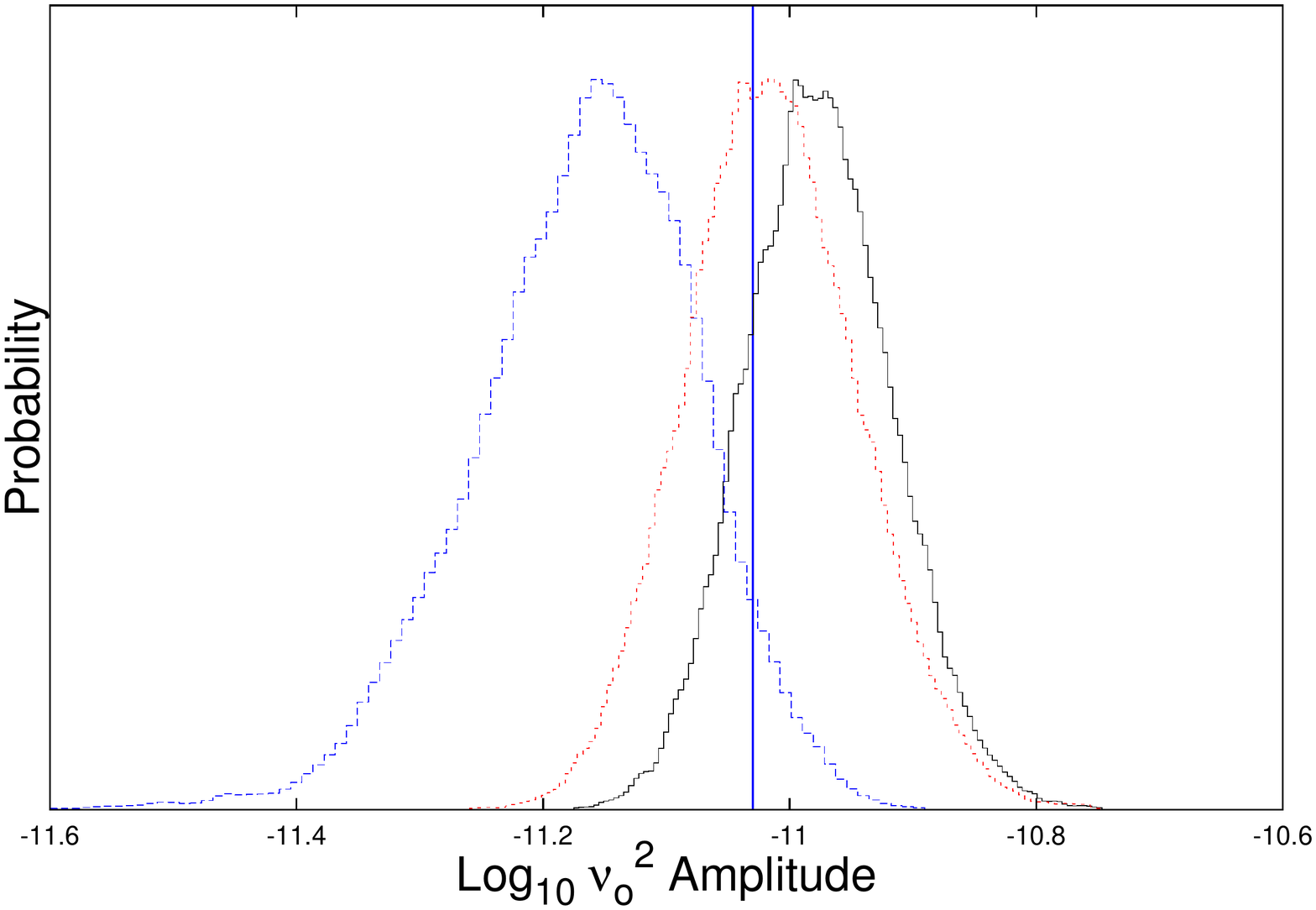} \\
\end{array}$
\end{center}
\caption{(Top left) Timing residuals for PSR J1643$-$1224 after subtracting the maximum-likelihood timing model from our optimal model.  Colours represent: 690-730~MHz PPTA (black points), 780-884~MHz NANOGrav (red points) and $1000-2000$~MHz data from all PTAs (blue points). For clarity, both the 780-884~MHz NANOGrav and the $1000-2000$~MHz data have been time-averaged over 10~day intervals.  (Top right) As for top-left panel, however the residuals have been scaling by $K\nu_o^2$, with $K$ as defined in Eq. \ref{Eq:DMScale}.   The lower frequency PPTA and NANOGrav data shows significantly more structure than the $1000-2000$~MHz data, even after scaling by $K\nu_o^2$.
(Middle) As for the top panels, however after subtracting the maximum-likelihood DM-variations signal.  The 1000-2000~MHz data shows no significant residual timing noise, however there is still significant structure in the lower frequency data.
(Bottom) One-dimensional posterior distributions for the amplitude of the power-law band-noise processes obtained for the PPTA 690-730~MHz data (black line), and NANOGrav 780-884~MHz data(red line), both using priors uniform in the log of the amplitude parameter, and for the 1000-2000~MHz data from all PTAs using a prior uniform in the amplitude (blue line).  We use the latter to obtain 95\% upper limits (blue vertical lines) both in units that scale as $\nu_o^0$ (bottom-left panel) and $\nu_o^2$ (bottom-right panel).}
\label{Fig:J1643BandNoise}
\end{figure*}

In Fig. \ref{Fig:J1643BandNoise} (top-left panel)  we show the timing residuals for PSR J1643$-$1224 after subtracting the maximum likelihood timing model from our optimal model.  For clarity, both the NANOGrav 780-884~MHz data and the 1000-2000~MHz data from all PTAs have been time-averaged over ten-day intervals.  In the top-right panel we show the same data after scaling the residuals by a factor $K\nu_o^2$, with $K$ as defined in Eq. \ref{Eq:DMScale}. The lower frequency PPTA and NANOGrav data show significantly more structure than the 1000-2000~MHz data, even after scaling by $K\nu_o^2$.  However, both the PPTA and NANOGrav data show coherent structure, indicating this is not simply an instrumental effect.

In the centre panels of Fig. \ref{Fig:J1643BandNoise} we show the DM-subtracted residuals, unscaled (centre left) and scaled by a factor $K\nu_o^2$ (centre right).  Both the PPTA and NANOGrav data clearly track each other across the time period for which both PTAs are present.  We stress that at no stage have we enforced any prior on the coherency of these two signals in the data.  The difference in the amplitude of the signals is also apparent by eye, with the lower frequency PPTA data showing larger fluctuations compared to the NANOGrav points.

We compare models that either assume the same amplitude in both bands, or allow the amplitude to take a different value in each band.  We also compare models that assume the same spectral exponent in each band, or that allow this parameter to vary between bands.  As for PSR J0437$-$4715 for models where the amplitude is described by a single value in both bands, we consider cases where this amplitude scales as $\nu_o^0$ as for a spin-noise process, and as $\nu_o^2$ for DM variations.  We find that the difference in $\log$ evidence for a model with separate power-law amplitudes and exponents in both the PPTA and NANOGrav low-frequency bands, compared to a model that fits a single coherent noise process with one amplitude and spectral exponent with a $\nu_o^0$ scaling, is 4.6, indicating that the data support the use of the more complex model in this case.  However, when assuming a $\nu_o^2$ scaling, the $\log$ evidence is 1.6 greater for the simpler two-parameter coherent-noise model.

That the evidence supports an additional coherent DM term could suggest that a power-law model for the DM variations is insufficient, and we are seeing residuals from that fit in our analysis.  If this is the case we would expect to see the same effect in the 1000-2000~MHz band, provided the data are sensitive enough for the signal to be observed in that band.

\begin{figure*}
\begin{center}$
\begin{array}{cc}
\hspace{-1.1cm}
\includegraphics[trim = 50 30 400 50, clip, width=80mm]{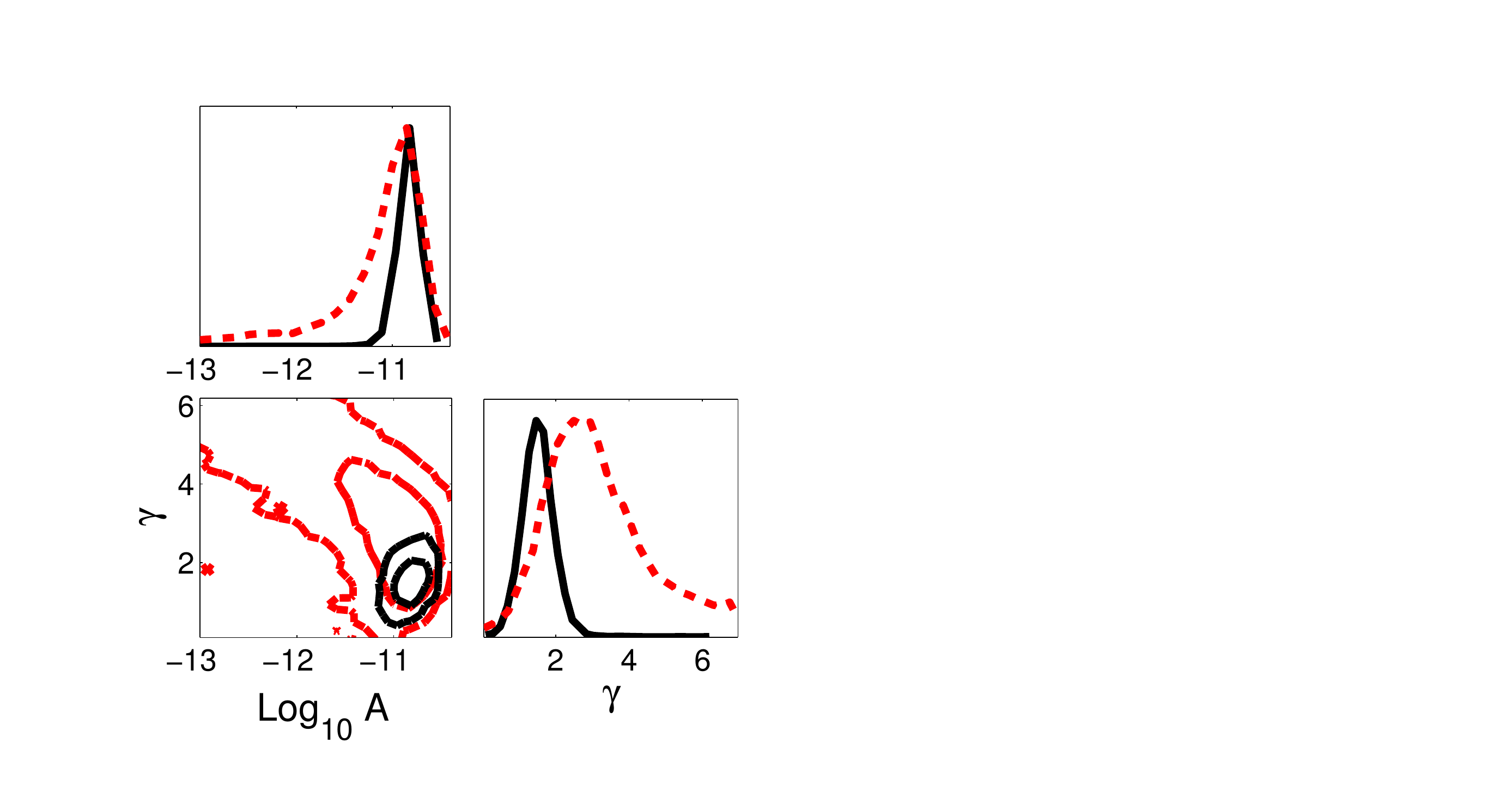} &
\includegraphics[trim = 0 20 0 40, clip, width=90mm]{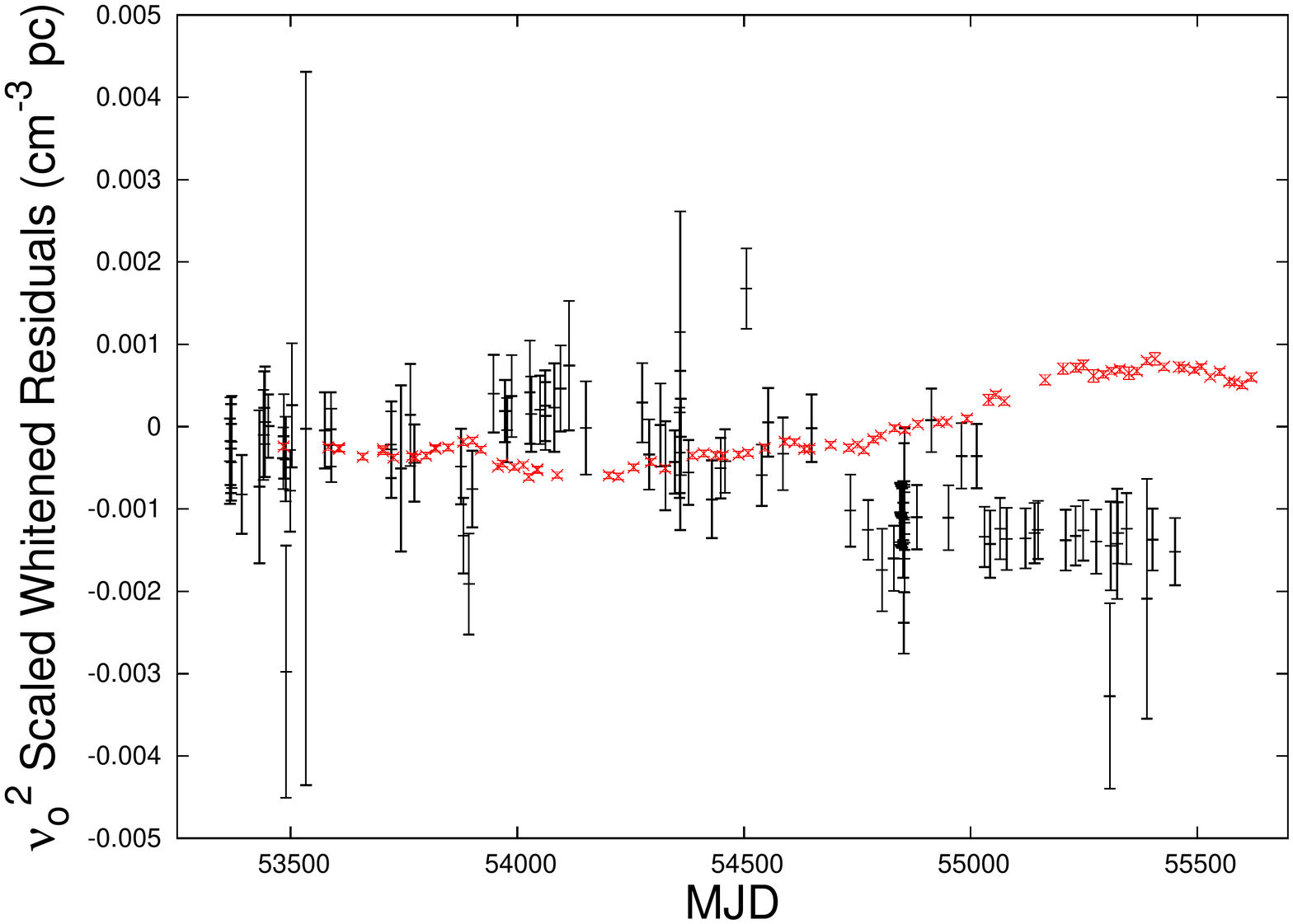} \\
\hspace{-1.1cm}
\includegraphics[ width=90mm]{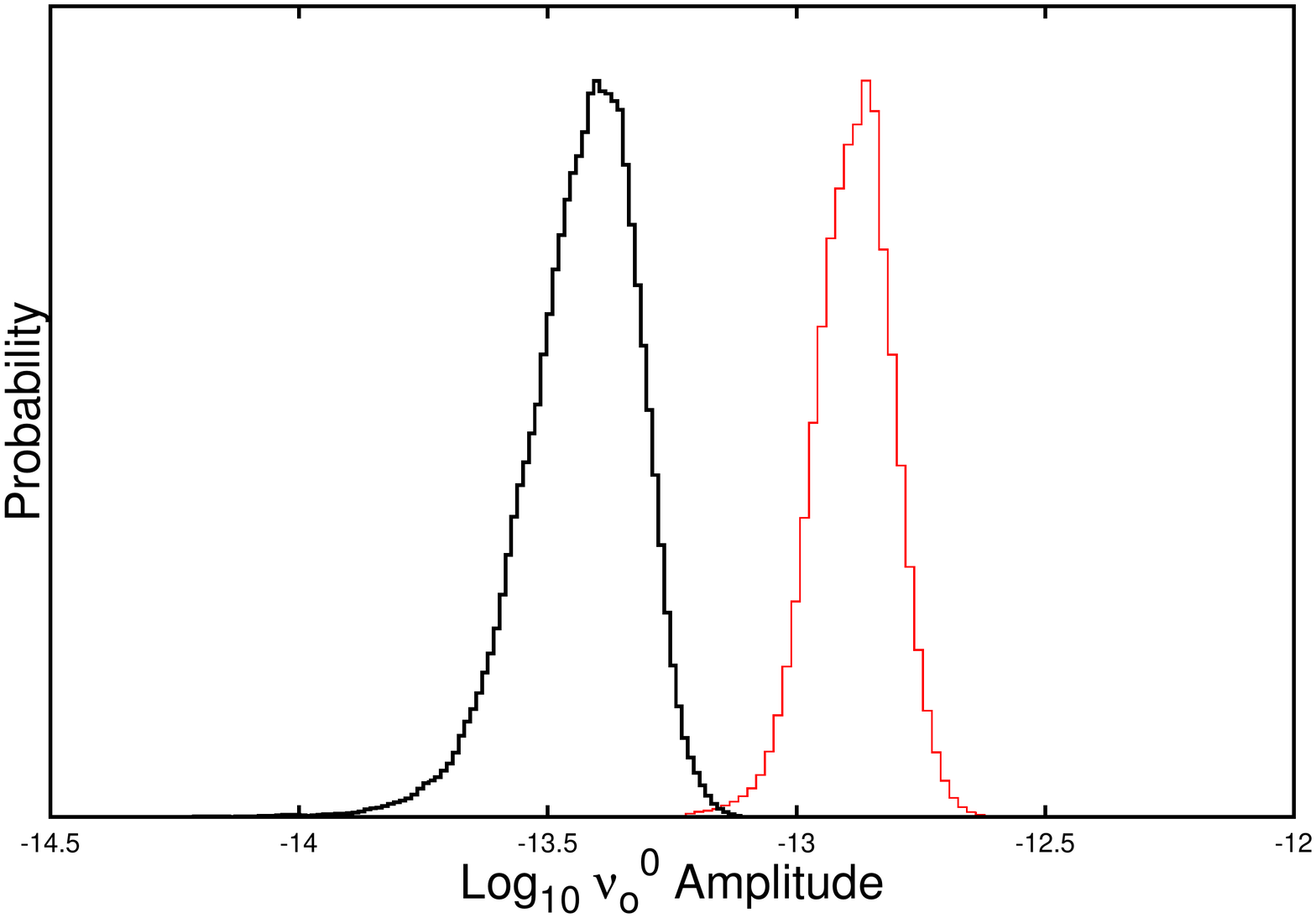} &
\includegraphics[width=90mm]{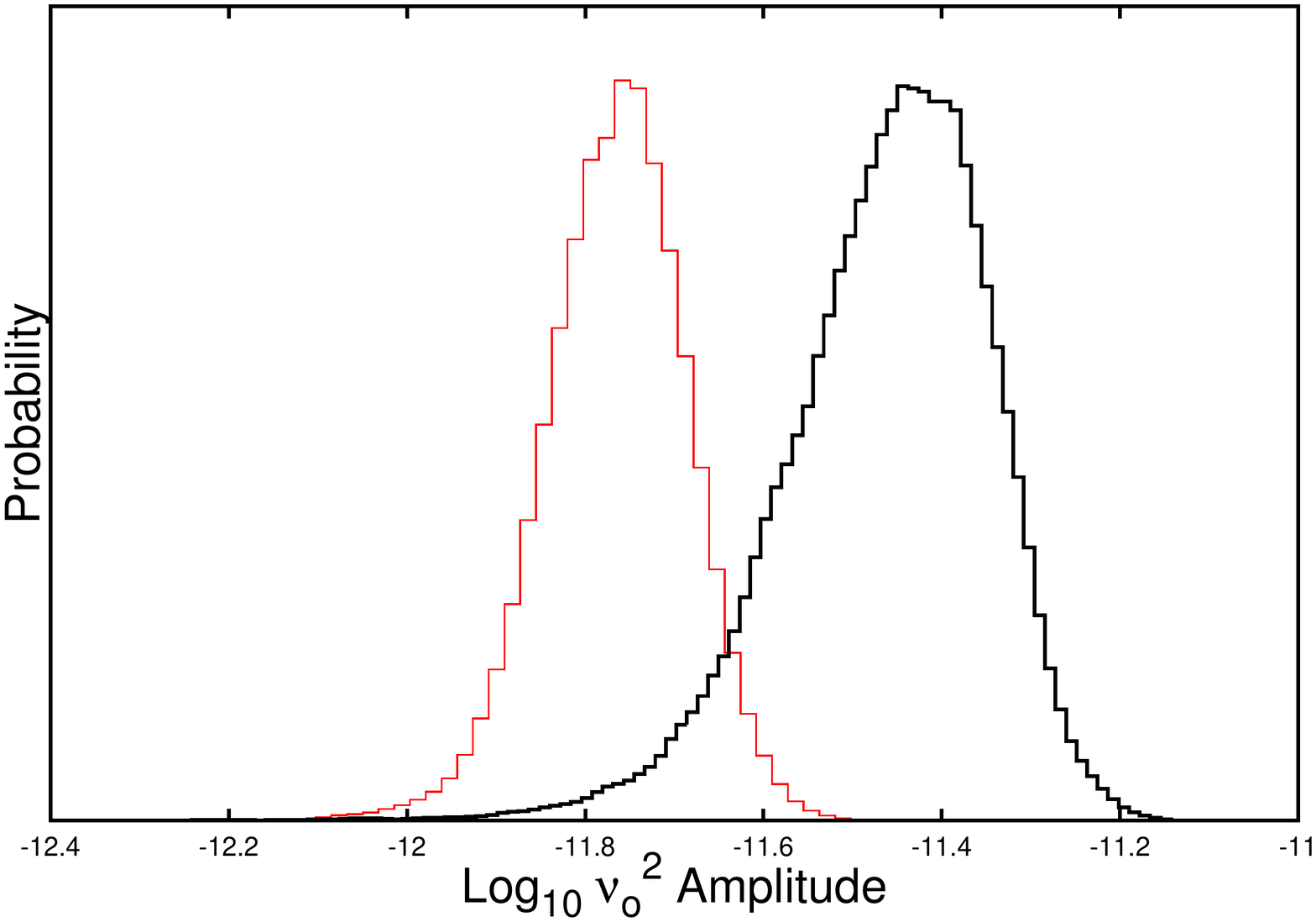} \\
\end{array}$
\end{center}
\caption{(Top Left) One- and two-dimensional marginalised posterior distributions for the amplitude and spectral exponent of the system-noise terms present in the Parkes 680-740~MHz CPSR2 system (red dashed lines) and PDFB3/APSR combined system group (black solid lines), obtained simultaneously with the additional model parameters included in the optimal model for PSR J1939+2134.  We find the parameter estimates for both terms are consistent, and that the evidence supports a single noise process across all Parkes 680-740~MHz data.
(Top right ) A subset of the timing residuals, scaled by $K\nu_o^2$, with $K$ as defined in Eq. \ref{Eq:DMScale}  for PSR J1939+2134 after subtracting the maximum likelihood timing model, as well as the maximum-likelihood DM variations and spin-noise signal realisations from our optimal model.  Colours represent: 680-740~MHz PPTA (red points)  and $2000-2200$~MHz data from the EPTA (black points).  While both frequency bands show a similar amount of structure, the two signals are not coherent across the time period where they overlap.
(Bottom panels) One-dimensional posterior parameter estimates for the amplitude of the power-law noise processes obtained for the 680-740~MHz band (red lines), and 2.0-2.5~GHz bands  (black lines).  Posteriors are shown when fitting for a single spectral exponent across all band-noise terms, but allowing the amplitude in each band to vary.  We perform the analysis in units that scale as $\nu_o^0$ (bottom-left panel) and as $\nu_o^2$ (bottom-right panel). }
\label{Fig:J1939BandNoise}
\end{figure*}

We quantify this possibility in the bottom panels of Fig. \ref{Fig:J1643BandNoise}.  Here we show the one-dimensional posterior distributions for the amplitude of the power-law band-noise processes obtained for the PPTA 690-730~MHz data, and NANOGrav 780-884~MHz data, both using priors uniform in the log of the amplitude parameter, and additionally, for the 1000-2000~MHz data from all PTAs using a prior that is uniform in the amplitude, which we use to obtain a 95\% upper limit.  We obtain the upper limit fitting for a single spectral exponent across all these band noise terms, but allowing the amplitude in each band to vary, and perform the analysis both in units that scale as $\nu_o^0$ (bottom-left panel) and $\nu_o^2$ (bottom-right panel).  We find that the 1000-2000~MHz data rules out the possibility of a $\nu_o^0$ process that has the same characteristics as the excess lower frequency noise with a probability of $> 99\%$, while a $\nu_o^2$ process is ruled out at $> 95\%$.  It is therefore unlikely that there are simply mis-modelled $\nu_o^2$ DM variations.  Instead, it is suggestive of a noise process that originates in the IISM, but with a steeper dependence on observing frequency, such as scattering which is expected to have $\nu_o^4$  \citep{2010arXiv1010.3785C}.

As for PSR J1600$-$3053 we use the 3100~MHz PPTA data to estimate the scintillation bandwidth, and find $\Delta \nu_d$ and  $\tau_d$  to be $\sim$ 2MHz, and 80ns at 3100~MHz, corresponding to $\tau_d$ of 25~$\mu$s in the 690-730~MHz band.  In Fig \ref{Fig:J1643BandNoise} (middle left panel) we see the peak to peak fluctuations of the DM-subtracted 690-730~MHz band noise are $\sim$ 25~$\mu$s, consistent with our estimate of the amplitude of the scattering from the high-frequency data.

\subsection{PSR J1939+2134 }

The final pulsar in the IPTA data set for which the data support band noise is PSR J1939+2134.  In this case we find that the optimal model supports time-correlated noise in both the 500-1000~MHz, and 2000-2500~MHz bands, in addition to spin noise, DM noise and system noise terms.  We find that neither the 500-1000~MHz nor the 3100~MHz PPTA data provide significant constraints on the frequency dependence of the additional band noise, with $95\%$  upper limits greater than the amplitude expected for a process that scales as $\nu_o^0$ from the 500-1000~MHz band noise.

\begin{table*}
\caption{Properties of the power-law spin noise and DM noise.  We denote the integrated power in the each model as $\sigma_\mathrm{SN}$, and $\sigma_\mathrm{DM}$ for the spin-noise, and DM-noise processes respectively.}
\begin{tabular}{ccccccc}
\hline\hline
Pulsar Name & \multicolumn{3}{c}{Spin Noise} & \multicolumn{3}{c}{DM Noise} \\
	    & Log$_{10}$ A$_\mathrm{SN}$ & $\gamma_\mathrm{SN}$ & Log$_{10}~ \sigma_\mathrm{SN}$ & Log$_{10}$ A$_\mathrm{DM}$ & $\gamma_\mathrm{DM}$& Log$_{10} ~\sigma_\mathrm{DM}$ \\
\hline\hline
J0218+4232	    & 	 - 		 &  - 		  &  - 		     & -11.15  $\pm$  0.05  &  2.1  $\pm$  0.3    &  -2.89  $\pm$  0.15   \\
J0437$-$4715	    & -			 & 	-	  & 	-	     & -11.90  $\pm$  0.07  &  2.9  $\pm$  0.3    &  -3.26  $\pm$  0.13   \\
J0613$-$0200          & -14.4 $\pm$ 0.5    & 5.0 $\pm$ 1.0  & -5.88 $\pm$ 0.17 & -11.72  $\pm$  0.04  &  1.8  $\pm$  0.2    &  -3.62  $\pm$  0.11	  \\
J0621+1002          & -11.91 $\pm$ 0.08  & 1.9 $\pm$ 0.3  & -4.81 $\pm$ 0.14 & 	 - 		    &  - 		  &  -	\\
J0711$-$6830	    & 	 - 		 &  - 		  &  - 		     &-12.3  $\pm$  0.5     &  3.7  $\pm$  1.2    &  -3.1  $\pm$  0.3     \\
J1012+5307          & -13.18 $\pm$ 0.09  & 1.5 $\pm$ 0.3  & -6.23 $\pm$ 0.09 & 	 - 		    &  - 		  &  -	\\
J1022+1001	    & 	 - 		 &  -		  &  - 		     & -11.53  $\pm$  0.05  &  0.9  $\pm$  0.3    &  -3.67  $\pm$  0.08   \\
J1024$-$0719          & -13.9 $\pm$ 0.2    &  5.4 $\pm$ 0.6 & -4.82 $\pm$ 0.16 & 	 - 		    &  - 		  &  -	\\
J1045$-$4509	    & 	 - 		 &  - 		  &  - 		     & -10.72  $\pm$  0.05  &  3.1  $\pm$  0.2    &  -1.90  $\pm$  0.13   \\
J1600$-$3053	    & 	 - 		 &  - 		  &  - 		     & -11.57  $\pm$  0.05  &  1.8  $\pm$  0.2    &  -3.52  $\pm$  0.11   \\
J1603$-$7202	    & 	 - 		 &  - 		  &  - 		     &-12.4  $\pm$  0.5     &  4.4 $\pm$  1.1     &  -2.9  $\pm$  0.2     \\
J1643$-$1224	    & 	 - 		 &  - 		  &  - 		     &-11.40  $\pm$  0.16   &  3.3  $\pm$  0.5    &  -2.42  $\pm$  0.19   \\
J1713+0747          & -14.0 $\pm$ 0.2    & 3.1 $\pm$ 0.6  & -6.14 $\pm$ 0.17 & -12.01  $\pm$  0.04  &  1.7  $\pm$  0.2    &  -3.90  $\pm$  0.12	  \\
J1732$-$5049	    & 	 - 		 &  - 		  &  - 		     & -11.7  $\pm$  0.5    &  3.9  $\pm$  1.3    &  -2.9  $\pm$  0.2     \\
J1824$-$2452A         & -12.73 $\pm$ 0.19  & 3.0 $\pm$ 1.0  & -5.6 $\pm$ 0.2   & -10.80  $\pm$  0.07  &  2.7  $\pm$  0.4    &  -2.59  $\pm$  0.16   \\
J1857+0943	    & 	 - 		 &  - 		  &  - 		     & -11.78  $\pm$  0.06  &  2.62  $\pm$  0.16  &  -3.07  $\pm$  0.08   \\
J1909$-$3744	    & 	 - 		 &  - 		  &  -		     & -12.14  $\pm$  0.03  &  1.69  $\pm$  0.17  &  -4.13  $\pm$  0.07   \\
J1939+2134          & -14.2 $\pm$ 0.2    &  6.0 $\pm$ 0.5 & -4.09 $\pm$ 0.16 & -11.35  $\pm$  0.03  &  2.73  $\pm$  0.12  &  -2.56  $\pm$  0.09	  \\
J2145$-$0750          & -12.98 $\pm$ 0.05  & 0.6 $\pm$ 0.2  & -6.18 $\pm$ 0.05 & -12.1  $\pm$  0.4    &  4.4  $\pm$  0.9    &  -2.52  $\pm$  0.16   \\
J2317+1439	    & 	 - 		 &  - 		  &  - 		     & -11.76  $\pm$  0.09  &  3.0  $\pm$  0.5    &  -3.1  $\pm$  0.2     \\
\hline
\end{tabular}
\label{Table:NoiseParams}

\end{table*}

In order to check whether this additional noise is consistent over all data in these bands, or is simply present in a single observing system, we fit two independent power-law noise processes simultaneously to separate Parkes systems that observe in the 680-740~MHz band.  In particular we take i) the CPSR2 data which extends from the year 2005 until 2010, and ii) data from both the APSR and PDFB3 systems which were in use from the year 2010 onwards.  As in the previous examples we perform this analysis jointly with the rest of the IPTA PSR J1939+2134 data set, including the spin noise, DM noise, system noise, and  2000-2500~MHz band noise from the optimal model. The one- and two-dimensional marginalised posterior distributions for the amplitude and spectral exponent for these two noise processes are shown in Fig. \ref{Fig:J1939BandNoise} (top-left) for the CPSR2 and APSR/PDFB3 groups. We find that there is significant excess noise detected in both systems, and that the parameter estimates are consistent with one another, with the evidence supporting a single-band process compared to separate system-noise processes.  We also check the 820-860~MHz WSRT data for the excess noise, however we find no evidence for such a process, but obtain an upper limit that is consistent with the levels observed in the Parkes data.

We perform a similar test for the 2000-2500~MHz band, dividing the band into two and fitting separate power-law processes to the 2.0-2.3~GHz EPTA data,  and the 2.4~GHz data from \cite{1994ApJ...428..713K}.  We find that, while there is still significant evidence for excess noise in the 2.0-2.3~GHz EPTA data, there is only minimal evidence ($\log \mathcal{Z} \sim 1$) for an additional noise process in the 2.4~GHz data.  The peak in the posterior however  is consistent with the amplitude observed in the EPTA 2.0-2.3~GHz data.

In Fig. \ref{Fig:J1939BandNoise} (top-right) we show the  $K\nu_o^2$ scaled residuals for a subset of the PSR J1939+2134 data set which has overlapping data from both the 680-740~MHz PPTA and 2.0-2.2~GHz EPTA data, after subtracting both the maximum-likelihood DM variations and spin-noise signal realisations.  While both show a similar degree of structure, they are clearly not coherent across the MJD range where they overlap, and we find the evidence does not support a coherent noise process with either $\nu_o^0$ or $\nu_o^2$ scaling.

Finally, we test models that enforce a single spectral exponent or amplitude for the different band-noise terms assuming either a $\nu_o^0$ or $\nu_o^2$ scaling of the amplitude between the different bands.  We find that in both cases a single spectral exponent is supported by the data.  In Fig. \ref{Fig:J1939BandNoise} (bottom panels) we show the one-dimensional posterior parameter estimates for the amplitude of the power-law noise processes obtained for the 680-740~MHz band (red lines), and 200-230~MHz band  (black lines) when fitting for a single spectral exponent across both band noise terms, but allowing the amplitude in each band to vary.  We find that the evidence does not support a single amplitude in the $\nu_o^0$ scaling case (bottom-left panel) with $>$ 99\% probability, however we find a single amplitude is sufficient to describe the data when assuming a  $\nu_o^2$ scaling between the bands (bottom-right panel).

In summary, we find that the evidence supports  incoherent noise processes in the two bands (i.e., the time-domain signals are not consistent with one another), in addition to the other noise processes mentioned previously.  The spectral exponents for these two band terms are however consistent, and further more we find the power-spectrum amplitudes scale with $\nu_o^2$.   As for J0437$-$4715,  this could therefore be indicative of different sampling of the IISM by the different wavelengths emitted by the pulsar.  This is consistent with the observations made at  both 1400~MHz and 2400~MHz presented in \cite{1990ApJ...349..245C} and later in \cite{2006ApJ...645..303R}.  As before, however, this could also result from telescope-dependent effects.  Without observations that overlap in time from different telescopes at the same frequency, it is not possible to differentiate between these possibilities.

\section{DM Variations}
\label{Section:DMEvents}

\begin{figure}
\begin{center}$
\begin{array}{c}
\hspace{-1.0cm}
\includegraphics[width=90mm]{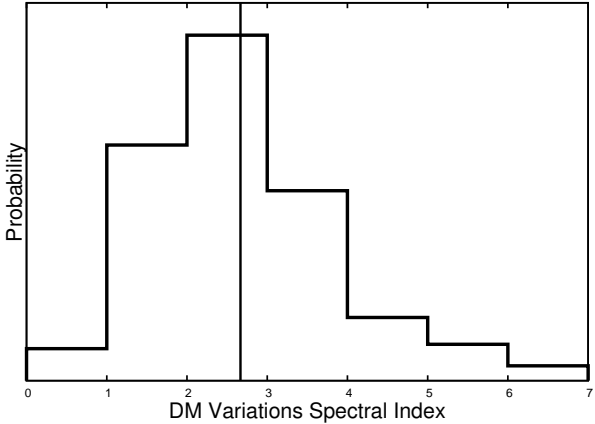} \\
\end{array}$
\end{center}
\caption{Sum of the one-dimensional marginalised posterior distributions for the spectral exponents of the power-law DM-noise for the 17 pulsars for which the data support this term in their optimal model.  The vertical line shows the spectral exponent expected for Kolmogorov turbulence.}
\label{Fig:DMHist}
\end{figure}

We find that 17 pulsars from the IPTA data set support power-law DM noise that can be clearly distinguished from spin noise and other system- or band-dependent effects.  We list the mean log amplitudes, spectral exponents and the total integrated power for these models in Table \ref{Table:NoiseParams}, along with their 1-$\sigma$ confidence intervals.  In Fig.  \ref{Fig:DMHist} we show a histogram of the sum of the 17 one-dimensional marginalised posteriors for the spectral exponent of the power-law DM noise model.  We find that, while there is non-zero probability for all spectral exponent bins, the peak is consistent with a value of $\gamma_\mathrm{DM} = 8/3$, as expected from Kolmogorov turbulence.

\begin{figure*}
\begin{center}$
\begin{array}{cc}
\hspace{-1cm}
\includegraphics[trim = 60 50 30 65, clip,width=90mm]{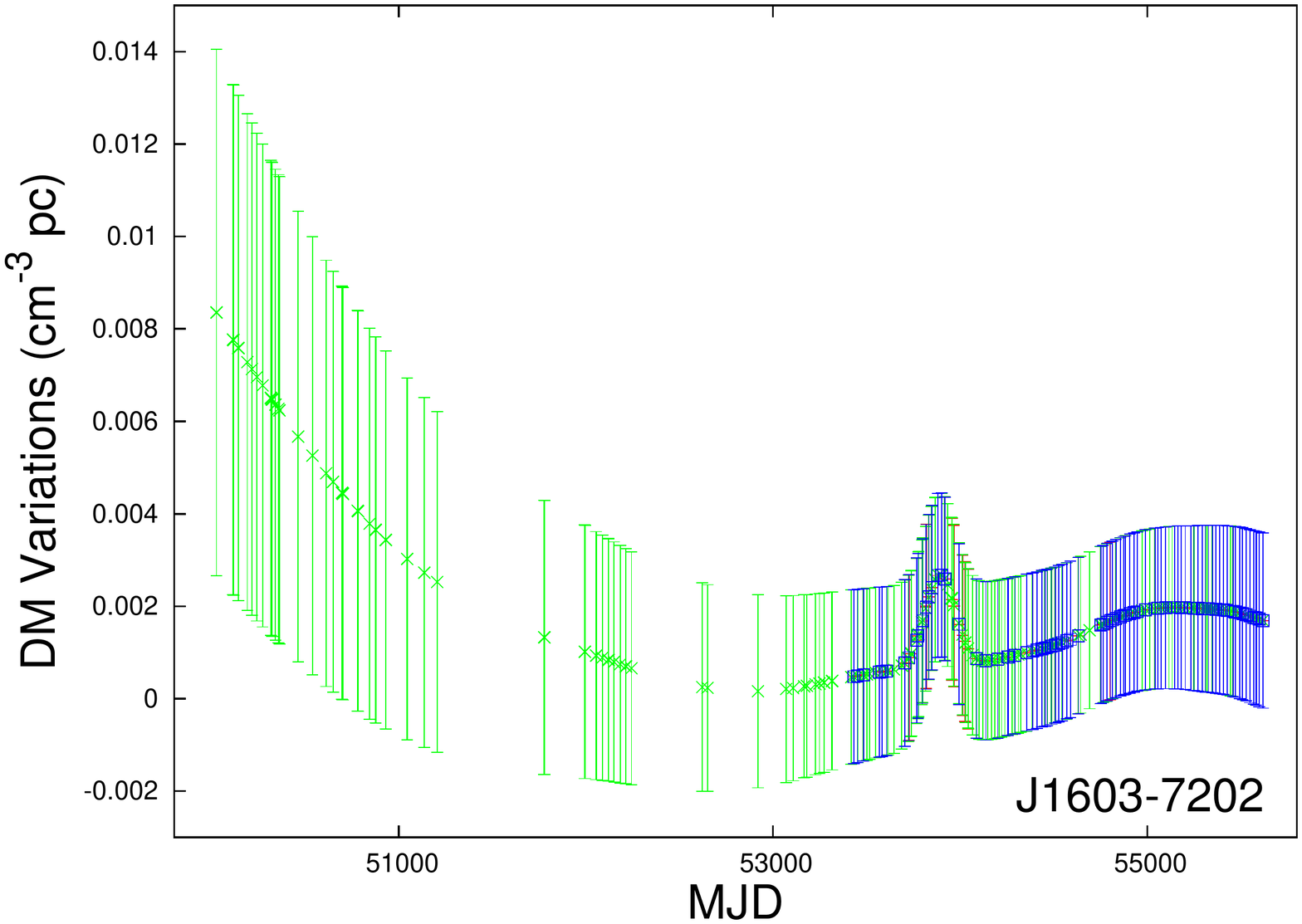}&
\hspace{0cm}
\includegraphics[trim = 60 50 30 65, clip,width=90mm]{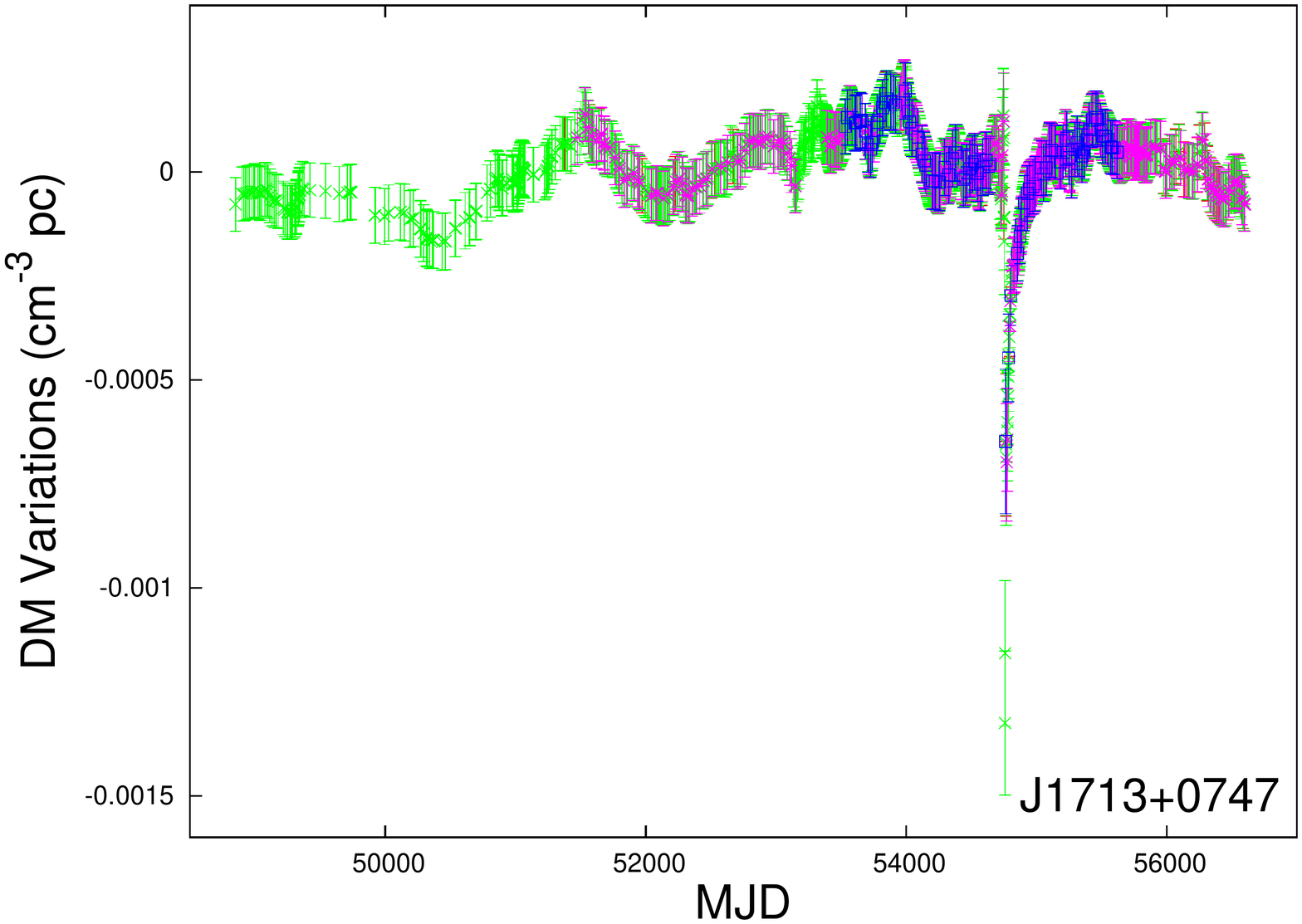} \\
\hspace{-1cm}
\includegraphics[width=100mm]{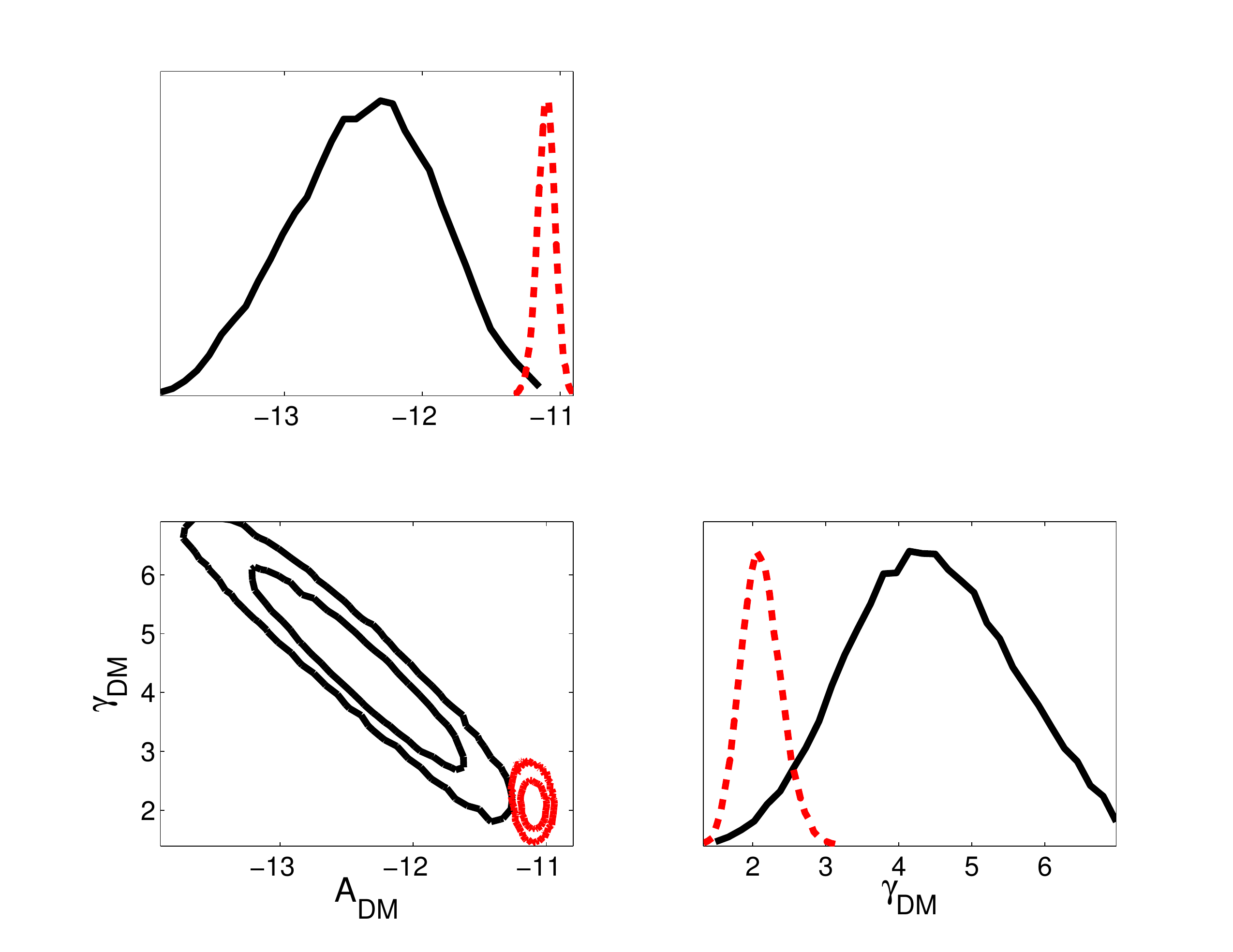}&
\hspace{-1cm}
\includegraphics[width=100mm]{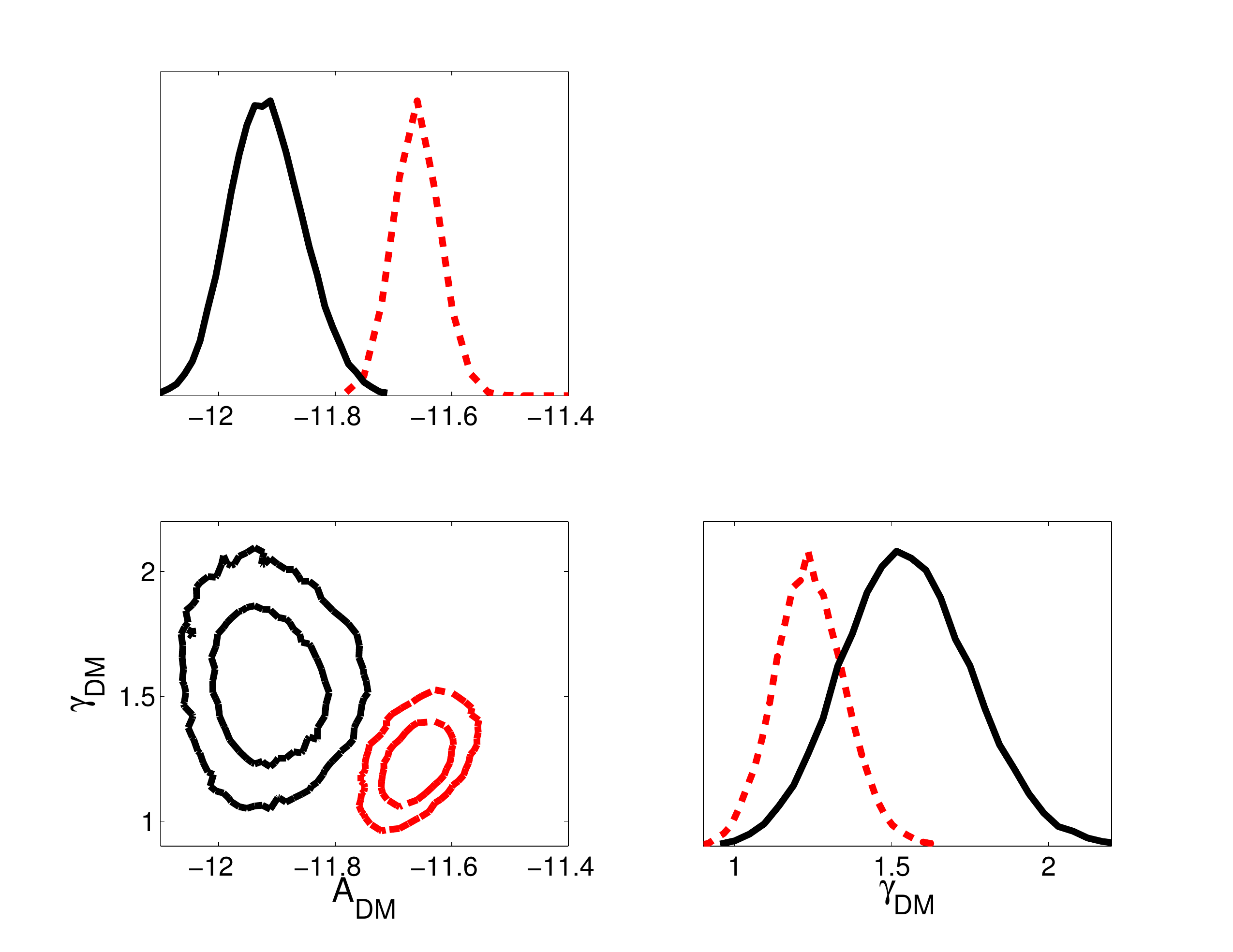} \\
\end{array}$
\end{center}
\caption{Maximum-likelihood signal realisations for the power-law DM noise model with the inclusion of a non-stationary DM event for PSRs J1603$-$7202 (top left) and  J1713+0747 (top right).  Colours represent the ToA observation frequency: $>$ 3~GHz (blue), 2-3~GHz (magenta), 1-2~GHz (green) and $<$ 1~GHz (red).  The properties of the time-stationary power-law component of the DM variations change significantly when including the DM-event model for both pulsars. The bottom two plots show the one- and two-dimensional marginalised posteriors for the amplitude and spectral exponent of the power-law DM variations when including the non-stationary DM event (black solid lines) or not (dashed red lines) for PSRs J1603$-$7202 (bottom left) and  J1713+0747 (bottom right).  In both cases when including the DM event the power-law component of the DM variations is consistent with smaller amplitudes and steeper spectral exponents.}
\label{Fig:J603DMEvent}
\end{figure*}

\begin{figure}
\begin{center}$
\begin{array}{c}
\hspace{-1cm}
\includegraphics[width=100mm]{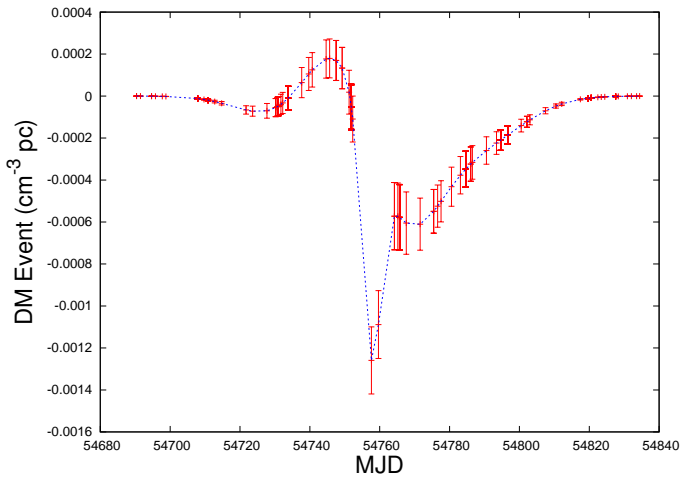} \\
\end{array}$
\end{center}
\caption{Maximum-likelihood model and 1-$\sigma$ confidence intervals for the PSR J1713+0747 DM event.  We find that the event model supports a shapelet model with five components and extends across an interval of $\sim 100$ days with a maximum decrease in DM of $\sim 1.3\times10^{-3}$~cm$^{-3}$~pc.}
\label{Fig:J1713DMEvent}
\end{figure}

We find that only for PSR J1603$-$7202 and PSR J1713+0747 do the data support the inclusion of non-stationary DM events in their stochastic model. In Fig. \ref{Fig:J603DMEvent} we show the maximum-likelihood DM signal realisations for these two pulsars when including the shapelet model for the DM event.  In PSR J1603$-$7202 the event corresponds to an increase in the electron density along the line of sight.  We find that only the lowest-order shapelet model is supported by the data, corresponding to a simple Gaussian model for the DM event, with centroid at MJD $53890\pm60$, full width at half maximum (FWHM) of $190\pm50$ days, and amplitude $3\pm0.5\times 10^{-3}$~cm$^{-3}$~pc, consistent with values previously reported in  \cite{2013MNRAS.429.2161K}.  In  PSR J1713+0747 we find a more complex model for the DM event is supported by the data, with five shapelet components.    We show the maximum likelihood model and 1-$\sigma$ confidence intervals for the PSR J1713+0747 DM event in Fig. \ref{Fig:J1713DMEvent}.  In the EPTA data alone only a single Gaussian component was found to be supported by the data for the DM event (Desvignes et al. submitted).  In the IPTA data set the additional NANOGrav and PPTA data improve both the constraints placed on the stationary component of the DM variations, and also improve the sampling of the DM event itself, warranting additional components in the model.  We find the event extends across a period of $\sim 100$ days with a maximum decrease in the DM over this period of $\sim 1.3\times10^{-3}$~cm$^{-3}$~pc at $\sim$ MJD~54757(see e.g. \cite{2015arXiv150607948C}, Desvignes et al. submitted).

In the bottom two panels of Fig. \ref{Fig:J603DMEvent} we compare the one- and two-dimensional posteriors for the amplitude and spectral exponent of the power-law DM noise model for both PSRs J1603$-$7202 (bottom left) and J1713+0747 (bottom right) when including, or not, the non stationary DM-event model.  In both cases we find that not including the model for the non-stationarity in the DM leads to a significantly higher estimate of the amplitude, with a flatter spectrum.

\begin{figure}
\begin{center}$
\begin{array}{c}
\hspace{-0.5cm}
\includegraphics[width=90mm]{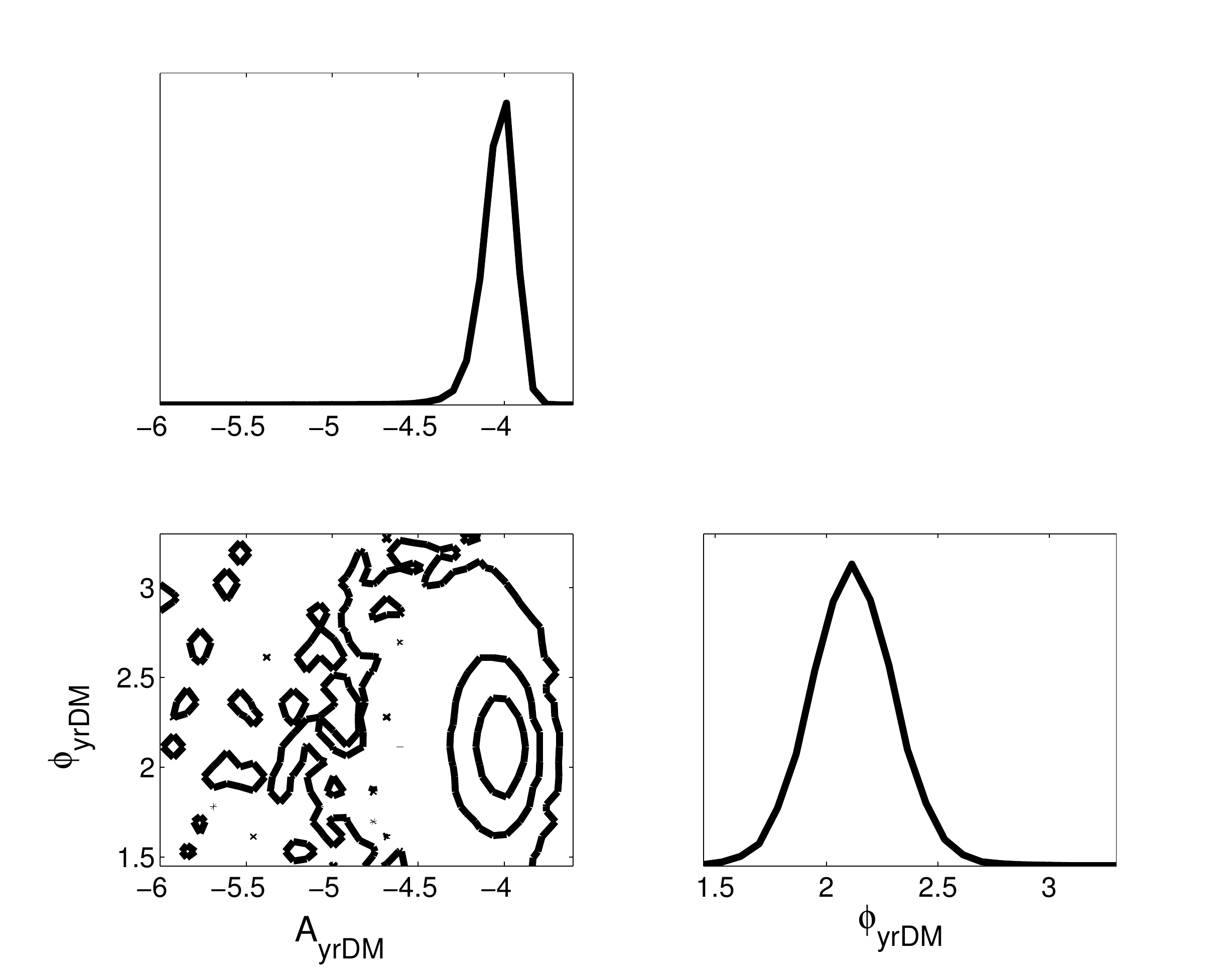} \\
\hspace{-1.0cm}
\includegraphics[width=90mm]{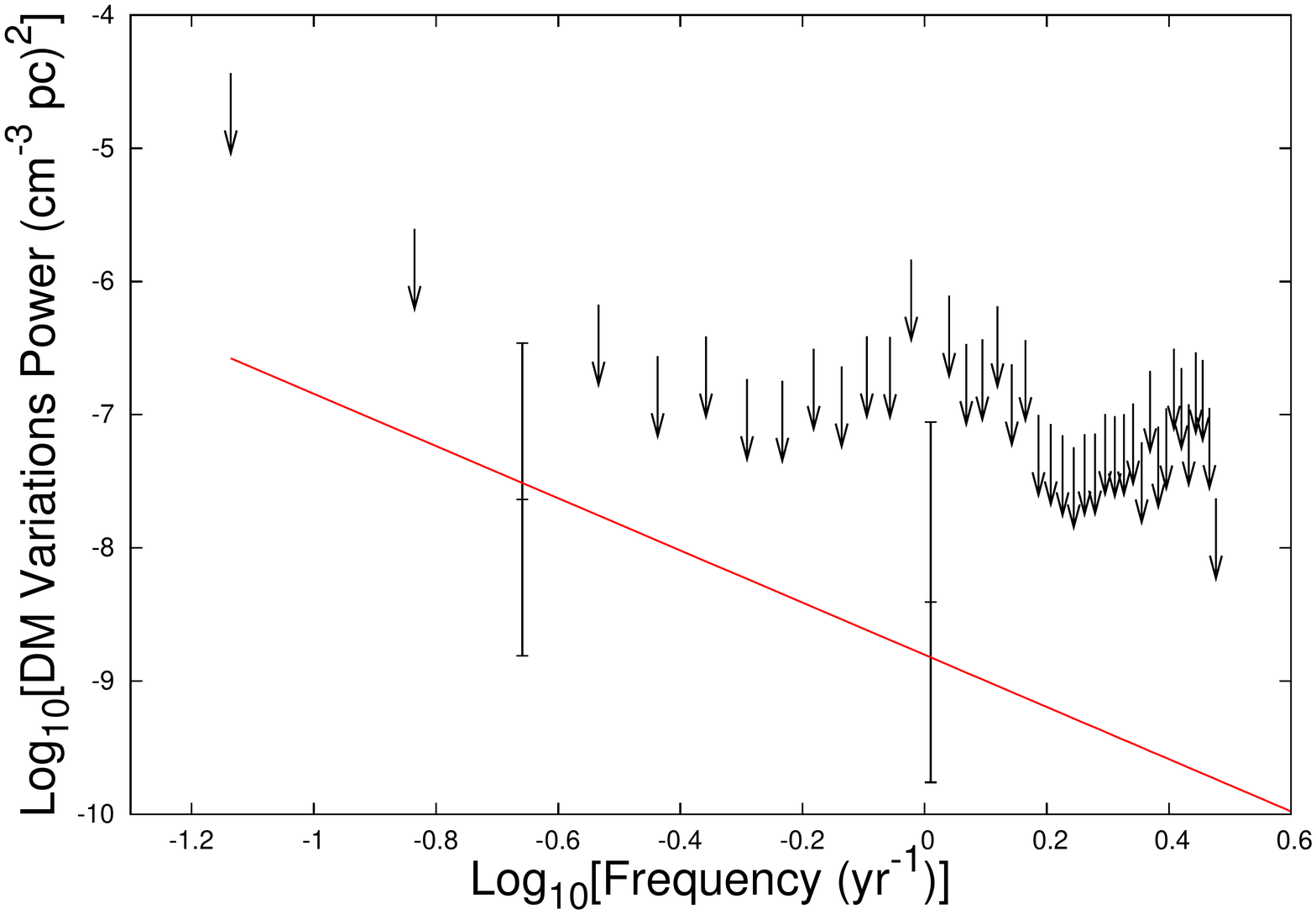} \\
\end{array}$
\end{center}
\caption{(Top) One- and two-dimensional marginalised posterior distributions for the log amplitude and phase of the yearly DM model solved for simultaneously with the optimal model  which includes power-law spin noise and DM noise, and additional system noise for the Nan{\c c}ay 1400~MHz data.  (Bottom) 95\% upper limits (arrows), and significant detections (points with 1-$\sigma$ error bars) for the power spectrum of the DM noise in PSR J0613-0200.  The straight line indicates the maximum-likelihood DM power law from the optimal model.   While there is a clear detection of power at a frequency of 1~yr$^{-1}$, a simple power-law model is still sufficient to describe the data.  This is reflected in the evidence, which increased by only $\sim 1$ compared to the optimal model when including the yearly variations.}
\label{Fig:J0613YrDM}
\end{figure}

We find that no pulsars support the addition of a yearly variation in DM in the IPTA data sets we have analysed.  This seemingly contradicts results presented in \cite{2013MNRAS.429.2161K} which observed significant yearly variations in PSR J0613$-$0200 in particular.  In Fig. \ref{Fig:J0613YrDM} (top) we show the one- and two-dimensional posteriors for the log amplitude and phase of the yearly DM model included in our analysis simultaneously with the optimal model from Table \ref{Table:evidenceValues} which includes power-law spin noise and DM noise, and additional system noise for the Nan{\c c}ay 1400~MHz data.  We find the increase in the evidence when including the yearly DM variations is $\sim 1$, which is not significant enough to warrant its
inclusion in the model.

To better explain this result, we perform an additional analysis on the PSR J0613-0200 data set where, rather than parameterising the DM noise as a power law, we allow the power at each Fourier frequency included in the model to vary as a free parameter.  In Fig. \ref{Fig:J0613YrDM} (bottom) we show the 95\% upper limits (arrows), and significant detections (points with 1-$\sigma$ uncertainties) for the power spectrum of the DM noise obtained from this analysis.  We consider significant detections in this case to be those frequencies for which less than $5\%$ of the posterior distribution is consistent with power of less than $10^{-18} (\mathrm{cm}^{-3}~\mathrm{pc})^2$, which we take to be effectively zero.   The straight line indicates the maximum likelihood DM noise power-law from the optimal analysis.   While there is a clear detection of power at a frequency of 1~yr$^{-1}$, a simple power-law model is still sufficient to describe the data.  This can be understood in terms of the differences in the DM models fitted in this work, compared to \cite{2013MNRAS.429.2161K}.  Here our stationary power-law model for the DM noise already includes power at a period of 1 yr, thus our model for yearly DM variations represents an excess with respect to the level already included in the power-law model.  In comparison, in \cite{2013MNRAS.429.2161K} a piecewise time-series model is used for the DM variations which does not explicitly include power at any periodicity.  We therefore are not claiming here that there is no power at a period of 1 year in the DM variations, simply that the power at that period is consistent with a simple power-law model for all pulsars in the IPTA data set.

\section{Spin Noise}
\label{Section:SpinNoise}

\begin{figure*}
\begin{center}$
\begin{array}{cc}
\hspace{-1cm}
\includegraphics[trim = 60 50 30 65, clip,width=80mm]{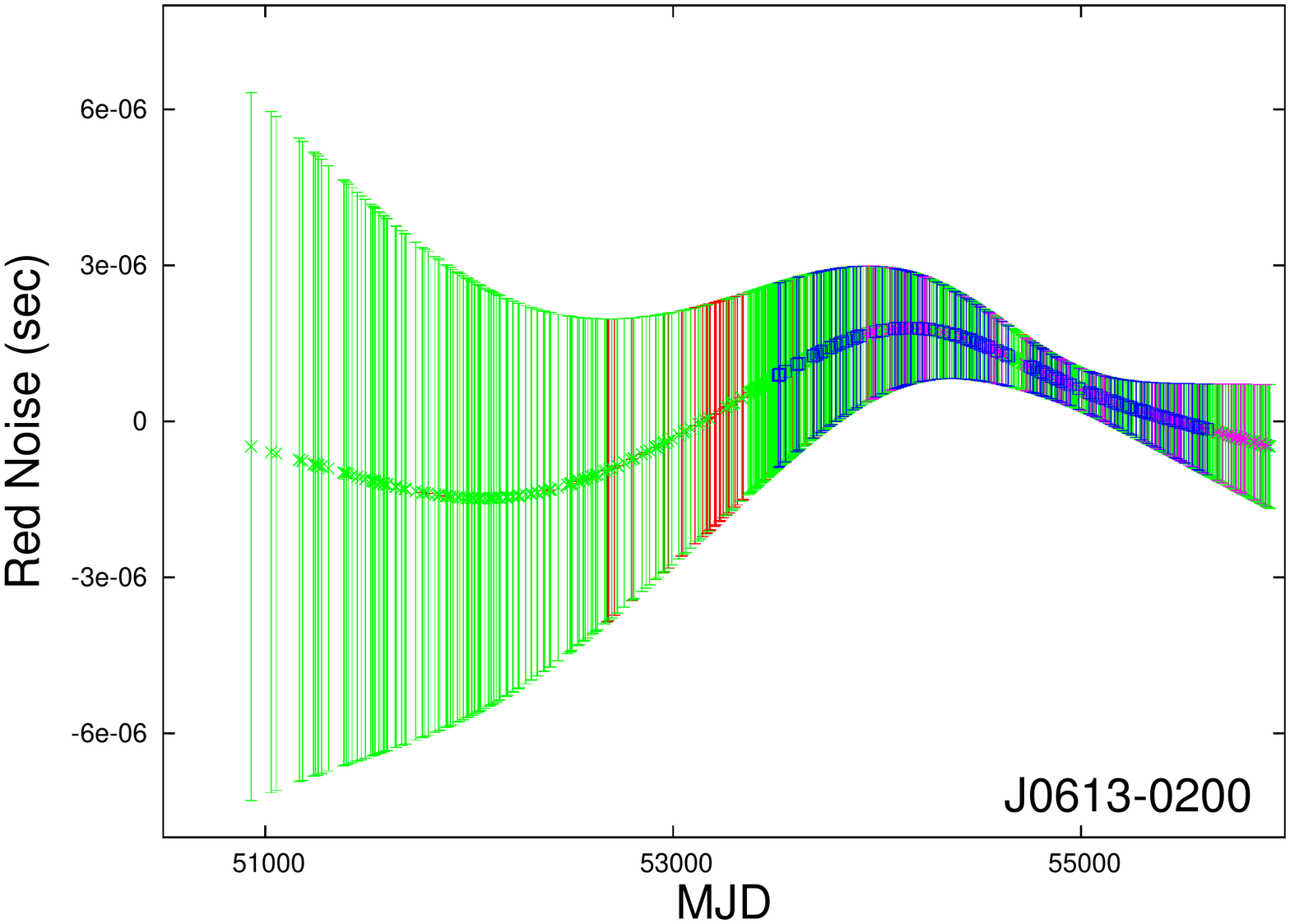}&
\hspace{0cm}
\includegraphics[trim = 60 50 30 65, clip,width=80mm]{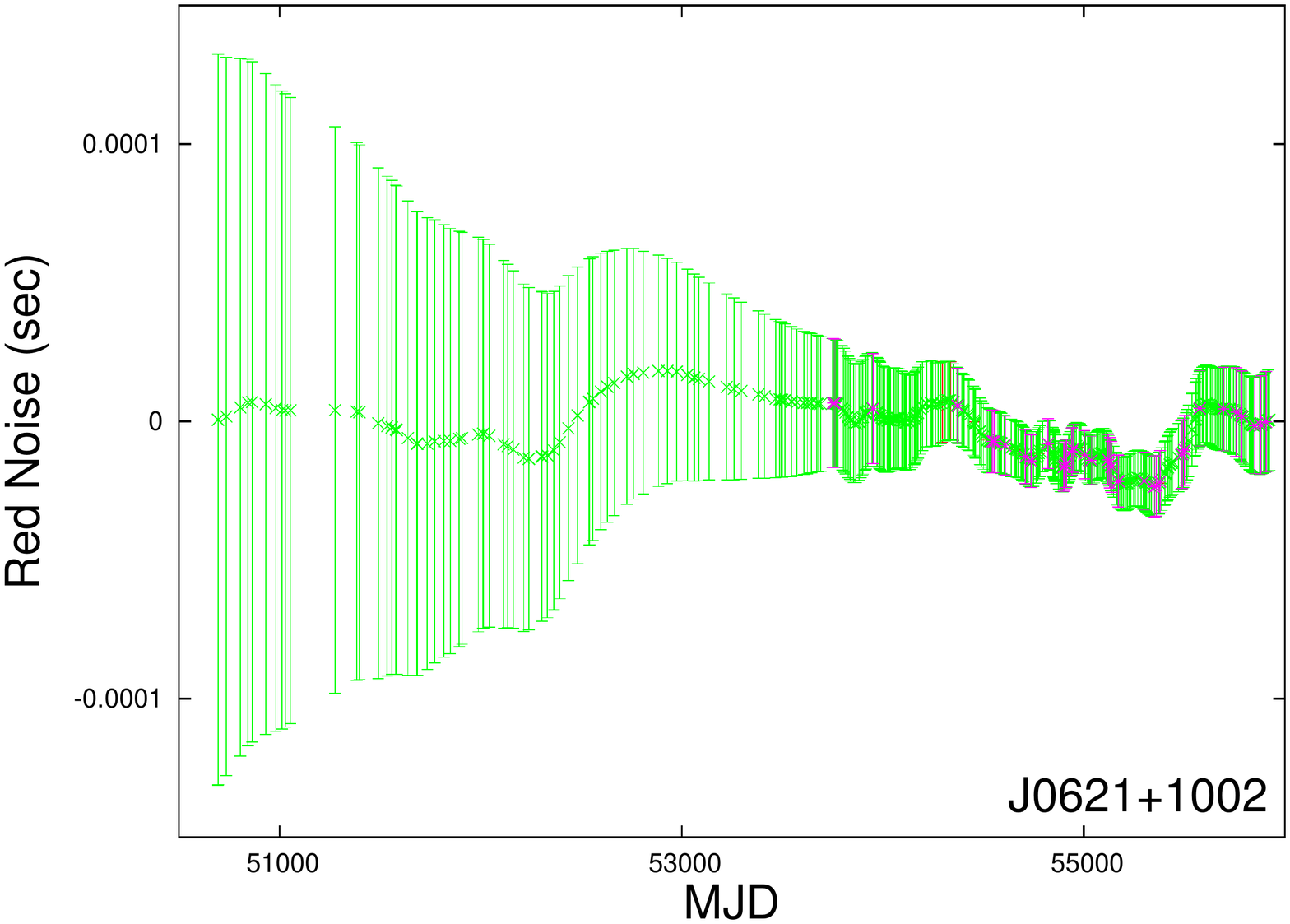} \\
\hspace{-1cm}
\includegraphics[trim = 60 50 30 65, clip,width=80mm]{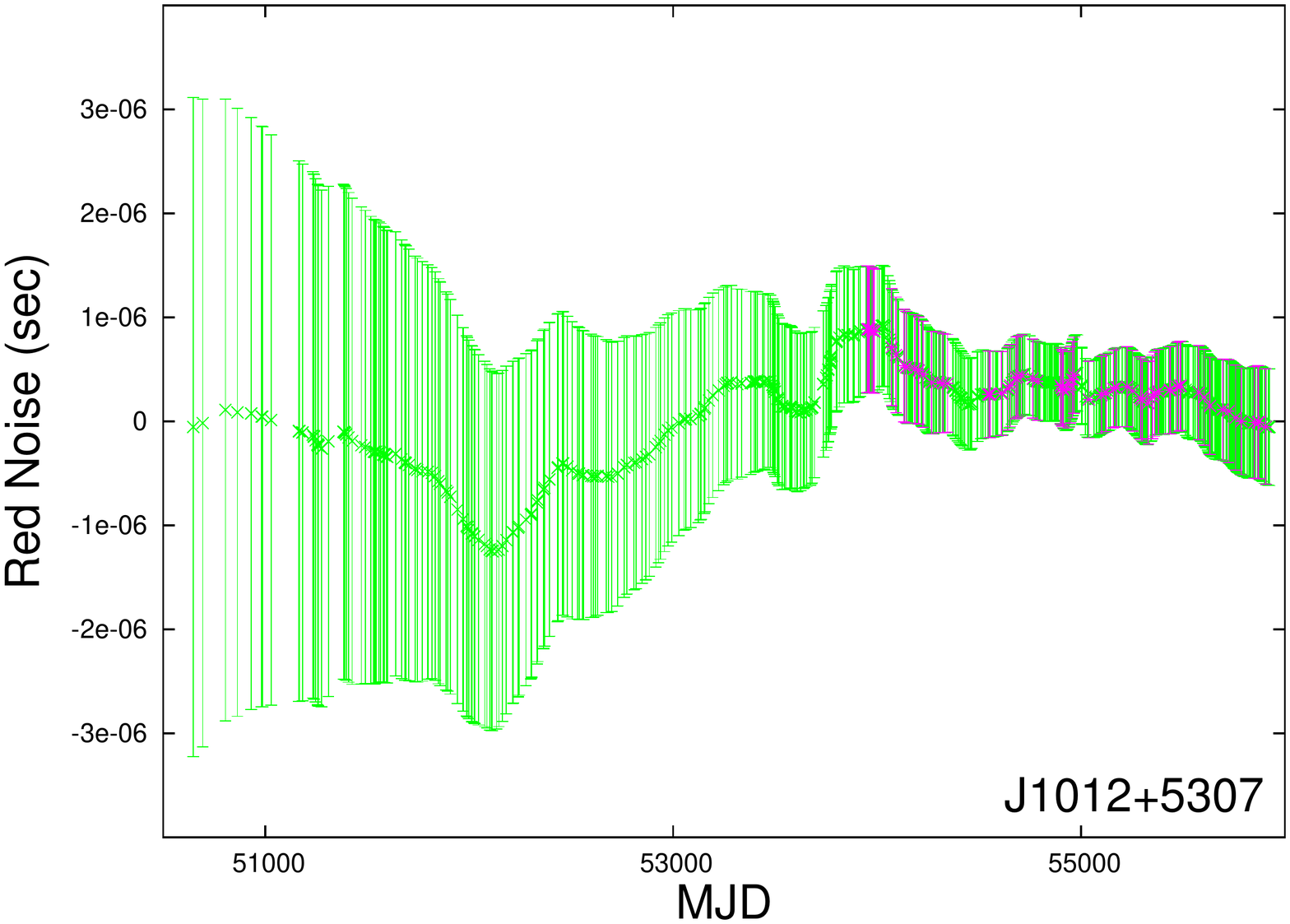}&
\hspace{0cm}
\includegraphics[trim = 60 50 30 65, clip,width=80mm]{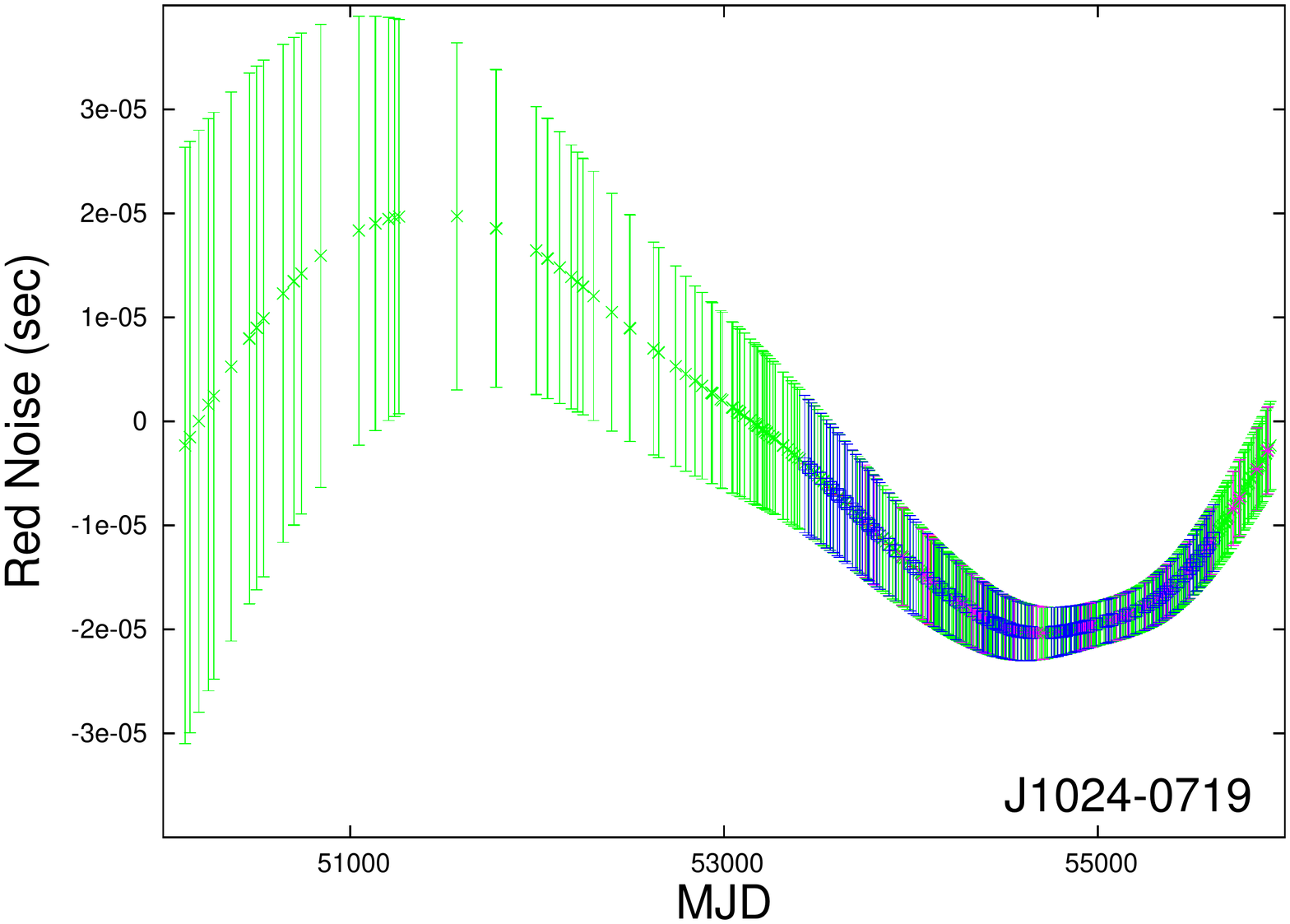} \\
\hspace{-1cm}
\includegraphics[trim = 60 50 30 65, clip,width=80mm]{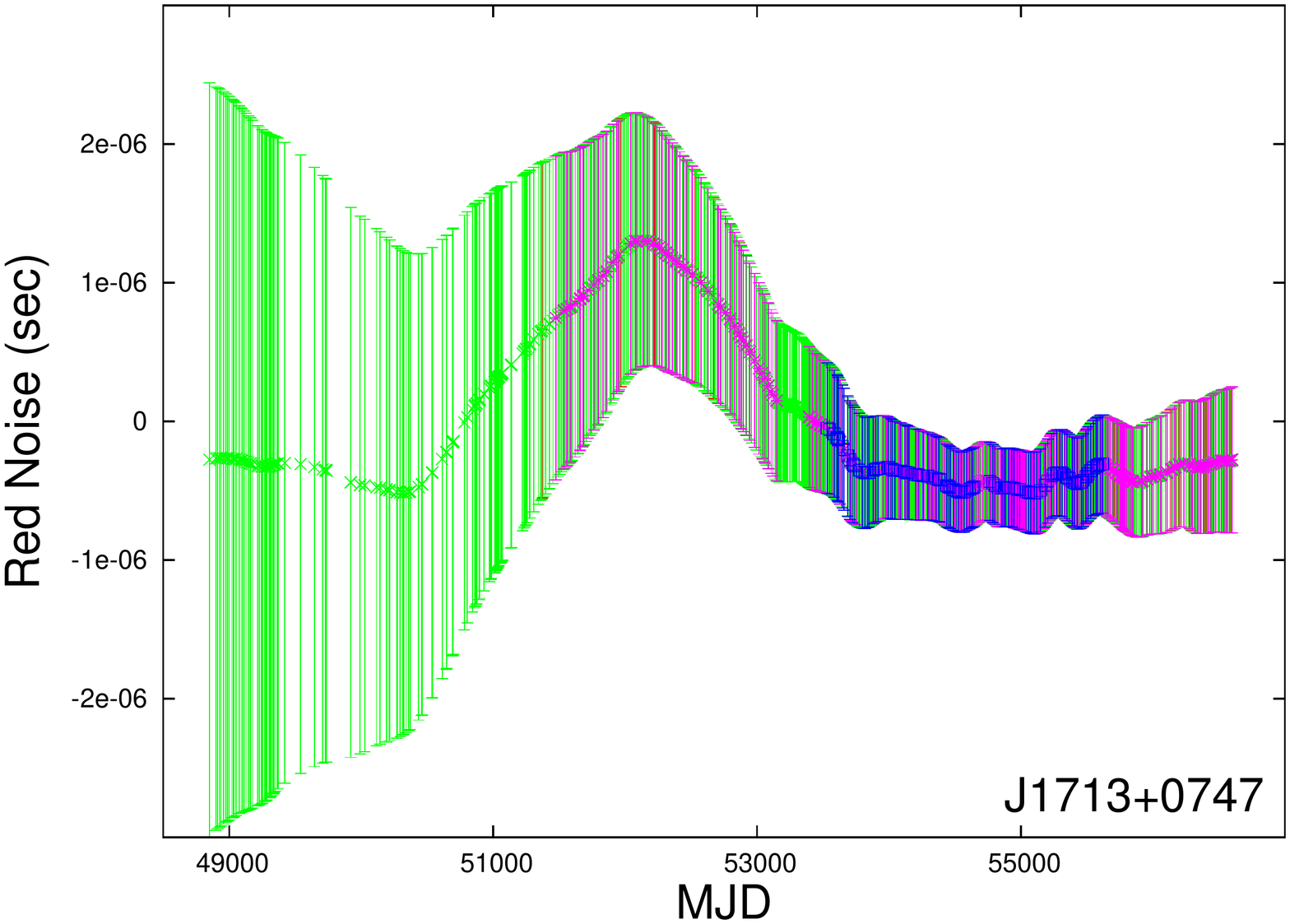}&
\hspace{0cm}
\includegraphics[trim = 60 50 30 65, clip,width=80mm]{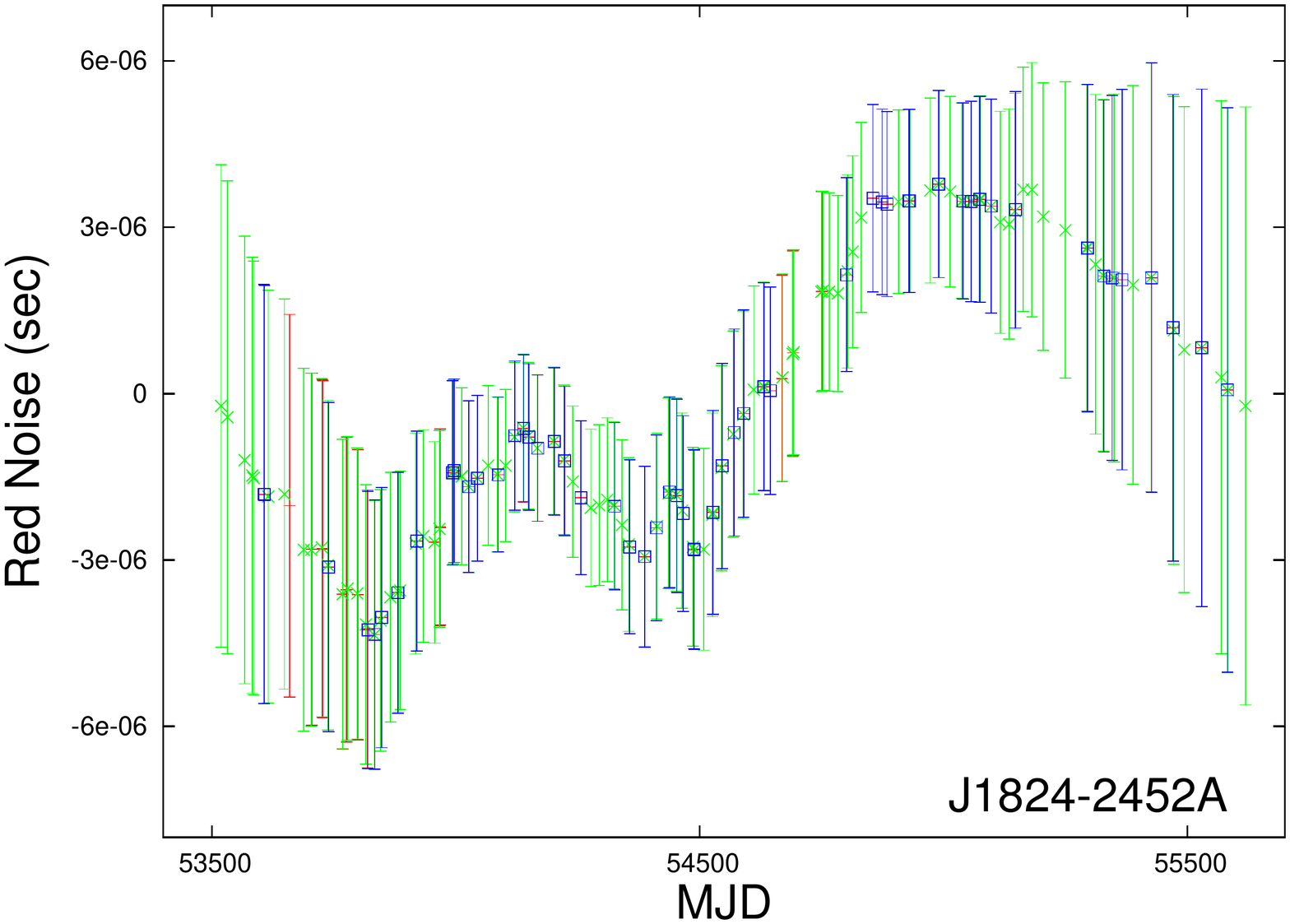} \\
\hspace{-1cm}
\includegraphics[trim = 60 50 30 65, clip,width=80mm]{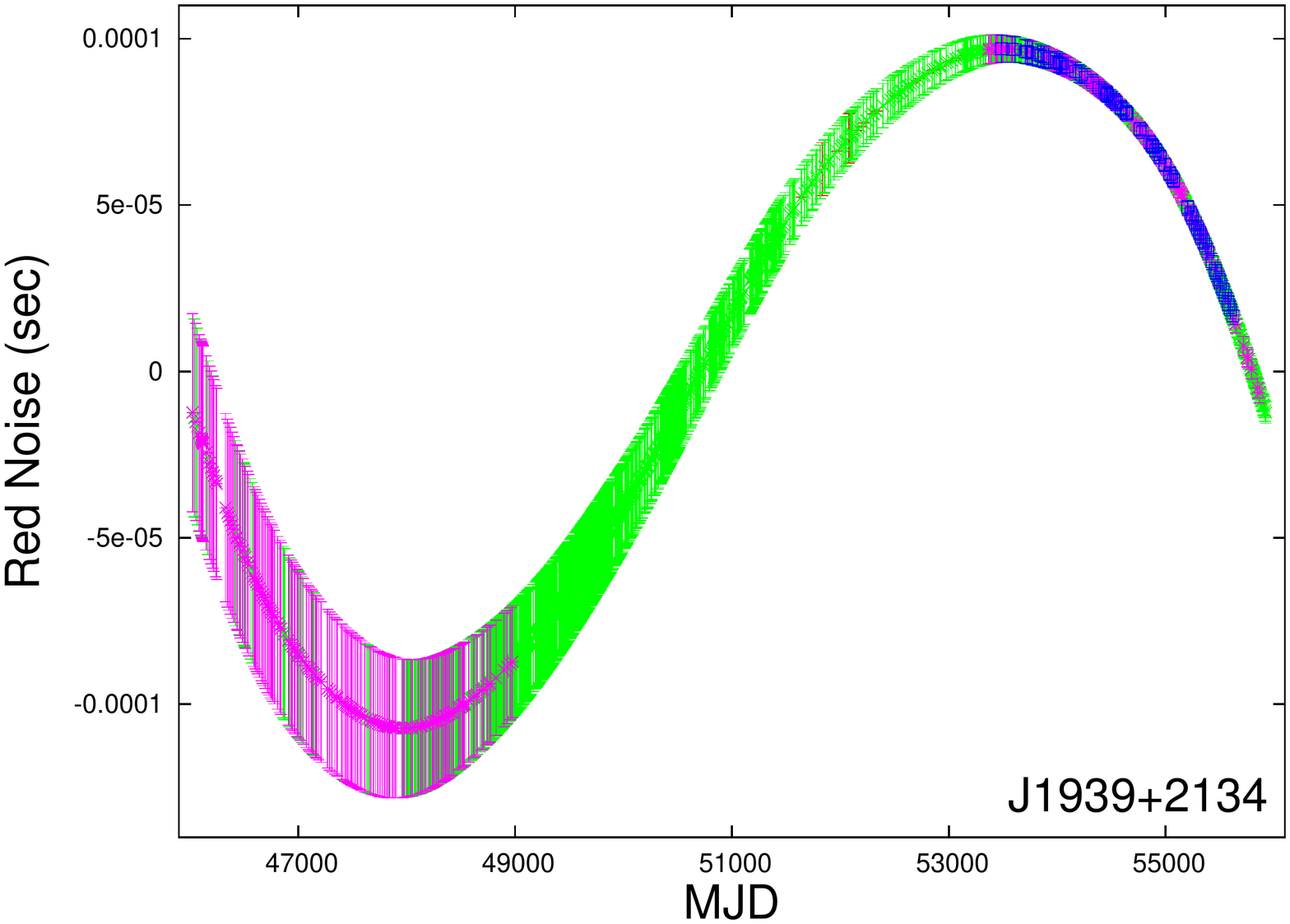}&
\hspace{0cm}
\includegraphics[trim = 60 50 30 65, clip,width=80mm]{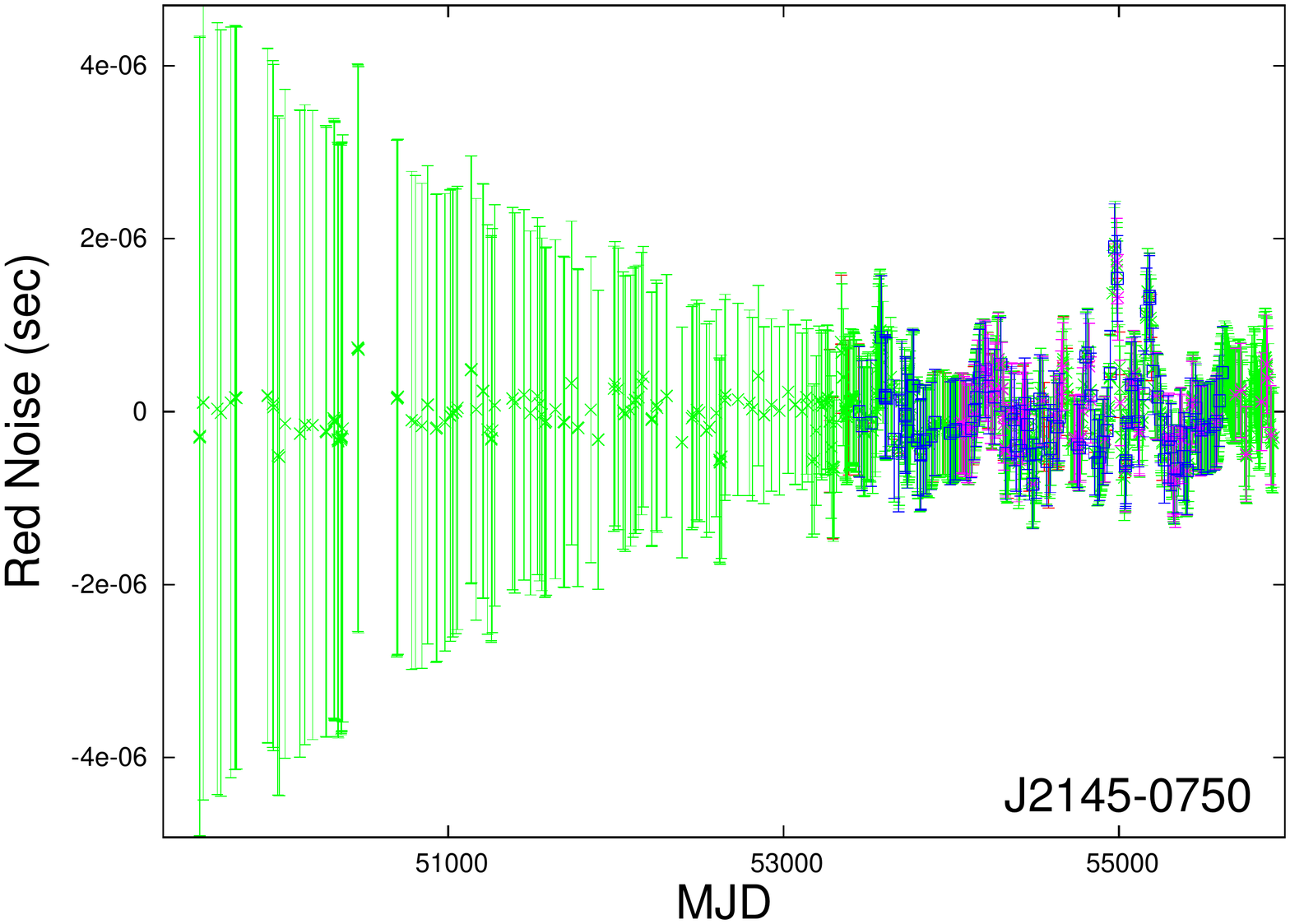} \\
\end{array}$
\end{center}
\caption{Maximum-likelihood signal realisations with 1-$\sigma$ uncertainties for the power-law spin-noise models for PSRs J0613$-$0200, J0621+1002,  J1012+5307, J1024$-$0719, J1713+0747, J1824$-$2452A, J1939+2134, and J2145$-$0750.  Colours represent the ToA observation frequency: $>$ 3~GHz (blue), 2-3~GHz (magenta), 1-2~GHz (green) and $<$ 1~GHz (red).}
\label{Fig:SpinNoiseModels}
\end{figure*}

We find that for eight pulsars in the IPTA data set the data support spin-noise processes that can be clearly distinguished from DM noise processes in the stochastic model. We list the mean log amplitudes, spectral exponents, and the total integrated power for these models in Table \ref{Table:NoiseParams}, along with their 1-$\sigma$ confidence intervals. In Fig. \ref{Fig:SpinNoiseModels} we show the maximum likelihood signal realisations for the spin noise in each of the eight pulsars: PSRs J0613$-$0200, J0621+1002,  J1012+5307, J1024$-$0719, J1713+0747, J1824$-$2452A, J1939+2134, and J2145$-$0750.  These cover a broad range of spectral exponents, from $0.6\pm0.2$ for  PSR J2145$-$0750 up to  $6.0\pm0.5$ for PSR  J1939+2134.

Given the large amplitude and steep spectral exponent observed in the timing noise for  PSR  J1939+2134 we also perform an evidence comparison for spin-noise models that include frequencies below 1/$T$, with $T$ the length of the data set.  We do this using the `log low frequency' model advocated in \cite{2015MNRAS.446.1170V}.   However we find that the posterior parameter estimates are consistent with those derived above and that the evidence does not support the addition of the  parameter describing the low-frequency cut off. This indicates that the spin-down quadratic included in the timing model is sufficient to model the low-frequency variations in the spin noise.

In the recent NANOGrav 9-yr data release \citep{2015arXiv150507540A} (henceforth A15) six pulsars were found to have timing noise with significance such that the evidence factor for models with timing noise was greater than three compared to a model without timing noise\footnote{A15 actually list ten pulsars as having possible spin-noise, but four of these are under our evidence threshold of three and hence are excluded here.}.  These were PSRs J0030+0451, J0613-0200, J1012+5307, J1643$-$1224, J1910+1256, J1939+2134.  We find that for all six pulsars, the IPTA data sets support time-correlated stochastic signals.  For PSRs J0030+0451 and J1910+1256 the IPTA data set lacks the multi-frequency coverage to distinguish between DM noise and spin noise, however for PSRs J0613-0200 and J1012+5307, we observed significant spin noise, with spectral exponents of $\gamma_{\mathrm{SN}} = 5.0\pm1.0$ , and $1.5\pm0.3$ respectively, roughly consistent with the values quoted in the NANOGrav data release of $\gamma=$ 2.9 and 1.7.  For PSR  J1643$-$1224 as discussed in Section \ref{Section:BandNoise} we found that there was significant band noise, inconsistent with either spin noise or DM variations, which if improperly modelled will manifest itself as spin noise.  This is the likely origin of the timing noise observed in the NANOGrav analysis.

\begin{figure}
\begin{center}$
\begin{array}{c}
\hspace{-1cm}
\includegraphics[width=100mm]{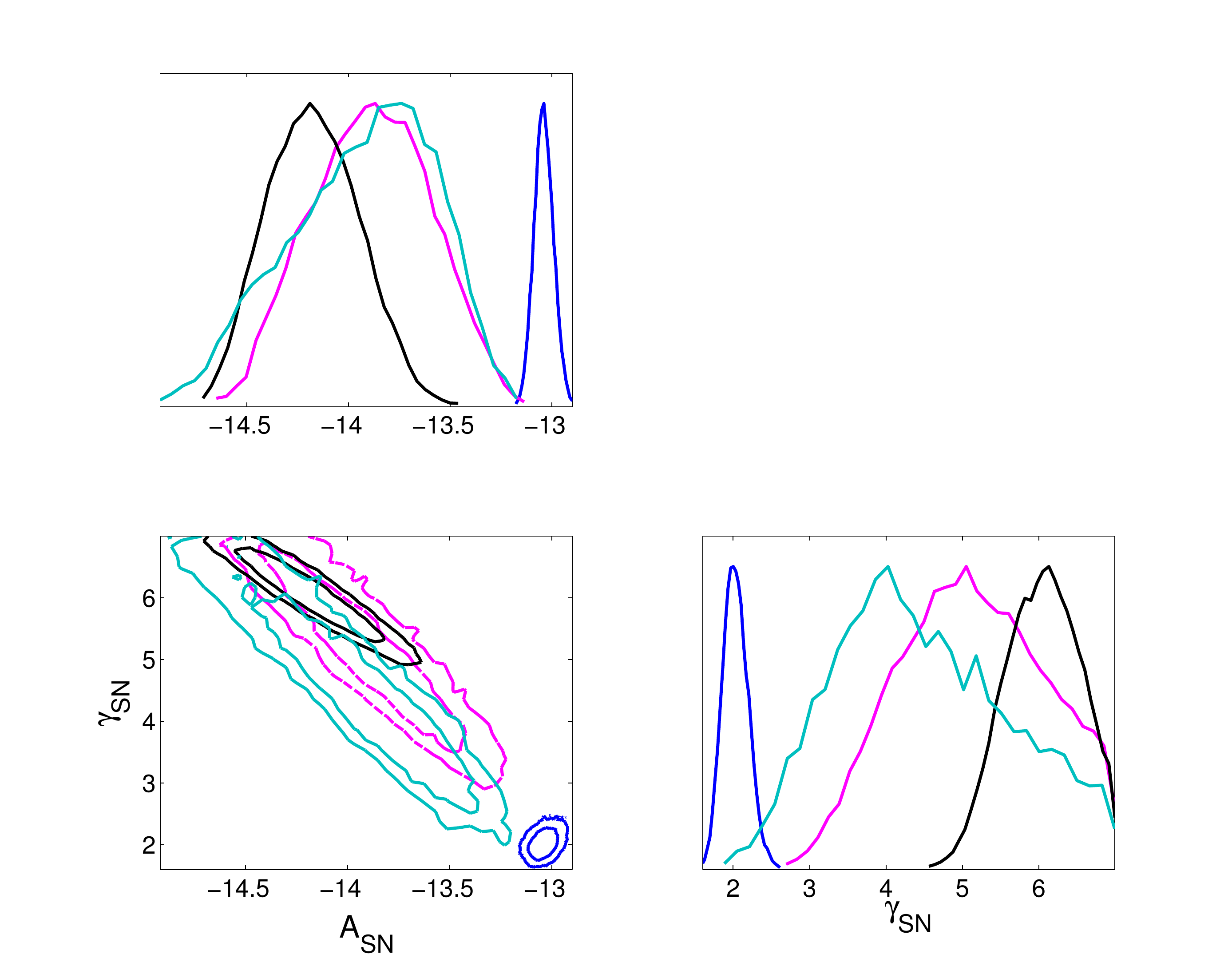} \\
\end{array}$
\end{center}
\caption{One- and two-dimensional marginalised posteriors for the log amplitude and spectral exponent of the spin-noise component of the PSR J1939+2134 for the optimal model from the IPTA data set (black lines), the 8-yr Kaspi et. al. (1994) subset of the IPTA data set (pink lines), the recent 9-yr NANOGrav data release \citep{2015arXiv150507540A} (blue lines), and a 9-yr PPTA data set consisting of 10~cm and 20~cm observations that extend over the same MJD range as the 9-yr NANOGrav data release (cyan lines). }
\label{Fig:J1939NoiseComps}
\end{figure}

We find that our analysis of the IPTA  PSR  J1939+2134 data set is incompatible with that presented in A15 where a timing-noise process with a spectral exponent of 2.4 was observed.    In  A15 the discrepancy with previously published analysis (e.g., \cite{1994ApJ...428..713K, 2010ApJ...725.1607S, 2014MNRAS.437.3004L} ) was attributed to either un-modelled IISM effects, or non-stationary timing noise.  To test the long-term stationarity of the timing noise in this pulsar we construct a 9-yr data set from 10~cm and 20~cm PPTA observations that covers the same MJD range as the data set presented in A15.  In Fig.  \ref{Fig:J1939NoiseComps} we compare one- and two-dimensional marginalised posteriors for the amplitude and spectral exponent of a spin-noise process from the optimal model for the IPTA data set (black lines), the 8-yr Kaspi et. al. (1994) subset of the IPTA data set (pink lines), our analysis of the data set presented in A15 (blue lines), and the 9-yr PPTA data set (light green lines).

We find that the analyses of the IPTA data set, the 8-yr Kaspi data set, and the 9-yr PPTA data set are all consistent with each other, indicating steep-spectrum spin noise ($\gamma_{\mathrm{SN}} = 6.0 \pm 0.5$) in the pulsar.  Our analysis of the data set presented in A15, where we include power-law spin noise, DM variations, and EFAC, EQUAD and ECORR parameters for each system is consistent with the analysis presented in A15, with a shallow spin-noise spectrum with $\gamma_{\mathrm{SN}} = 2.0 \pm 0.2$.  However this is inconsistent with the other PSR J1939+2134 data sets analysed.  We note that the IPTA data set for PSR J1939+2134 contains no contribution from NANOGrav, and so can be considered a completely independent data set from that presented in A15.  While these results suggest that the long-term behaviour of the spin noise in this pulsar is stationary over the observed time-span, it does not rule out either band noise due to un-modelled IISM effects, or system noise in the A15 data set, either of which could result in a flatter spectrum.

While A15 do not observe timing noise in their PSR J1713+0747 data set, we find our analysis is consistent with that presented in \cite{2015ApJ...809...41Z}, in which a timing noise process with a spectral exponent of $\gamma = 3.6 \pm1.4$ was observed, compared to our value $3.1\pm0.6$.  A15 also did not observe timing noise for PSR J1024$-$0719.  For these two pulsars our results show that the spin noise processes present have steep spectra.  The IPTA data sets are considerably longer (21- and 16-yr for PSRs J1713+0747 and J1024$-$0719, respectively) than the 9-yr data sets analysed in A15, implying that the shorter A15 data set is simply not yet sensitive to these processes.

Finally, in \cite{2015arXiv151009194C} an analysis of timing noise in an extended EPTA data set compared to that included in the first IPTA data release is presented.  We find that for those pulsars that have no detectable system noise in the EPTA data we obtain consistent parameter estimates for the properties of the spin noise. In addition, several pulsars that we identify as supporting system noise in the Nan{\c c}ay 1400~MHz data (likely due to polarisation calibration errors, see Section \ref{Section:SystemNoise}), such as PSRs J1022+1001, J1600$-$3053, and J1744$-$1134 are found to have significant timing noise in the extended EPTA data set.

\section{Conclusions}
\label{Section:Conclusions}

In this paper we have presented an analysis of the stochastic timing properties of the 49 pulsars included in first IPTA data release.  We performed model selection using the Bayesian evidence to determine the optimal model for the time-correlated signals present in each pulsar.  In addition to power-law spin noise and DM noise, these models could include system noise, present in a single observing system, and band noise present in all observing systems within some frequency band.

In total we found that for 19 pulsars the data support no time-correlated timing noise components, of which notably PSRs J1640+2224, J2124$-$3358, and J2129$-$5721 all have time spans of greater than 15 years and $\sigma_\mathrm{w}$ of less than 3~$\mu$s.

We find for 17 pulsars the data support power-law DM noise,  for eight pulsars the data support a model for intrinsic spin noise, for ten pulsars the data support system noise and finally for four pulsars the data support a model with additional band noise.

We showed that the improved frequency coverage, and the wealth of overlapping data from different telescopes analysed using different data reduction pipelines in the IPTA data set enables us to separate out system- and band-dependent effects with much greater efficacy than just using the individual PTA data sets.  Additionally, we showed that failing to model these effects appropriately can dramatically alter the interpretation of the signals observed in the residuals, in the most extreme cases revealing a significant detection of spin-noise, as a purely systemic effect.

For example, for PSR J1643$-$1224, we showed that the data set has, in addition to DM variations, further frequency-dependent noise that is coherent between different frequency bands observed by different PTAs, but does not scale as either $\nu_\mathrm{o}^0$ or $\nu_\mathrm{o}^2$ as would be expected for spin noise or DM variations respectively.  This allows us to interpret this timing noise as likely being the result of chromatic IISM effects, such as refraction and scattering, as opposed to being due to spin noise intrinsic to the pulsar.

One of the primary goals of the IPTA is to detect GWs.  The signal induced by a GW background will be highly correlated with the intrinsic timing noise present in each pulsar in the data set. Therefore the strength and properties of this spin noise will certainly affect the timeline for detection of GWs using a PTA.  We showed in the context of the PSR J0437$-$4715 dataset, that by more optimally modelling the different components of the stochastic signals present in the data set, the sensitivity to a GWB could be improved by $\sim 60\%$ compared to a model that includes only DM variations and spin noise.  This clearly demonstrates the importance of performing a comprehensive analysis  of the combined data sets - such as that presented here - in order to best exploit the potential of pulsar timing arrays to detect GWs.

It is clear that it will be critical for future GW-detection efforts to ensure that multiple telescopes continue to observe the same pulsars, at the same frequencies, in order to robustly identify system-dependent noise.  Thus, even when future telescopes such as the Square Kilometre Array come online, it will be necessary to continue observing with as many large telescopes as possible, in order to characterise systemic effects in the data sets from these new instruments.

It has also become clear that band-dependent effects can have a significant impact on the sensitivity attainable by a PTA at the level required to detect GWs. At low radio frequencies, variable interstellar scattering can result in non-$\nu^{-2}$ delays that induce erroneous DM corrections \citep{2015arXiv150308491C}, but other band-dependent effects are also possible. As demonstrated by the most recent PPTA GWB limit of $1\times 10^{-15}$, the most sensitive limit to date, which was derived using only 10-cm ($\sim 3$~GHz) data \citep{2015Sci...349.1522S}, such effects can be mitigated by observing at as high a frequency as possible. However, the problem is complex since different pulsars have different amounts of scattering, different flux densities and different radio spectral indices. For some pulsars with limited scatter-broadening, low-frequency observations using e.g., LOFAR \citep{2011A&A...530A..80S, 2015arXiv150802948K}, will help to identify band-dependent effects. Such observations, together with those obtained using the ultra-wideband receivers that are currently installed or under construction at several observatories, will greatly assist in achieving optimal sensitivity for GW detection by PTAs.

\bibliographystyle{mn2e}
\bibliography{references}

\section{Author Affiliations}

$^{1}$ Astrophysics Group, Cavendish Laboratory, JJ Thomson Avenue, Cambridge, CB3 0HE, UK \\
$^{2}$ CSIRO Astronomy and Space Science, Australia Telescope National Facility, Box 76, Epping NSW 1710 \\
$^{3}$ International Centre for Radio Astronomy Research, Curtin University, Bentley, WA 6102, Australia \\
$^{4}$ ECE Department, University of California at San Diego, La Jolla, CA, 92093-0407, USA  \\
$^{5}$ Fakult\"{a}t f\"{u}r Physik, Universit\"{a}t Bielefeld, Postfach 100131, 33501 Bielefeld, Germany \\
$^{6}$ Max-Planck-Institut f{\"u}r Radioastronomie, Auf dem H{\"u}gel 69, D-53121 Bonn, Germany \\
$^{7}$ Jet Propulsion Laboratory, California Institute of Technology, Pasadena, California 91109, USA \\
$^{8}$ Center for Research and Exploration in Space Science and Technology/USRA and X-Ray Astrophysics Laboratory, \\
NASA Goddard Space Flight Center, Code 662, Greenbelt, MD 20771, USA \\
$^{9}$ MPI for Gravitational Physics (Albert Einstein Institute), Golm-Potsdam 14476, Germany \\
$^{10}$ ASTRON, the Netherlands Institute for Radio Astronomy, Postbus 2, 7900 AA Dwingeloo, the Netherlands \\
$^{11}$ INAF - Osservatorio Astronomico di Cagliari, via della Scienza 5, 09047 Selargius (CA), Italy \\
$^{12}$ National Radio Astronomy Observatory, PO Box O Socorro NM 87801, USA \\
$^{13}$ Cornell Center for Astrophysics and Planetary Science, Cornell University, Ithaca, NY 14853, USA \\
$^{14}$ Laboratoire de Physique et Chimie de l'Environnement et de l'Espace LPC2E CNRS-Universit{\'e} d'Orl{\'e}ans,\\
F-45071 Orl{\'e}ans, France \\
$^{15}$ Station de radioastronomie de Nan{\c c}ay, Observatoire de Paris, CNRS/INSU F-18330 Nan{\c c}ay, France \\
$^{16}$ Department of Astronomy, School of Physics, Peking University, Beijing, 100871, China  \\
$^{17}$ Department of Physics, Hillsdale College, 33 E. College Street, Hillsdale, Michigan 49242, USA \\
$^{18}$ McGill University, Department of Physics, Rutherford Physics Building, 3600 University Street, Montreal, QC,\\
H3A 2T8, Canada \\
$^{19}$ Department of Physics and Astronomy, University of British Columbia, 6224 Agricultural Road, Vancouver, BC V6T 1Z1 Canada \\
$^{20}$ School of Mathematics, University of Edinburgh, King's Buildings, Edinburgh EH9 3JZ, UK \\
$^{21}$ Vancouver Coastal Health, Department of Nuclear Medicine, 899 W 12th Ave, Vancouver, BC, V5Z 1M9, Canada \\
$^{22}$ Anton Pannekoek Institute for Astronomy, University of Amsterdam, Science Park 904, 1098 XH Amsterdam, The\\
Netherlands \\
$^{23}$ NCSA, University of Illinois at Urbana-Champaign, Urbana, Illinois 61801, USA \\
$^{24}$ Department of Physics, Columbia University, New York, NY 10027, USA \\
$^{25}$ Jodrell Bank Centre for Astrophysics, University of Manchester, Manchester, M13 9PL, United Kingdom \\
$^{26}$ Monash Centre for Astrophysics (MoCA), School of Physics and Astronomy, Monash University, Victoria 3800, Australia  \\
$^{27}$ Kavli institute for astronomy and astrophysics, Peking University, Beijing 100871, P.R.China \\
$^{28}$ Department of Physics and Astronomy, West Virginia University, Morgantown, WV 26506, USA \\
$^{29}$ National Radio Astronomy Observatory, P.O. Box 2, Green Bank, WV, 24944, USA \\
$^{30}$ National Radio Astronomy Observatory, 520 Edgemont Rd., Charlottesville, VA 22903, USA \\
$^{31}$ TAPIR, California Institute of Technology, MC 350-17, Pasadena, CA 91125 USA \\
$^{32}$ Physics Department, Lafayette College, Easton, PA 18042 USA \\
$^{33}$ University of Virginia, Department of Astronomy, P.O. Box 400325 Charlottesville, VA 22904-4325, USA \\
$^{34}$ Universit\'e Paris-Diderot-Paris7 APC - UFR de Physique, Batiment Condorcet, 10 rue Alice Domont et L\'eonie\\
Duquet 75205 PARIS CEDEX 13, France \\
$^{35}$ Centre for Astrophysics \& Supercomputing, Swinburne University of Technology, PO Box 218, Hawthorn VIC 3122,\\
Australia \\
$^{36}$ School of Physics and Astronomy, University of Birmingham, Edgbaston, Birmingham, B15 2TT, United Kingdom \\
$^{37}$ Center for Gravitation, Cosmology and Astrophysics, Department of Physics, University of Wisconsin-Milwaukee, P.O. Box 413, Milwaukee, WI 53201, USA \\
$^{38}$ Physics and Astronomy Dept., Oberlin College, Oberlin OH 44074, USA \\
$^{39}$ Department of Physics and Astronomy, University of New Mexico, Albuquerque, NM 87131, USA	 \\
$^{40}$ Laboratoire Univers et Th[\'e]ories LUTh, Observatoire de Paris, CNRS/INSU, Universit{\'e} Paris Diderot, 5\\
place Jules Janssen, 92190 Meudon, France  \\
$^{41}$ Xinjiang Astronomical Observatory, Chinese Academy of Science, 150 Science 1-Street, Urumqi, Xinjiang 830011,\\
China \\
$^{42}$ School of Physics, Huazhong University of Science and Technology, Wuhan, Hubei Province 430074, China \\
$^{43}$ School of Physical Science and Technology, Southwest University, Chongqing 400715, China \\
$^{44}$ School of Physics, University of Western Australia, Crawley WA 6009, Australia \\

\section{Acknowledgements}

The NANOGrav project receives support from National Science Foundation (NSF) PIRE program award number 0968296 and NSF Physics Frontier Center award number 1430284.
The National Radio Astronomy Observatory is a facility of the NSF operated under cooperative agreement by Associated Universities, Inc.
The Arecibo Observatory is operated by SRI International under a cooperative agreement with the NSF (AST-1100968), and in alliance with Ana G. M\'{e}ndez-Universidad Metropolitana, and the Universities Space Research Association.
The Westerbork Synthesis Radio Telescope is operated by the Netherlands Institute for Radio Astronomy (ASTRON) with support from The Netherlands Foundation for Scientific Research NWO.
The 100-m Effelsberg Radio Telescope is operated by the Max-Planck-Institut fur Radioastronomie at Effelsberg. Some of the work reported in this paper was supported by the ERC Advanced Grant `LEAP', Grant Agreement Number 227947 (PI M. Kramer).
Pulsar research at the Jodrell Bank Centre for Astrophysics is supported by a consolidated grant from STFC.
The Parkes radio telescope is part of the Australia Telescope National Facility which is funded by the Commonwealth of Australia for operation as a National Facility managed by the Commonwealth Scientific and Industrial Research Organisation.
LL was supported by a Junior Research Fellowship at Trinity Hall College, Cambridge University.
CGB acknowledges support from the European Research Council under the European Union's Seventh Framework Programme (FP/2007-2013) / ERC Grant Agreement nr. 337062 (DRAGNET; PI Jason Hessels)
NDRB is supported by a Curtin Research Fellowship.
RNC acknowledges the support of the International Max Planck Research School Bonn/Cologne and the Bonn-Cologne Graduate School.
JG's work is supported by the Royal Society.	
MEG was partly funded by an NSERC PDF award.
JAE acknowledges support by NASA through Einstein Fellowship grant PF4-150120.
JWTH acknowledges funding from an NWO Vidi fellowship and ERC Starting Grant `DRAGNET' (337062).
GH is supported by an ARC Future Fellowship grant.
RK acknowledges the support of the ERC Advanced Grant “LEAP” (Number 227947, PI M. Kramer).
PDL is supported by the Australian Research Council Discovery Project DP140102578.
PL acknowledges the support of IMPRS Bonn/Cologne	
KJL gratefully acknowledges support from National Basic Research Program of China, 973 Program, 2015CB857101 and NSFC 11373011.
KL acknowledges the support of the ERC Advanced Grant “LEAP” (Number 227947, PI M. Kramer). KL acknowledges the financial support by the European Research Council for the ERC Synergy Grant BlackHoleCam under contract no. 610058.
CMFM was supported by a Marie Curie International Outgoing Fellowship within the European Union Seventh Framework Programme.
SO is supported by the Alexander von Humboldt Foundation.
PAR is supported by the Australian Research Council Discovery Project DP140102578.
SAS acknowledges support from a NWO Vidi fellowship (PI: J.W.T. Hessels).
AS is supported by a University Research Fellowship of the Royal Society.
RMS acknowledges travel support through a John Philip early career research award from CSIRO.
Pulsar research at UBC is supported by an NSERC Discovery Grant and Discovery Accelerator Supplement and by the Canadian Institute for Advanced Research.
SRT is supported by an appointment to the NASA Postdoctoral Program at the Jet Propulsion Laboratory, administered by Oak Ridge Associated Universities through a contract with NASA.
MV acknowledges support from the JPL RTD program. Portions of this research were carried out at the Jet Propulsion Laboratory, California Institute of Technology, under a contract with the National Aeronautics and Space Administration.
RvH acknowledges support by NASA through Einstein Fellowship grant PF3-140116.
JBW is supported by West Light Foundation of CAS XBBS201322 and NSFC project No.11403086.	
YW was supported by the National Science Fundation of China (NSFC) award number 11503007.
XPY acknowledges support by NNSF of China (U1231120) and FRFCU (XDJK2015B012).

\end{document}